\documentclass[prd,aps,twocolumn,nofootinbib,superscriptaddress,eqsecnum,floatfix,preprintnumbers,amsmath,amssymb,
longbibliography]{revtex4-1}

\usepackage{graphicx}
\usepackage[dvipsnames]{xcolor}
\usepackage[utf8]{inputenc}
\usepackage{stmaryrd}
\usepackage{mathrsfs}
\usepackage{mathalfa}
\usepackage{accents}
\usepackage{enumitem}
\usepackage[normalem]{ulem}

\DeclareMathOperator\arccot{arccot}

\newcommand{\realn}{\mathbb{R}}

\newcommand{\OO}{\mathcal{O}}

\newcommand{\eps}{\varepsilon}
\newcommand{\ph}{\varphi}
\newcommand{\tht}{\vartheta}
\newcommand{\ts}[1]{{#1}}

\newcommand{\pa}{\partial}

\newcommand{\lap}{\triangle}

\newcommand{\abs}[1]{{\left|#1\right|}}

\newcommand{\be}{\begin{equation}}
\newcommand{\ee}{\end{equation}}
\newcommand{\bea}{\begin{eqnarray}}
\newcommand{\eea}{\end{eqnarray}}
\newcommand{\ba}{\begin{equation}\begin{aligned}}
\newcommand{\ea}{\end{aligned}\end{equation}}
\newcommand{\beg}{\begin{gather*}}
\newcommand{\eng}{\end{gather*}}
\newcommand{\hh}{,\hspace{0.5cm}}
\newcommand{\hhh}{,\hspace{0.2cm}}

\newcommand{\n}[1]{\label{#1}}

\newcommand{\Z}{{\mathcal{Z}}}
\newcommand{\Zo}{{\mathcal{Z}}}
\newcommand{\Zp}{{\mathcal{Z}_+}}
\newcommand{\Zm}{{\mathcal{Z}_-}}
\newcommand{\Zpm}{{\mathcal{Z}_\pm}}

\newcommand{\const}{\mbox{const}}
\newcommand{\oix}{\mathrm{o}}
\newcommand{\zm}{\mathrm{zm}}
\newcommand{\sh}{\mathrm{sh}}
\newcommand{\sht}{\mathrm{sh\,t}}
\newcommand{\shi}{\mathrm{sh\,i}}
\newcommand{\hm}{\mathrm{hom}}
\newcommand{\wh}{\mathrm{wh}}
\newcommand{\trn}{\mathrm{t}}
\newcommand{\ind}{\mathrm{i}}

\newcommand{\rwhspc}{\widetilde{R}}
\newcommand{\pot}{U}
\newcommand{\delU}{{\Delta\pot}}
\newcommand{\hmf}{{w_\oix}}
\newcommand{\mzm}{{m_\infty}}

\begin{document}

\title{Ring wormholes and time machines}

\author{Valeri P. Frolov}
\email{vfrolov@ualberta.ca}
\affiliation{Theoretical Physics Institute, University of Alberta, Edmonton, Alberta, Canada T6G 2E1}
\author{Pavel Krtou\v{s}}
\email{Pavel.Krtous@utf.mff.cuni.cz}
\affiliation{Institute of Theoretical Physics,
Faculty of Mathematics and Physics, Charles University,
V~Hole\v{s}ovi\v{c}k\'ach 2, Prague, Czech Republic}
\author{Andrei Zelnikov}
\email{zelnikov@ualberta.ca}
\affiliation{Theoretical Physics Institute, University of Alberta, Edmonton, Alberta, Canada T6G 2E1}


\begin{abstract}
In the present paper we discuss properties of a model of a ring wormhole, recently proposed by Gibbons and Volkov \cite{Gibbons:2017djb,Gibbons:2017jzk,Gibbons:2016bok}. Such a  wormhole connects two flat spacetimes which are glued through discs of the radius $a$ bounded by the string with negative angle deficit $-2\pi$. The presence of the string's matter violating null energy condition makes the wormhole static and  traversable. We study gravitational field of static sources in such a spacetime in the weak field approximation. In particular, we discuss how a field of an oblate thin massive shell surrounding one of the wormhole's mouth is modified by its presence. We also obtain a solution of a similar problem when both mouths of the wormhole are located in the same space. This approximate solution if found for the case when the distance $L$ between these mouths is much larger than the radius $a$ of the ring. We demonstrate that the corresponding locally static  gravitational field in such a multiply connected space is non-potential. As a result of this, the proper time gap for the clock's synchronization linearly grows with time and closed timelike curves are formed. This process inevitably transforms such a traversable ring wormhole into a time machine. We estimate the time scale of this process.
\medskip

\hfill {\scriptsize Alberta Thy 2-23}
\end{abstract}

\maketitle

\section{Introduction}
\label{sc:intro}

Even in the absence of matter the space topology in the general relativity can be very non-trivial. A well known example is a famous Einstein-Rosen bridge connecting two asymptotically flat spaces. Examples of solutions of the initial value problem describing wormholes in a three dimensional space were constructed by Misner \cite{Misner:1960zz} and Brill and Lindquist \cite{Brill:1963yv}.  Wormholes and their properties were described in detail by Wheeler in his famous book on the geometrodynamics \cite{Wheeler}. A generic property  of vacuum wormhole solutions of the Einstein equations is that as a result of their time evolution they shrink and form a singularity so fast that one can not use them to `travel in space'.  Gannon \cite{Gannon:1975} proved that any asymptotically flat spacetime with a nonsimply connected Cauchy surface has singular time evolution if it satisfies the weak energy condition (see also \cite{Lee:1976,Galloway:1983,Friedman:1993}).

A special class of wormholes, called traversable, attracted in the past a lot of attention. Their characteristic property is that a region with the non-trivial topology is located inside a compact spatial domain and the particles and light can penetrate through a topological handle  and return to the exterior region without meeting a singularity \cite{Morris:1988cz,Morris:1988tu,Visser_1989,Hochberg:1997wp}. As it was demonstrated by Thorne and collaborators \cite{Morris:1988tu,Friedman:1990xc} a relative motion of the traversable wormhole's   mouths can generate closed timelike curves, so that such a wormhole becomes a `time machine'. A similar effect of `time machine' creation from a traversable wormhole can be achieved when the wormhole's mouths are not moving, but one of them is surrounded by some mass distribution \cite{Frolov:1990si}. In the presence of closed timelike curves a problem of self-consistency of the standard physics becomes highly non-trivial \cite{Friedman:1990xc,Echeverria:1991nk,Carlini:1995st,Dolansky:2010nr}. For example, infinite amplification of zero-point vacuum fluctuation during the creation of the time machine may result in (formally) infinite growth of the value of the renormalized vacuum stress-energy tensor, that indicates that its backreaction would become important \cite{Kim:1991mc,Frolov:1991nv}. Hawking formulated this problem as a chronology protection conjecture \cite{Hawking:1991nk}. For general discussion of wormholes and a time machine problem see e.g. books \cite{Visser:1995cc,Frolov:1998wf,FrolovZelnikov:2011}.

The simplest model of a traversable wormhole connecting two asymptotically flat spaces has a spherically-symmetric geometry \cite{Morris:1988tu,Visser_1989,Hochberg:1997wp}. For such a wormhole there exists a two-sphere of the minimal area in the wormhole throat and the null-energy condition should be violated in its vicinity (see e.g. \cite{Visser_1989}). Recently, an interesting model of a traversable wormhole was proposed by Gibbons and Volkov \cite{Gibbons:2017djb,Gibbons:2017jzk,Gibbons:2016bok}.
In this model two flat spacetimes are connected through a disk such that its boundary is a circular ring
where there exists a conical curvature singularity with the angle deficit $-2\pi$. This model of a traversable wormhole is a particular implementation of a loop-based wormhole discussed by Visser \cite{Visser:1995cc}.
One can identify this ring with a cosmic string with the corresponding negative energy distribution. Namely the presence of this matter violates the null energy condition and makes it possible an existence of  a static traversable wormhole. The geometry of the spacetime everywhere outside the ring (including the disk where two flat spacetime geometries are glued) is flat. The radius of the ring $a$ can be arbitrary large and a particle propagating through the disc from one flat space to the other meets neither negative energy density regions, nor a strong gravitational field. We refer to this type of geometries as to the ring wormhole.  A main purpose of this paper is to demonstrate how such a ring wormhole can be transformed into a  time machine.

For this goal we discuss a gravitational field created by static sources in the presence of a ring wormhole. We consider two cases: (i) The ring wormhole connects two different flat spaces, and (ii) such a wormhole connects two spatially separated regions of the same spacetime. In the latter case such a space is multiply connected. We assume that the gravitational field is weak and use the linearized gravity equations. We also assume that source of the gravitational field is an oblate massive thin shell spheroid surrounding a mouth of the ring wormhole confocal to the string ring. An exact solution for the gravitational field of such a shell for the ring wormhole connecting two flat space is obtained.

For the second problem (the wormhole in a single space) an approximate solution is found for the case when the distance $L$ between the mouths is much larger than the ring's size $a$. We demonstrate that in the second case the locally static gravitational field is non-potential and as a result of this closed timelike curves are created some time  after the massive source surrounding one of the mouths is `switched on'. We also estimate the corresponding time required for this as a function of mass of the shell, its size and the distance between the mouths.

The paper is organized as follows. In the next section we remind the reader properties of the ring wormholes.
In the section~\ref{sc:weakgrav} we explain the meaning of an approximation of a weak gravitational field in application to the wormhole spacetime. Gravitational field in the presence of a ring wormhole connecting two flat spaces is discussed in section~\ref{sc:twoasympWHspc}.
A similar problem for a string wormhole connecting two separated regions of a single space is considered in section~\ref{sc:oneasympWHspc}. Closed timelike paths formation, i.e., appearance of a time machine, in the space of a wormhole with a non-potential locally static gravitational field is discussed in section~\ref{sec:TMformation}. In particular,
it contains estimation of the time required for a closed timelike curves formation. In Section~\ref{S6} we discuss the obtained results and their possible consequences. Two appendices collect information concerning the static gravitational field in the weak field approximation and general properties of locally static non-potential gravitational fields.

In this paper we use sign conventions adopted in \cite{Misner:1973prb}.

\section{Geometry of a ring wormhole}
\label{sc:ringyWH}

\subsection{A ring wormhole connecting two asymptotically flat spaces}

\begin{figure}
    \centering
      \includegraphics[width=0.4\textwidth]{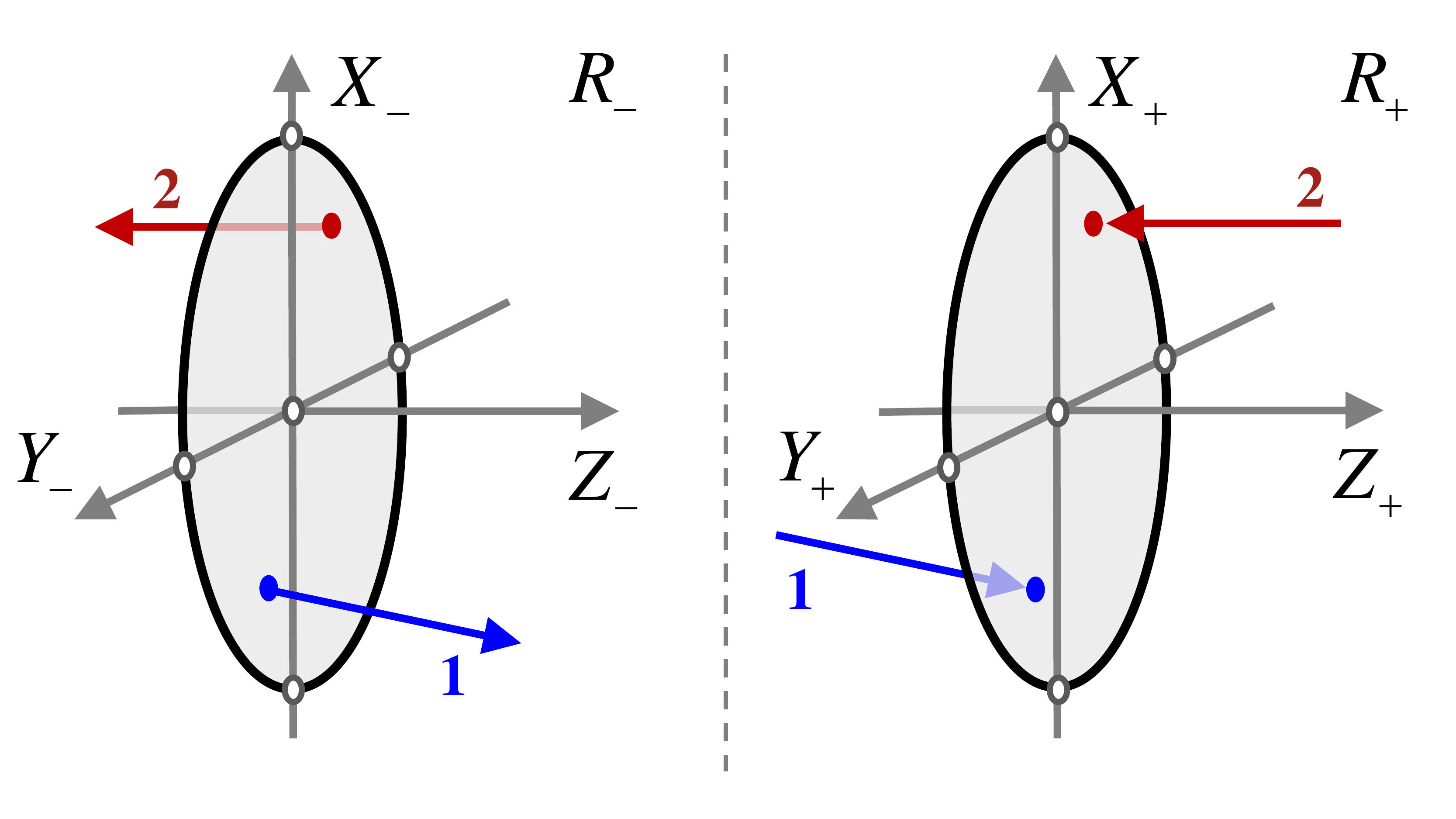} 
    \caption{\label{fig:WH3D}A ring wormhole connecting two flat spaces $R_+$ and~$R_-$.}
\end{figure}

Let us discuss first a spatial configuration of the wormhole spacetimes. We begin with a case of a ring wormhole connecting two flat spaces. Such a wormhole can be obtained as follows. Consider two copies of a three-dimensional flat space. Denote by $X^i=(X_{\pm},Y_{\pm},Z_{\pm})$ standard Cartesian coordinates in these spaces. Let  $D_{\pm}$  be  discs in $R_{\pm}$ of radius $a$ in the planes $Z_{\pm}=0$ with the center at the origin of these spaces (see  Fig.~\ref{fig:WH3D}).

To join the spaces $R_-$ and $R_+$ with a wormhole, we identify surfaces of the discs $D_-$ and $D_+$. This identification is done as follows. We denote by $D_{\pm}^>$ and $D_{\pm}^<$ the `right' and `left' faces of the discs. Namely, if the disc $D_{\pm}$ is reached by a point with the positive value of $Z_{\pm}$ we say that it belongs to  $D_{\pm}^>$. In the opposite case when a point approaches the disc with ${Z_{\pm}<0}$ we say that it belongs to $D_{\pm}^<$. One identifies $D_+^<$ with $D_-^>$ and $D_+^>$ with $D_-^<$, without any rotation around $z$-axis.

This means that a free particle 1 moving in space $R_+$, which meets the disc $D_+^<$, enters space $R_-$ at $D_-^>$ with identical coordinates $X_-=X_+$, $Y_-=Y_+$, $Z_-=Z_+=0$, and continues its motion with increasing $Z_-$ coordinate. Similarly if a particle 2 enters $D_+^>$, it appears at $D_-^<$ part of the disc $D_-$ as illustrated in Fig.~\ref{fig:WH3D}.

In what follows we denote $\rwhspc$ the space obtained by unifying two spaces $R_+$ and $R_-$ with identification described above. We call the discs $D_-$ and  $D_+$ in $R_-$ and $R_+$ , respectively, the wormhole mouths, and their identification $D$ the wormhole throat.

The radius $a$ of the disc has the dimension of length. It is convenient to use it as a natural scale and write the metric in a flat space in the form
\begin{gather}
dL^2 = \delta_{ij}\, d X^i d X^j = a^2 dl^2\;,\\
dl^2 = \delta_{ij}\,dx^{i}dx^{j} = dx^2+dy^2+dz^2\;.
\end{gather}
In what follows we shall use the dimensionless metric $dl^2$ and dimensionless coordinates $x^{j}=X^{j}/a$.
We denote by ${(z,\,\rho,\,\ph)}$ dimensionless cylindrical coordinates
\be\label{metcylcoor}
\rho^2=x^2+y^2\;,\quad x=\rho\cos\ph\;,\quad y=\rho\sin\ph\, ,
\ee
with the metric taking the form
\be
dl^2=d\rho^2+dz^2+\rho^2 d\ph^2\; .
\ee
The ring equations are
\be
\rho=1\hh z=0\, .
\ee

It will be very useful to introduce also the oblate spheroidal coordinates $(\chi,\tht,\ph)$ related to the cylindrical coordinates as follows:
\be \n{rrzz}
 z=\sinh\chi\cos\tht\;,\quad \rho=\cosh\chi\sin\tht\;.
\ee
A two-dimensional surface ${\chi=\const}$ is an oblate spheroid, confocal with the ring, with $\cosh\chi$ being its larger semi-axis and $\sinh\chi$ being the smaller one. A surface $\tht=\const$ is a hyperboloid confocal also with the ring. Let us note that the relations \eqref{rrzz} can be written in the following complex form
\be
  z+i \rho= \sinh(\chi+i\tht)\, .
\ee
The flat metric in the oblate spheroidal coordinates is
\be\label{mtrcsphroidal}
dl^2= (\sinh^2\!\chi{+}\cos^2\!\tht)(d\chi^2+d\tht^2) +\cosh^2\!\chi \sin^2\!\tht\, d\ph^2\, .
\ee

For a single flat space $R$, the oblate spheroidal coordinates have a two-fold degeneracy:  $(\chi,\tht,\ph)$ and $(-\chi,\pi-\tht,\ph)$, correspond to the same point. For this reason, it would be sufficient to consider the ranges of the oblate spheroidal coordinates $\chi\in(0,\infty)$, $\tht\in (0,\pi)$ and $\ph\in (0,2\pi)$.

In the wormhole space $\rwhspc$, we introduce separate copies of flat ${(x_\pm,\,y_\pm,\,z_\pm)}$ and cylindrical  ${(z_\pm,\,\rho_\pm,\,\ph_\pm)}$ coordinates in spaces $R_-$ and $R_+$. However, we take advantage of the degeneracy of the oblate spheroidal coordinates and define just single coordinates ${(\chi,\,\tht,\,\ph)}$ covering the whole wormhole space, where  $\chi\in(-\infty,\infty)$, simply assuming that $\chi>0$ in $R_+$ and $\chi<0$ in $R_-$.

\begin{figure}
    \centering
      \includegraphics[width=0.5\textwidth]{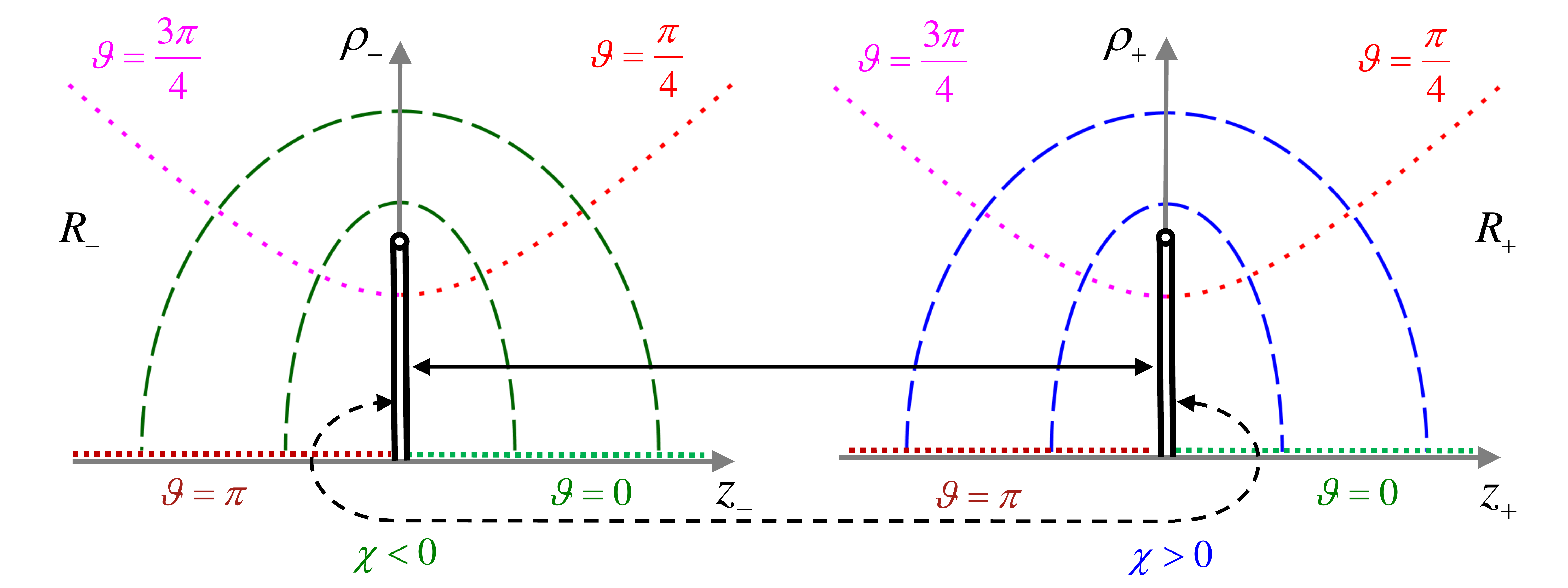} 
    \caption{\label{fig:WH2M2D}Ring wormhole in space $\rwhspc$ with two asymptotic regions $R_-$ and $R_+$. Each of the domains $R_\pm$ is covered by cylindrical coordinates ${\rho_\pm,\,\ph_\pm,\, z_\pm}$. Directions related to angular coordinate $\ph_\pm$ are not shown. They can be obtained by rotations around axes $\rho_\pm=0$. The whole space $\rwhspc$ is covered by one copy of the oblate spheroidal coordinates ${\chi\in(-\infty,\infty)}$, ${\tht\in(0,\pi)}$ and ${\ph\equiv\ph_\pm\in(-\pi,\pi)}$. Dashed lines corresponds to $\chi=\const$, dotted lines to $\tht=\const$. Double lines indicate the discs which are identified as indicated by arrows. They represent the mouths of the wormhole.}
\end{figure}

Fig.~\ref{fig:WH2M2D} shows space $\rwhspc$ in $(\chi,\tht)$ coordinates.
To specify a point, one needs to add an azimuthal angle $\ph$.
The coordinates $(\chi,\tht,\ph)$ cover the complete space and they are continuous at the discs representing the mouths of the ring wormhole. The left figure corresponds to $R_-$ domain where $\chi<0$, while the right one corresponds to $R_+$ where $\chi>0$.

Let us note that it is possible to generalize the described ring wormhole space to the case when there exist not only one but several discs in space $R_+$, each `connected' with its own version of flat space. We do not consider such multi-wormhole configurations in the present paper.

\subsection{A ring wormhole in a single space}

It is easy to use a simple procedure to construct a space that contains both mouths of a single wormhole. One just places two discs representing the wormhole's mouths into one flat space. Obviously, there is an ambiguity in the choice of positions and orientations of the discs. We consider the simplest case when the discs are orthogonal to a common axis and separated by a distance~$L$.

If we introduce global dimensionless flat and cylindrical coordinates ${(x,\,y,\,z)}$ and ${(z,\,\rho,\,\ph)}$, the centers of two discs $D_\pm$ of radius 1 are located along $z$-axis at $z=\pm\ell/2$, and the discs are orthogonal to the $z$-axis. Here, we denoted by $\ell=L/a$ the dimensionless version of the discs' distance.

\begin{figure}
    \centering
      \includegraphics[width=0.5\textwidth]{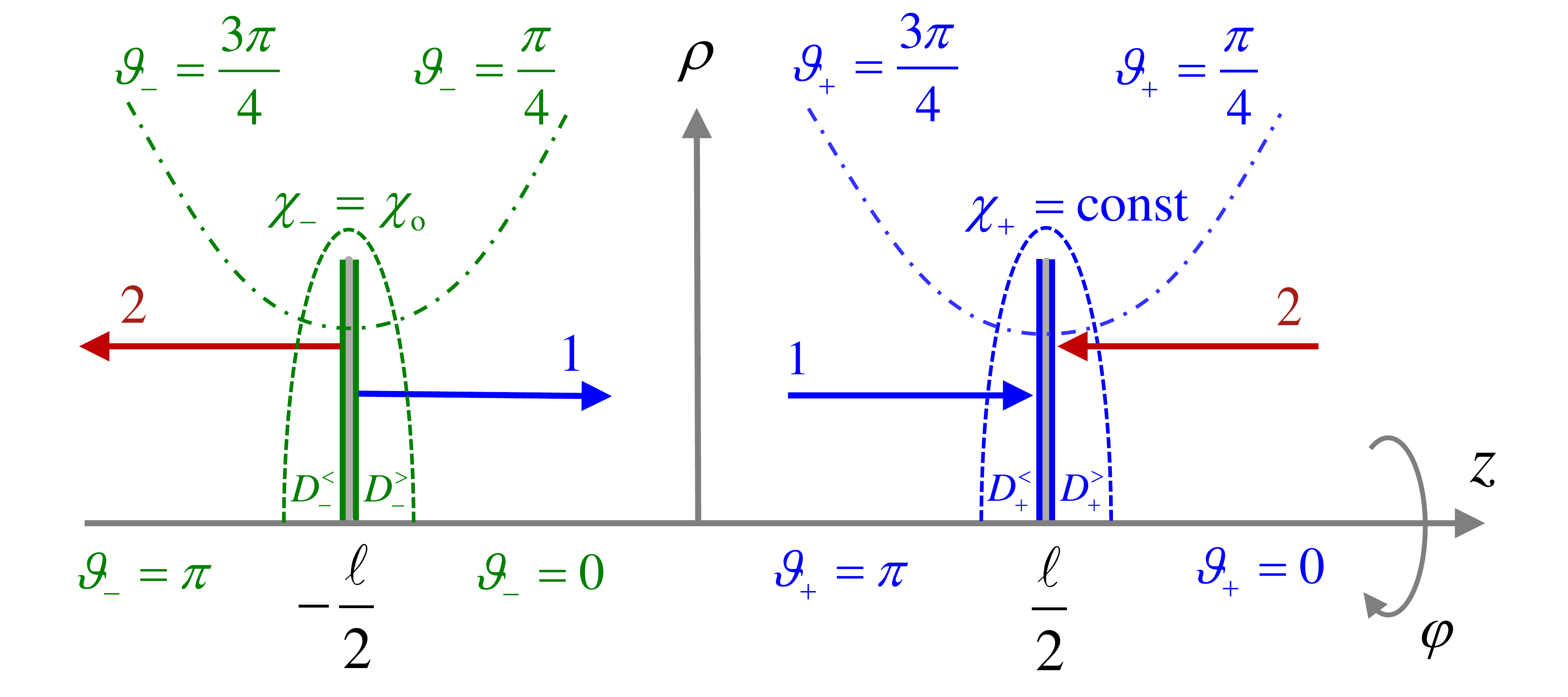}\\    
    \caption{\label{fig:WH1M2D}Ring wormhole space $R_\wh$ with one asymptotic region. We have global cylindrical coordinates ${\rho,\,\ph,\,z}$ and two copies of the oblate spheroidal coordinates ${\chi_\pm,\,\tht_\pm,\,\ph_\pm}$. The coordinate $\ph\equiv\ph_\pm$ is not shown. The discs $D_\pm$ representing the wormhole's mouths are indicated by double lines located at $z=\pm\ell/2$. The values of the coordinate lines indicated in the diagram corresponds to the domains $\bar{V}_\pm$ described in the text and Fig.~\ref{fig:domdef}. The coordinates in $\bar{V}_\pm$ are related to the opposite coordinates in the leaked domains $\hat{V}_\mp$ by $\chi_+=-\chi_-$ and $\tht_+=\pi-\tht_-$. Compare also Fig.~\ref{fig:whcoor}.
    }
\end{figure}

To form the wormhole, we identify  the discs $D_-$ and $D_+$ in a similar manner as we described above: Each of the discs has two faces which we denoted by $D_{\pm}^>$ and $D_{\pm}^<$.
One identifies $D_-^<$ with $D_+^>$ and $D_-^>$ with $D_+^<$, without any rotation around the $z$-axis. This identification is illustrated in Fig.~\ref{fig:WH1M2D}.

We call the resulting space~$R_\wh$. The two-faces discs $D_-$ and $D_+$ are called left and right mouths of the wormhole.
Identified, they form the throat of the wormhole. Strictly speaking, the standard coordinates ${(x,\,y,\,z)}$ and $(\rho,\ph,z)$, when understood as coordinate maps on $R_\wh$, are well defined only outside the throat.

Clearly, the wormhole spacetime $R_\wh$ has a different topology than the original empty flat space. Its first homology group is non-trivial, since we have non-contractible loops spanned through the wormhole.

In $R_\wh$, we introduce two sets of oblate spheroidal coordinates. One of them, $(\chi_+,\tht_+,\ph_+)$, is centered on the ring at $z=+\ell/2$, and the other one, $(\chi_-,\tht_-,\ph_-)$, is centered on the ring at $z=-\ell/2$.

For $\chi_\pm>0$, we restrict the coordinates $(\chi_+,\tht_+,\ph_+)$ to the domain where $z>-\tilde{\delta}$ and similarly, we restrict the coordinates  $(\chi_-,\tht_-,\ph_-)$  to the domain where $z<\tilde{\delta}$.

We assume that these coordinates are valid not only for positive values of $\chi_{\pm}$ but are extended to some negative value ${-\delta<\chi_\pm}$. This provide us with a `leakage' of these coordinates through wormhole mouths. Namely, coordinates $(\chi_+,\tht_+,\ph_+)$ cover not only the space outside discs $D_+$ but also some domain $-\delta<\chi_+$ in the vicinity of the disc $D_-$. Similarly, the coordinates $(\chi_-,\tht_-,\ph_-)$ cover exterior of the disc  $D_-$ and a domain $-\delta<\chi_-$  in the vicinity of~$D_+$.

\begin{figure}
    \centering
      \includegraphics[width=0.9\columnwidth]{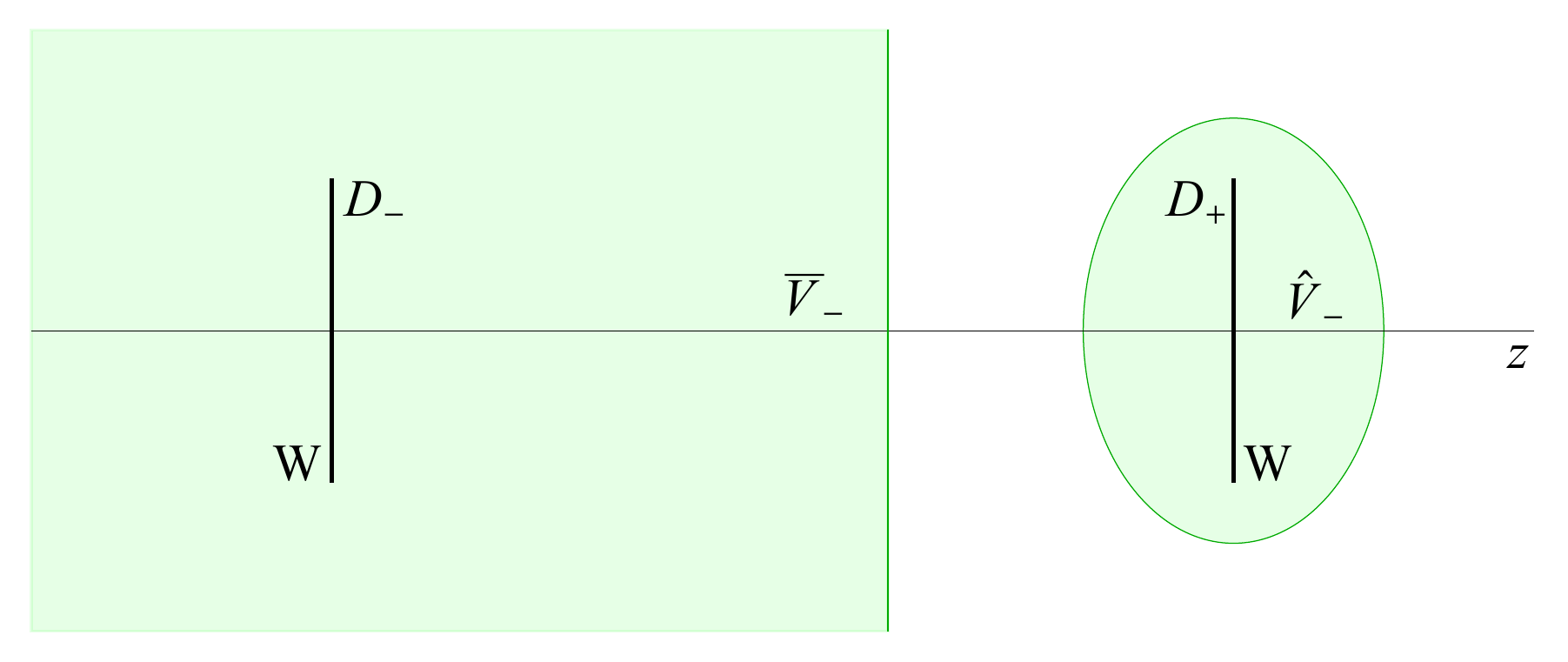}\\[-1ex]
      {\footnotesize Domain ${V_-=\bar{V}_-\cup\hat{V}_-}$.}\\[1ex]
      \includegraphics[width=0.9\columnwidth]{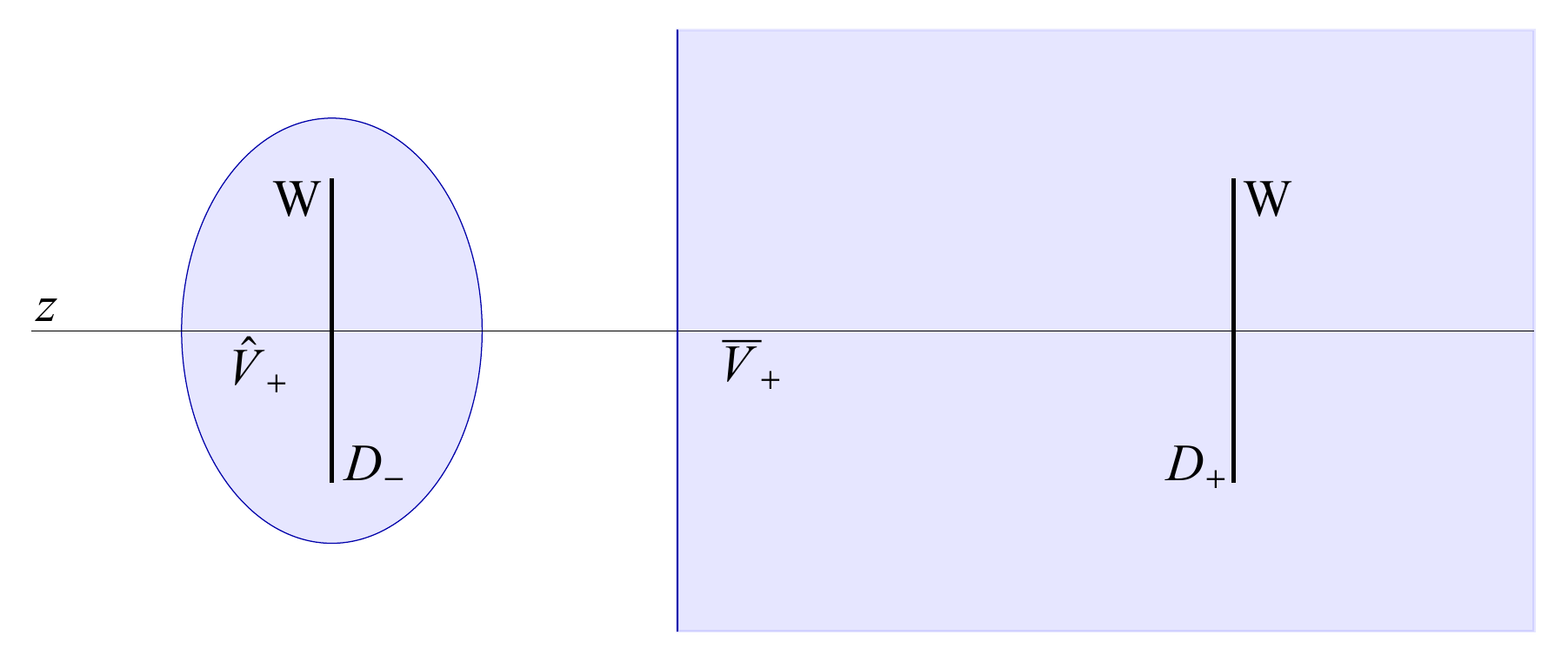}\\[-1ex]
      {\footnotesize Domain ${V_+=\bar{V}_+\cup\hat{V}_+}$.}\\[1ex]
      \includegraphics[width=0.9\columnwidth]{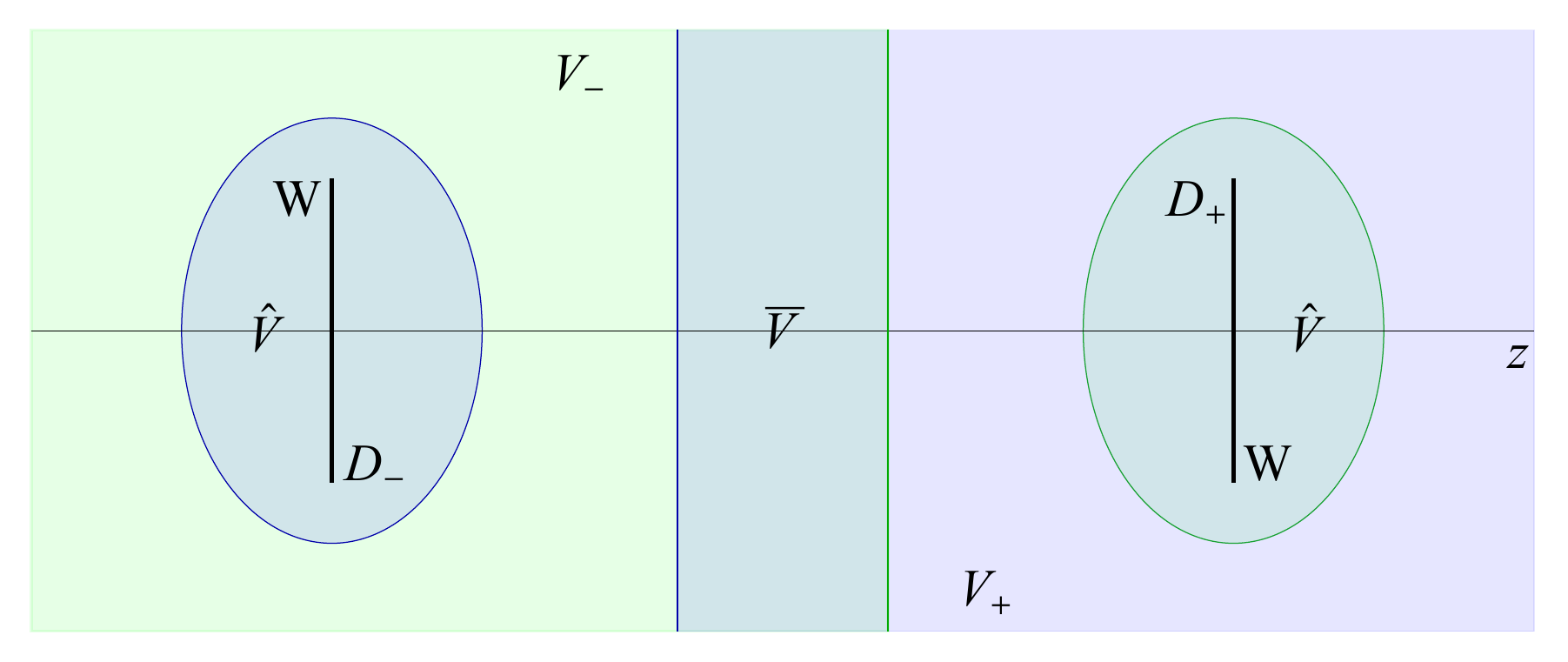}\\[-1ex]
      {\footnotesize Domains ${\bar{V}}$ and ${\hat{V}}$.}\\[1ex]
    \caption{\label{fig:domdef}Domains $V_-$ (top) and $V_+$ (middle) are covering together the whole space. They are composed by the `primary' parts $\bar{V}_\pm$ and `leaked' parts $\hat{V}_\pm$. The bottom diagram shows two intersections $\bar{V}$ and $\hat{V}$ of these two domains. Horizontal direction corresponds to the $z$ axis, vertical to $\rho$ direction. The mouths $D_\pm$ of the wormhole are represented by the thick vertical lines. The distance between mouths is~$\ell$.
    }
\end{figure}

We denote the corresponding domains of definitions of coordinates $(\chi_\pm,\tht_\pm,\ph_\pm)$ as $V_{\pm}$, respectively, see Fig.~\ref{fig:domdef}. These two domains intersect in a narrow strip $\bar{V}$ where $-\tilde{\delta}<z<\tilde{\delta}$, and in a spheroidal neighborhood $\hat{V}$ of the throat, given by $-\delta<\chi_\pm<\delta$, see Fig.~\ref{fig:domdef}. We assume $\delta$ and $\tilde{\delta}$ to be sufficiently small so that $\bar{V}$ and $\hat{V}$ do not overlap.

We denote the part of $V_+$ with $\chi_+>0$ as $\bar{V}_+$ and the `leaked' part of $V_+$, i.e., the part with $-\delta<\chi_+\le0$, as $\hat{V}_+$. We define $\bar{V}_-$ and $\hat{V}_-$ analogously. The domains are related
\begin{equation}\label{domrels}
\begin{gathered}
    V_- = \bar{V}_- \cup \hat{V}_-\;,\quad
    V_+ = \bar{V}_+ \cup \hat{V}_+\;,\\
    \bar{V} = \bar{V}_- \cap \bar{V}_+\;,\quad
    \hat{V} = \hat{V}_+ \cup \hat{V}_-\;.
\end{gathered}
\end{equation}

In domains $\bar{V}_+$ and $\bar{V}_-$, the oblate spheroidal coordinates are related to the global cylindrical coordinates as
\be \label{rrtt}
\begin{gathered}
z=\pm \frac\ell2+\sinh\chi_{\pm}\cos\tht_{\pm}\;,\\
\rho=\cosh\chi_{\pm}\sin\tht_{\pm}\;,\\
\chi_\pm>0\;,\quad\tht_\pm\in(0,\pi)\;.
\end{gathered}
\ee
These formulas also establish relations between coordinates $(\chi_+,\tht_+)$ and $(\chi_-,\tht_-)$ in $\bar{V}$,
\be\label{sphcoorpmrel}
\begin{gathered}
\cosh\chi_+\sin\tht_+=\cosh\chi_-\sin\tht_-\, ,\\
-\ell/2+\sinh \chi_-\cos\tht_-=\ell/2+\sinh \chi_+\cos\tht_+\, .
\end{gathered}
\ee

In the `leaked' domains $\hat{V}_-$ and $\hat{V}_+$, the oblate coordinates  $(\chi_+,\tht_+)$ and $(\chi_-,\tht_-)$ are related to the global cylindrical coordinates $(\rho,z)$ as
\be \label{rrttleak}
\begin{gathered}
z=\mp \frac\ell2+\sinh\chi_{\pm}\cos\tht_{\pm}\;,\\
\rho=\cosh\chi_{\pm}\sin\tht_{\pm}\;,\\
\chi_\pm<0\;,\quad\tht_\pm\in(0,\pi)\;.
\end{gathered}
\ee
As a consequence, we have in $\hat{V}$
\begin{equation}\label{sphcoorpmrel1}
  \chi_+=-\chi_-,\,\quad\tht_+=\pi-\tht_- \;.
\end{equation}

The relation of coordinates $\ph_\pm$ is trivial. We can consistently set $\ph=\ph_-=\ph_+$ on all overlaps of the domains of definition.

\begin{figure}
    \centering
      \includegraphics[width=\columnwidth]{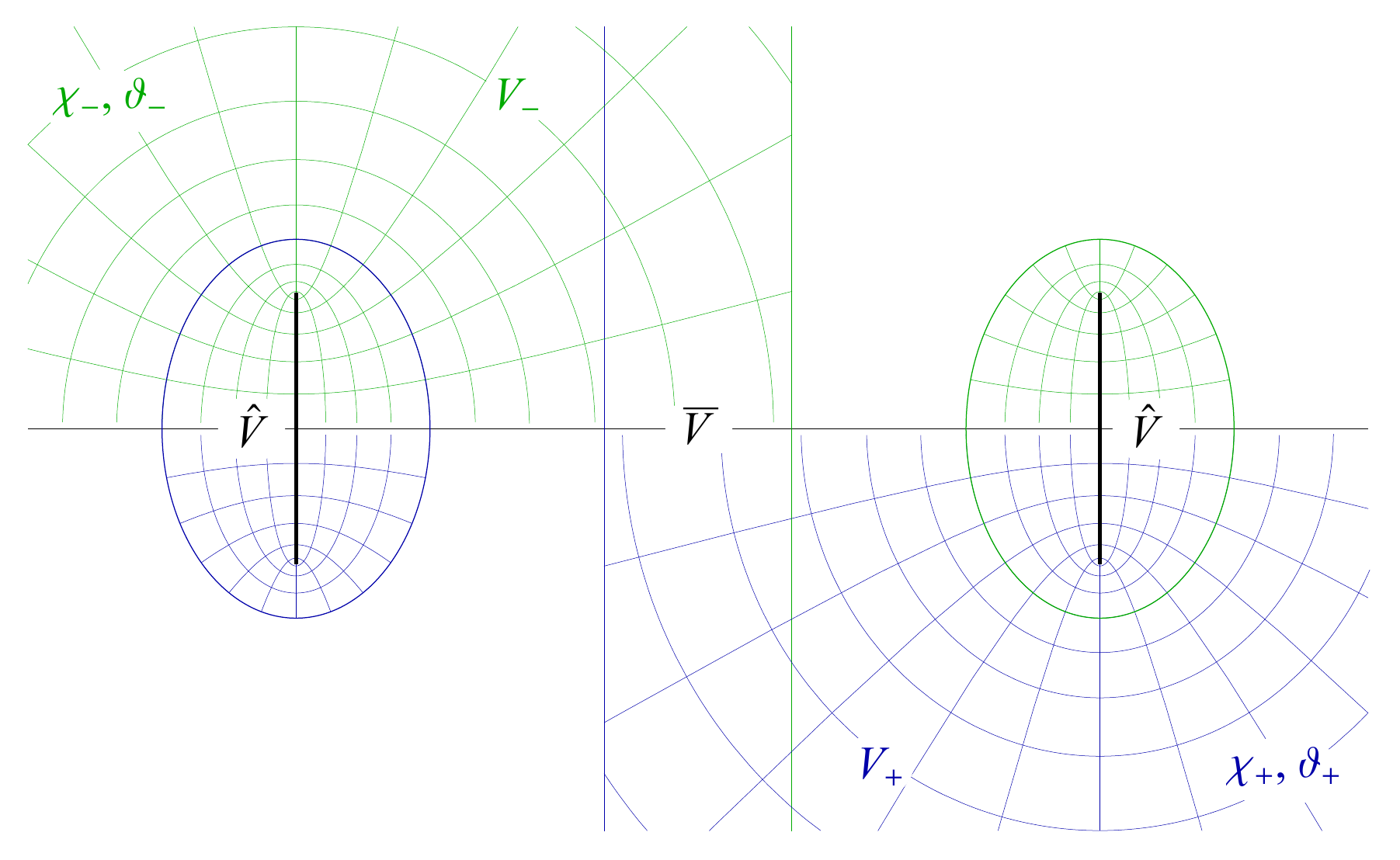}
    \caption{\label{fig:whcoor}Coordinate maps ${(\chi_-,\,\tht_-)}$ in the domain $V_-$ (top half) and ${(\chi_+,\,\tht_+)}$ in $V_+$ (bottom half).
    }
\end{figure}

\subsection{Static wormhole spacetimes}

Let us note that after fixing three-dimensional spatial geometry $R$, we have to fix also full spacetime geometry $M$. In this paper we will assume that the time direction is added to the given space in a trivial way, namely that the resulting spacetime will be locally static with time-slices locally equivalent to the chosen spatial geometry.

The simplest example is a globally static spacetime $M=R\times \realn$, with a global dimensionless coordinate $t$ independent of the spatial geometry, and with spacetime metric
\begin{equation}\label{staticst}
    ds^2 = - dt^2 + dl^2\;.
\end{equation}
We can generalize this case in several ways. First, we can admit a redshift factor
\begin{equation}\label{staticst}
    ds^2 = - e^{2U} dt^2 + dl^2\;,
\end{equation}
where $U$ is globally well-defined spatial function.

When the spatial manifold is topologically non-trivial, as the space $R_\wh$ discussed above, we can have more elaborate generalizations. First, we can identify the mouths of the wormhole at different times.

Namely, suppose we introduce time $t$ synchronized with clocks at infinity, defined everywhere except at the wormhole throat. It complements the global flat coordinates ${(x,\,y,\,z)}$ or cylindrical coordinates ${(\rho,\,z,\,\ph)}$.

Now, we do not identify the discs $D_\pm$ at the same values of time $t$, but we identify $D_-$ at $t=t_\oix$ with $D_+$ at $t=t_\oix+\Delta t$, with $\Delta t$ being a constant. We require that the spacetime geometry is still described by the metric
\begin{equation}\label{staticstDeltat}
    ds^2 = - dt^2 + dl^2\;.
\end{equation}
It is straightforward to see that the metric is continuous through the wormhole, since the constant shift $\Delta t$ disappears in the differential $dt$.

Consider a particle which enters the disc $D_-$ at time $t=t_\oix$. It appears from the disc $D_+$ at time $t=t_\oix+\Delta t$. We thus say that there exists a time gap $\Delta t$ associated with the wormhole. Since the proper time at both sides of the identified discs must be the same (the geometry is continuous through the wormhole), the time gap must remain the same. This means that if a second particle is sent through $D_-$ shortly after the first one at time $t=t_\oix+\delta t$ it will appear at $D_+$ at time $t=t_\oix+\delta t+\Delta t$.

However, if we put some matter around one of the mouths of the wormhole, then its gravitational field affects a proper time of static observers and the static spacetime metric needs further modification. In general, the time shift of coordinate $t$ will not remain constant. Our spacetime becomes only locally static as we will describe in the following sections.

\subsection{A nature of the ring}

Before going further let us make a following remark. Gibbons and Volkov \cite{Gibbons:2017djb,Gibbons:2017jzk,Gibbons:2016bok} demonstrated that in fact there exists a $\delta-$like singularity of the curvature at the location of the ring. The corresponding distribution of matter violates the null energy condition which makes it possible the very existence of a static traversable wormhole. This distribution of matter can be identified with an infinitely thin cosmic string (strut) with a negative angle deficit $-2\pi$. In order to exclude the $\delta-$like singularity at the ring one can smear the distribution of the matter of the corresponding cosmic string. This makes the spacetime smooth, but the equation of state of the corresponding matter is not very realistic. If $\lambda$ is the radius of the smeared string, we assume that the parameter $d=\lambda/a$ is small. This always can be achieved since the radius of the ring $a$ can be arbitrary large. In what follows we always assume that, if necessary, the corresponding smearing is done.

\section{Weak gravitational field}
\label{sc:weakgrav}

\subsection{Gravitational potential}

In the weak-field approximation, a solution of the Einstein equation for a static matter distribution can be written in the form
\be \n{MMUU}
 ds^2=-e^{2\pot} dt^2 +e^{-2\pot} dl^2\;.
\ee
where the  gravitational potential $\pot$  obeys the Poisson equation
\be\n{Poisson}
\lap \pot = 4\pi \mu\; .
\ee
Here $\mu$ is the mass density defined with respect to the flat metric $dl^2$.
The metric (\ref{MMUU}) is a perturbation of the flat background spacetime
\be \n{flatbackgr}
 ds^2_\oix =- dt^2 + dl^2\;.
\ee
For details and further discussion of static metrics in the weak field approximation see Appendix~\ref{apx:statST}.

We denote $\xi^\mu$ a time-like Killing vector and $u^{\mu}$ a normalized four-velocity of static observers moving along this Killing vector,
\begin{equation}\label{uxirel}
    u^\alpha = e^{-\pot} \xi^\alpha\;.
\end{equation}
Obviously,
\begin{equation}\label{udtrel}
    u_\alpha = -e^{\pot} t_{,\alpha}\;.
\end{equation}
The four-acceleration of these static observers,
\begin{equation}\label{wdef}
     w^{\mu}=u^{\mu}_{\ ;\nu}u^{\nu}\;,
\end{equation}
satisfies the relation $w_{[\alpha ,\beta]}=0$ and hence, at least locally, one has
\begin{equation}\label{wdurel}
    w_\alpha = \pot_{,\alpha}\;.
\end{equation}

We call $w^\alpha$ the acceleration field, and $-\pot_{,\alpha}$ the gravitational field strength. Since $w^\alpha$ is orthogonal to the time direction $u^\alpha$, it can be restricted to its spatial components. The field strength is a fictitious force that one assumes in the non-inertial static frame to explain a tendency of free observers to move with respect to the frame. Non-moving static observers have to `compensate' this force by a real force equal to $w_j$ per unit mass. Of course, in the full spacetime description, the static observers move along non-geodesic trajectories with four-acceleration $w^\alpha$ caused by the real force.

In static spacetimes we can effectively `ignore' the time coordinate. The weak-field equation \eqref{Poisson} can be understood as the equation on spatial section $t=\const$, or, better, on a factorized three-dimensional space, points of which are orbits of the Killing vector $\xi^\alpha$. We will see in a moment, that the latter approach is more general.

\subsection{Locally static spacetimes}
\label{ssc:locstatST}

The above description of the static spacetime, as well as of the weak field, was local. Let us make some remarks about global aspects. We start with a general static spacetime without any approximation.

First, we observe that metric \eqref{MMUU} is  invariant under a scaling transformation
\be \n{staticrescaling}
\pot=\hat{\pot}+\delU\, ,\quad  t=e^{-\delU} \hat{t}\,,\quad dl^2 = e^{2\delU} \hat{dl}^2\;,
\ee
parametrized by a constant\footnote{Please, notice a difference between character for Laplace operator, $\lap\pot$, and the letter Delta in the constant $\delU$.}  $\delU$. This rescaling tells us, that two potentials on the background spacetime which differ only by a constant are in some sense equivalent.

This opens a very important possibility for globally non-trivial spacetimes. Consider spacetime with a non-trivial first homology group, i.e., spacetime containing non-contractible loops. From the duality between homology and de~Rham cohomology we know that in such a spacetime closed 1-forms do not have to be exact,\footnote{Here, one should understand tensor indices as abstract indices, not coordinate ones. On a simple local coordinate map, any closed form is exact. But we speak here about tensor fields defined on the whole spacetime.}
\begin{equation}\label{nontriviality}
    \sigma_{[\mu,\nu]} = 0 \quad \nRightarrow \quad \sigma_\mu = \alpha_{,\mu}\;.
\end{equation}
In particular, we can have spacetime with global static observers given by globally well-defined four-velocity $u_\alpha$.
It gives a globally well-defined four-acceleration $w_\alpha$. Although it is closed,
\begin{equation}\label{rotiszero}
    w_{[\mu,\nu]} =0\;,
\end{equation}
it may be non-potential. Such $w_\alpha$ corresponds to a non-trivial element of the first cohomology group.

Of course, locally, on a topologically trivial domain $V$, we can always find  a potential $\pot$ such that \eqref{wdurel} holds. But in a general case such a potential does not have a global smooth extension to the whole spacetime.

If we have two such potentials $\pot$ and $\hat{\pot}$ on domains $V$ and $\hat{V}$, respectively, on their intersection $V\cap\hat{V}$ the potentials may differ only by a constant,
\begin{equation}\label{difpotconst}
    w_{\alpha} = \pot_{,\alpha} = \hat{\pot}_{,\alpha}\quad \Rightarrow\quad
    \delU = \pot-\hat{\pot} = \const\;.
\end{equation}

We can cover the whole manifold by simply connected domains with associated with them potentials. Next, we may shift the potentials by a constant on some of these domains to fit potentials in the neighborhood domains. However, in topologically non-trivial spacetime it may not be possible to do it consistently everywhere. It may happen that we find two domains $V$ and $\hat{V}$ with potentials which are already continuously extended along some path in the manifold. However, these potentials do not fit; they differ on the intersections of these domains by a constant $\delU$.

A maximal extension of the potential $\pot$ can be specified by choosing a point $p_\oix$ of vanishing potential (typically in an asymptotical region) and paths $\gamma_p$ from this point to each point $p$ in the spacetime. The potential is given by
\be \n{UINF}
\pot=\int_{\gamma_p} w_\alpha\, dx^\alpha\; .
\ee
For a continuous change of the path $\gamma_p$ the potential $U$ changes smoothly.
However, in a topologically non-trivial space some pathes heading to two close points in the intersection $V\cap\hat{V}$ cannot be smoothly deformed to each other. Then, such a potential may not be continuous.

Impossibility to extend the potential to a single-valued global smooth function implies that also Killing vector $\xi^\alpha$, given by \eqref{uxirel}, cannot be defined globally. It can be continuously extended to domains $V$ and $\hat{V}$, similarly to potential~$\pot$. But on the intersection $V\cap\hat{V}$, it will differ by rescaling
\begin{equation}\label{KVrescaling}
  \xi^\alpha = e^{\delU}\hat{\xi}^\alpha\;.
\end{equation}
The same applies to the time coordinate $t$ related to the four-velocity by \eqref{udtrel}. On the intersection of domains $V$ and $\hat{V}$ we find
\begin{equation}\label{trescaling}
  t = e^{-\delU}\hat{t}\;.
\end{equation}

Spacetime with such a structure is called a locally static \cite{Frolov:1990si}, in contrast to the globally static spacetimes. In globally static spacetimes all quantities $ds^2$, $dl^2$, $u^\alpha$, $w^\alpha$, $\xi^\alpha$, $\pot$, and $t$ are defined globally. In locally static spacetimes, $\xi^\alpha$, $\pot$, $dl^2$, and $t$ have only local meaning. See Appendix~\ref{apx:locstaticst} for more details.

\subsection{Locally static weak gravitational field}
\label{ssc:locstatwf}

Our goal now is to build a locally static spacetime describing the gravitational field in the presence of a ring wormhole in the weak field approximation.

We start with a globally static spacetime, in fact, with a flat spacetime which differs from the Minkowski spacetime just by a wormhole identifications described in Sec.~\ref{sc:ringyWH}. On this background we find a gravitational potential $\pot$ which satisfies locally Poisson equation \eqref{Poisson} with some physically reasonable sources. But we are interested in situations when the potential cannot be extended globally.

In such a case, the globally well-defined quantities are acceleration field $w_\alpha$, four-velocity $u^\alpha$, and, of course, the full perturbed metric $ds^2$.

The weak-field equation \eqref{Poisson} can be rewritten using globally defined quantity $w^j$. The potential relation \eqref{wdurel} is superseded by \eqref{rotiszero}. We get
\begin{gather}
    \nabla_{[i} w_{j]} = 0\;,\label{rotw}\\
    \nabla_{i} w^{i} = 4\pi\mu\;.\label{divw}
\end{gather}
These equation can be understood as the gravito-static equations on the three-dimensional factorized space mentioned above.

Taking into account \eqref{MMUU}, we conclude that inertial time $t$ and flat metric $dl^2$, as well as the implicitly defined cartesian coordinates $x^i$, depend on the choice of the potential. Their perturbed versions are not thus defined globally.

Similarly to the discussion above, if we find domains $V$ and $\hat{V}$ on which the potentials  differ by a constant $\delU$, expression \eqref{MMUU} for the metric on these domains define $(t,x^j)$ and $(\hat{t},\hat{j})$ which will be related by
\be \n{UUCC}
\begin{gathered}
\pot=\hat{\pot}+\delU\;,\\
t=e^{-\delU} \hat{t}\;,\quad
x^j = e^{\delU} \hat{x}^j\;,\\
dl^2=e^{2\delU} \hat{dl}^2\;.
\end{gathered}
\ee
Identification of the resulting spacetime with the original background with metric $ds^2_\oix$ thus cannot be done globally in the  coordinates $(t, x^j)$.

However, the inertial coordinates are very simple and offer a straightforward local identification of the perturbed spacetime with the background. We want to take advantage of them. Therefore, we modify our perturbation procedure slightly.

Namely, we do not require that our perturbed metric $ds^2$ is a global perturbation of one unperturbed background $ds^2_\oix$. To justify the validity of the weak-field equations, it is enough that the resulting spacetime is a perturbation of a background on local domains. The domains must cover whole resulting spacetime. But the background can be assumed for each domain separately.

In this approach we can easily identify background spacetimes with the resulting perturbed spacetime using locally defined inertial coordinates. However, we have to face a problem of the unequality of these coordinates on overlaps of the domains.

In the specific cases of the wormhole spacetimes, we will be able to extend the potential and related quantities to the full spacetime, except the wormhole throat, using the procedure \eqref{UINF}. However, similarly to the potential, the extended inertial coordinates will not be continuous through the wormhole and we will have to discuss, how they are related. We will demonstrate this procedure in Sec.~\ref{sc:oneasympWHspc}.

Non-existence of a global static time coordinate $t$ is also a reason why we cannot use a global spatial section ${t=\const}$. However, we can still introduce the three-dimensional factorized space of orbits of the Killing vector $\xi^\alpha$ and solve the weak-field equations \eqref{rotw} and \eqref{divw} on it.

Finally, let us note that rescaling \eqref{UUCC} can be used also for identification of locally inertial coordinates near an arbitrary point $p$. By a suitable choice of $\delU$ one can always put $\hat{\pot}(p)=0$. For this gauge the coordinate $\hat{t}$ at $p$ coincides with the proper time at this point and locally Cartesian coordinates $\hat{x}^j$ near $p$ coincide with proper lengths along the axes.

\subsection{Gravity field flux}
\label{ssc:gravflux}

Let $S$ be a closed two-dimensional surface embedded in the spatial section $t=\const$. We define a gravity field flux over this surface as
\be\n{mS}
\Pi_S=\frac1{4\pi}\oint_S w^{a} n_a\, d^2\!S\, .
\ee
Here $n^a$ is an outward-pointing unit normal to $S$ and $d^2\!S$ is a two-dimensional surface area element on $S$.

Let $S_1$ and $S_2$ be two homological closed surfaces and $V$ is a three-dimensional volume restricted by them
\be
\pa V=S_2-S_1\, .
\ee
Then the Gauss theorem applied to volume $V$ and the weak-field equation \eqref{divw} implies that
\be \n{PPMM}
\Pi_{S_2}-\Pi_{S_1}=\frac1{4\pi}\int \nabla_{i} w^i\; d^3\! V= \int \mu\, d^3\! V\;.
\ee
It has  important consequences:
\begin{enumerate}[label=(\roman*)]
\item If there is no matter in the space between $S_1$ and $S_2$, the fluxes through these surfaces coincide.
\end{enumerate}
Applying \eqref{PPMM} on a boundary $\partial V$ of a compact volume $V$, we have:
\begin{enumerate}[resume,label=(\roman*)]
\item Consider adiabatic change of the distribution of mass. It does not change the flux through a two-dimensional surface $S=\partial V$ enclosing a three-dimensional volume $V$ until the matter crosses $S$.
\end{enumerate}
In the case of space with a non-trivial topology or with several asymptotic regions, one can formulate a more general proposition:
\begin{enumerate}[resume,label=(\roman*)]
\item The flux through any closed two-dimensional surface $S$ does not change under an adiabatic change of a mass distribution until the matter crosses the surface $S$.
\end{enumerate}
This last proposition has a direct counterpart in a well known result of the Maxwell electrodynamics, connecting the change of the flux over a closed surface with the current crossing it. The result can be easily proved by using the Stokes theorem. In the case of a weak gravitational field, the proposition does not follow just from properties of the Poisson equation. One would need to specify laws governing the dynamics. We will not discuss these laws here. Let us just mention, that there exists a well known analogy between the weak gravitational field and the Maxwell equations, see, e.g., \cite{Mashhoon:2003ax,Clark:2000ff}.

\section{Gravitational field near the wormhole connecting two asymptotic domains}
\label{sc:twoasympWHspc}

In this and the following sections, we work in a three-dimensional flat space with a non-trivial identification given by the wormhole. We will discuss some solutions of the field equation \eqref{Poisson}, or \eqref{rotw} and \eqref{divw}, respectively. As earlier, in Sec.~\ref{sc:ringyWH} we use a radius $a$ of the ring to define dimensional metric and coordinates. We will return to the spacetime description in Sec.~\ref{sec:TMformation}.

\subsection{Gravitational field of a massive shell}
\label{ssc:thickshell}

It is instructive to discuss first a gravitational field of a thick homogeneous ellipsoidal shell in a simple flat space which does not contain any wormhole. Let us consider a shell bounded by two similar ellipsoids filled by the matter of constant density. It will be sufficient for us to observe that according to famous Newton's theorem, cf.~\cite{Chandra:1969}, the gravitational force $-\nabla\pot$ inside the shell (inside the inner ellipsoid) vanishes.\footnote{Let us stress that Newton's theorem requires the bounding ellipsoids being \emph{similar} and, therefore, they are not \emph{confocal}. Thus, even when the ellipsoids are oblate spheroids, they cannot be specified just by two values of spheroidal coordinate $\chi$.} The potential in the cavity is thus constant $\pot=\pot_\oix$. Here, we normalize the potential $\pot$ so that it decreases at the infinity outside the outer ellipsoid.

The field both outside and in the shell's interior (between both ellipsoids) can also be written explicitly \cite{lan84,Chandra:1969}. However, these expressions are rather complicated. Instead of this, we will consider below a much simpler model in which the matter surrounding a throat is taken in the form of a thin spheroidal mass shell. This will allow us to obtain simple analytical expressions for the gravitational field.

Using the described field in the shell's cavity, it is  easy to construct a solution of the equation \eqref{Poisson} in the space~$\rwhspc$ where the ring wormhole connects two flat spaces. We assume that the disc associated with the ring wormhole in $R_+$ is inside the cavity. In this case the corresponding solution inside the shell and out of it remains the same as in a case without the ring wormhole, while in the cavity of the shell and everywhere in space $R_-$ the potential $\pot$ is constant equal to $\pot_\oix$.

In a special case when all the axes of the ellipsoidal shell are equal, it takes the form of the sphere. For such a case, this conclusion  remains valid even if the gravitational field of the shell is not week. Namely, one can use a massive thin shell approach by Israel \cite{Israel:1966rt} to glue the Schwarzschild metric outside the shell with the inner flat solution. By imposing the ring wormhole inside the shell does not change the gravitational field around.

Let us emphasize that the presence of a ring wormhole has non-trivial consequences. We can point out two aspects:
\begin{enumerate}[label=(\roman*)]
\item Even in the case of a ring wormhole connecting two flat spaces, the above described solution of the Poisson equation \eqref{Poisson} with the described source is not unique.
\item In the presence of a ring wormhole in a single space, the statement that the gravitating force vanishes in the cavity of the shell is not correct.
\end{enumerate}
We will discuss these issues below.

\subsection{Zero modes}

The space $\rwhspc$ with the wormhole connecting two asymptotic regions has the first cohomology group trivial and thus, we do not have to face problems with non-existence of a global potential. However, it has two asymptotic region and it allows ambiguity in the potential, even for trivial sources. Let us illustrate this point.

\begin{figure}
    \centering
      \includegraphics[width=0.5\columnwidth]{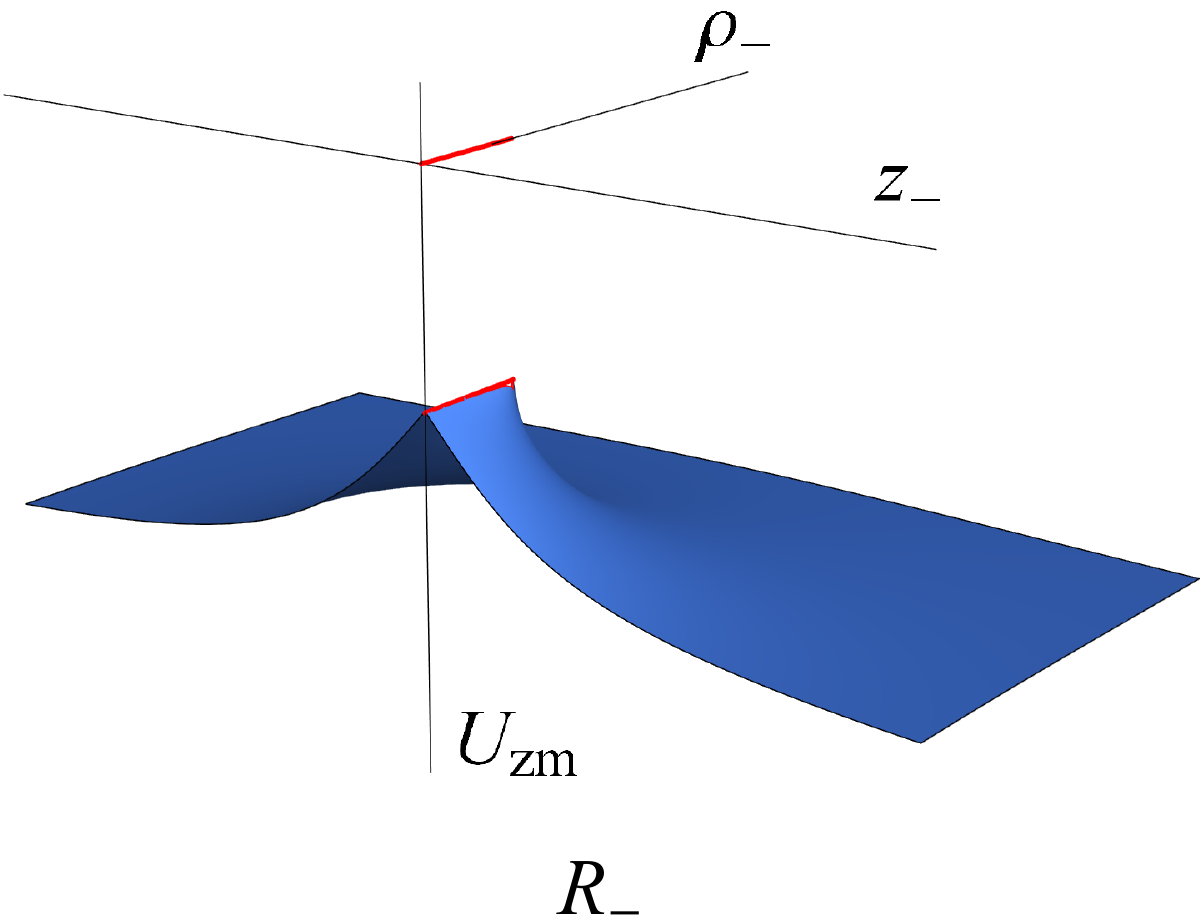}\includegraphics[width=0.5\columnwidth]{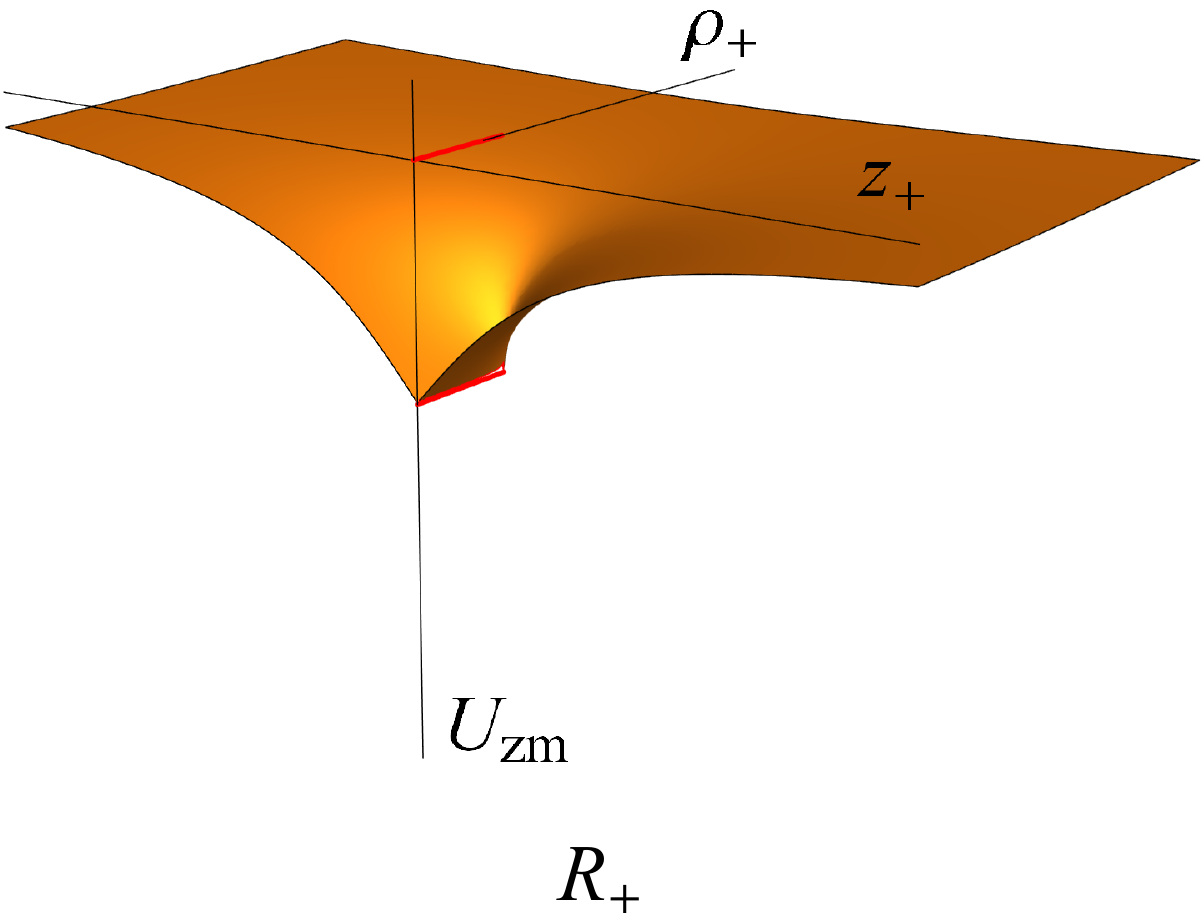}\\[3ex]
      \includegraphics[width=0.5\columnwidth]{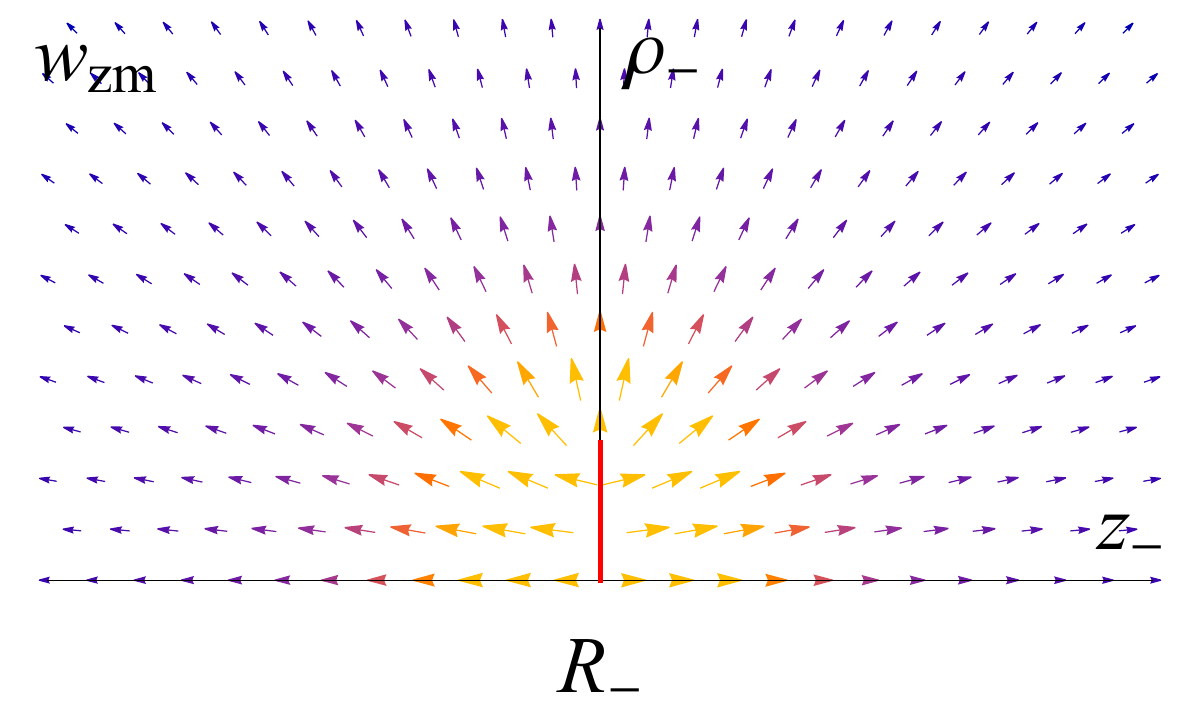}\includegraphics[width=0.5\columnwidth]{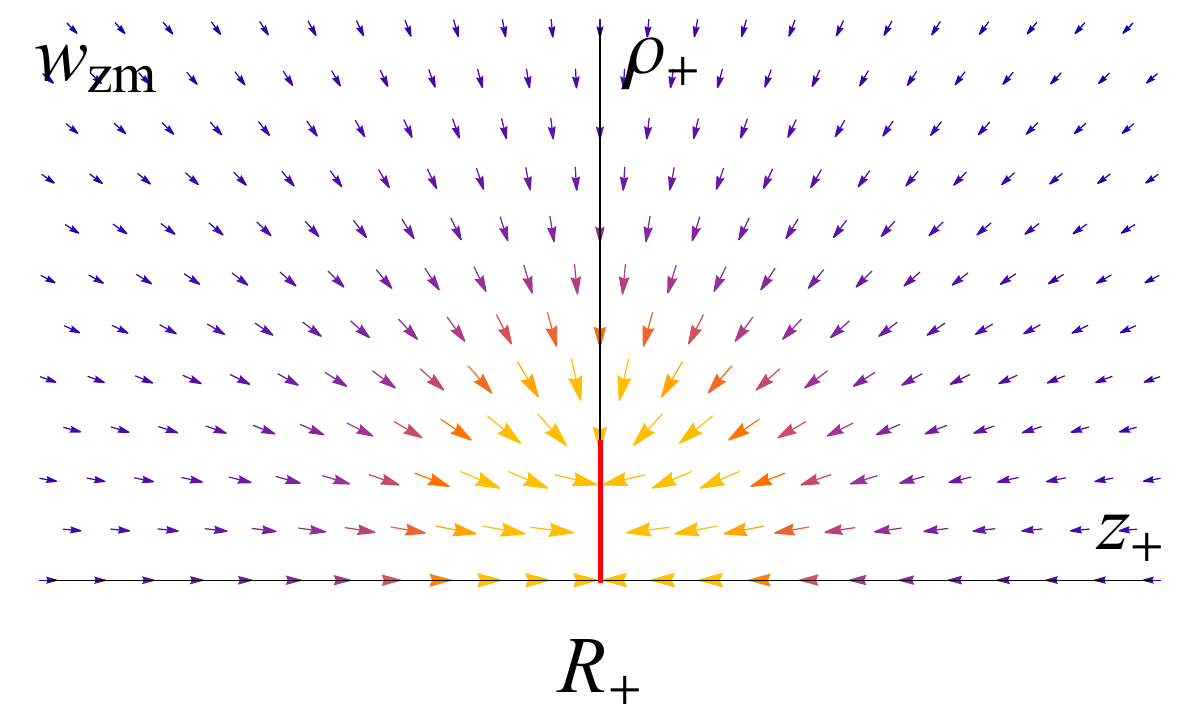}\\[2ex]
    \caption{\label{fig:zmUW}Potential $\pot_\zm$ (top) and field strength $w_\zm^j$ (bottom) of the zero mode. The left diagram shows the field in space $R_-$,  and the right one in $R_+$. Thick segments on the $\rho_\pm$ axes represent the mouths of the wormhole. Angular coordinate $\ph$ is ignored and can be recovered by a rotation around the axes $\rho_\pm=0$.}
\end{figure}

Consider the function\footnote{The function ${\Zo}$ is related to Legendre function ${\mathsf{Q}_0}$ evaluated on the imaginary axis with a branch cut chosen such that it is analytical on the whole imaginary axis.}
\be\label{Z0def}
\Zo(\chi)=-\arccot(\sinh\chi)
\,,
\ee
with $\arccot s$ taking values in the interval $(0,\pi)$. It has the following properties
\be\label{Z0prop}
\begin{gathered}
\Zo'(\chi)=
  \frac1{\cosh \chi}\, ,\\
\Zo(-\infty)=-\pi\hhh
\Zo(0)=-\frac\pi2 \hhh
\Zo(+\infty)=0\, .
\end{gathered}
\ee
\begin{figure}[h]
    \centering
      \includegraphics[width=0.6\columnwidth]{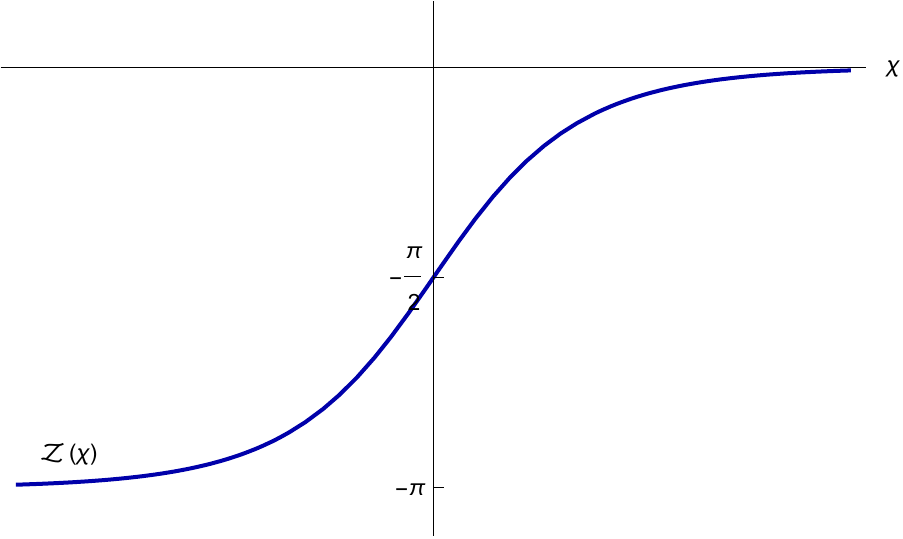}
    \caption{\label{FigZ}Functions $\Zo(\chi)$.}
\end{figure}
Let $\mzm$ and $C$ be two constants, then it is easy to check that
\be\label{zmpotC}
\pot_\zm(\chi)=\mzm\,\Zo(\chi)+C
\ee
is a solution of the homogeneous Laplace equation
\be\label{Laplace}
\lap \pot_\zm=0\, .
\ee
We call such a solution a zero mode. It is convenient to use the standard ambiguity in the choice of the potential and to put $C=0$, then
\be\label{zminf}
\pot_\zm|_{\chi=+\infty}=0\hh U_\zm|_{\chi=-\infty}=-\mzm\pi\, .
\ee

In the absence of the ring wormhole, one should put $\mzm=0$ as well, in order to avoid a discontinuity in the first derivative at $\chi=0$. For nonvanishing $\mzm$, a jump in the field strength across the disc $\chi=0$ would correspond to a mass distribution on the disc, and the potential \eqref{zmpotC} would not be a solution of source-free \eqref{Laplace}.

In the presence of the ring wormhole connecting spaces $R_+$ and $R_-$, the situation is quite different. Even if one
imposes a condition that  the potential vanishes at infinity of $R_+$ by putting $C=0$, there exists a one-parameter family of homogeneous solutions parameterized by a constant $\mzm$. The difference of potentials at the infinities of the spaces $R_+$ and $R_-$ is
\be\label{zmpotdif}
\delU=\pot_\zm|_{\chi=+\infty}-\pot_\zm|_{\chi=-\infty}=\mzm\pi\, .
\ee

In the presence of a zero-mode potential, there exists a non-zero force acting on a unit-mass particle equal to
\be
-\nabla \pot_\zm = -\frac{\mzm}{\cosh\chi}\,\nabla\chi\, .
\ee
For positive $\mzm$ this force is directed towards the decreasing $\chi$.
In $R_+$  this force describes an attraction of the particle by the wormhole, while in $R_-$ it is repulsive from the wormhole.

It is easy to calculate a flux $\Pi_D$ through the wormhole associated with the zero mode. Consider a closed two-dimensional surface $S$ surrounding the throat $D$ of the wormhole, with normal oriented towards $R_+$ asymptotic. Since the zero mode has vanishing sources, the flux does not depends on exact shape and position of the surface, cf.\ argument \eqref{PPMM}. The calculation for spheroid ${\chi=\const}$ gives
\be\label{fluxzm}
\Pi_S= \mzm\, ,
\ee
where we used
\begin{equation}\label{ndS}
\begin{gathered}
    n^\alpha = \frac{1}{\sqrt{\sinh^2\chi+\cos^2\tht}}\,
    \delta_\chi^\alpha\;,\\
    dS = \sqrt{\sinh^2\chi+\cos^2\tht}\, \cosh\chi \sin\tht\, d\tht d\ph\;.
\end{gathered}
\end{equation}
For surface $D$ at $\chi=0$, representing the identified discs, we obtain the flux $\Pi_D$ through the wormhole throat.

A zero-mode field can be created in the following process. Let us start with empty space with vanishing potential. Consider a particle of mass $m$ at the infinity $R_+$ and move it adiabatically to bring it close to the throat. The flux $\Pi_D$ through the surface $D$ at the throat is not changed in such a process until the mass crosses $D$. After crossing $D$ the flux becomes $\Pi_D=m$ and it remains such if the mass stays in $R_-$. When the mass $m$ is brought to infinity in the space $R_-$, one is left with a space without any matter, but with a non-vanishing flux $\Pi_D$ through the throat. It will be exactly the zero mode described above.

\subsection{A field of a massive thin shell around the wormhole}

\begin{figure}
    \centering
      \includegraphics[width=0.5\columnwidth]{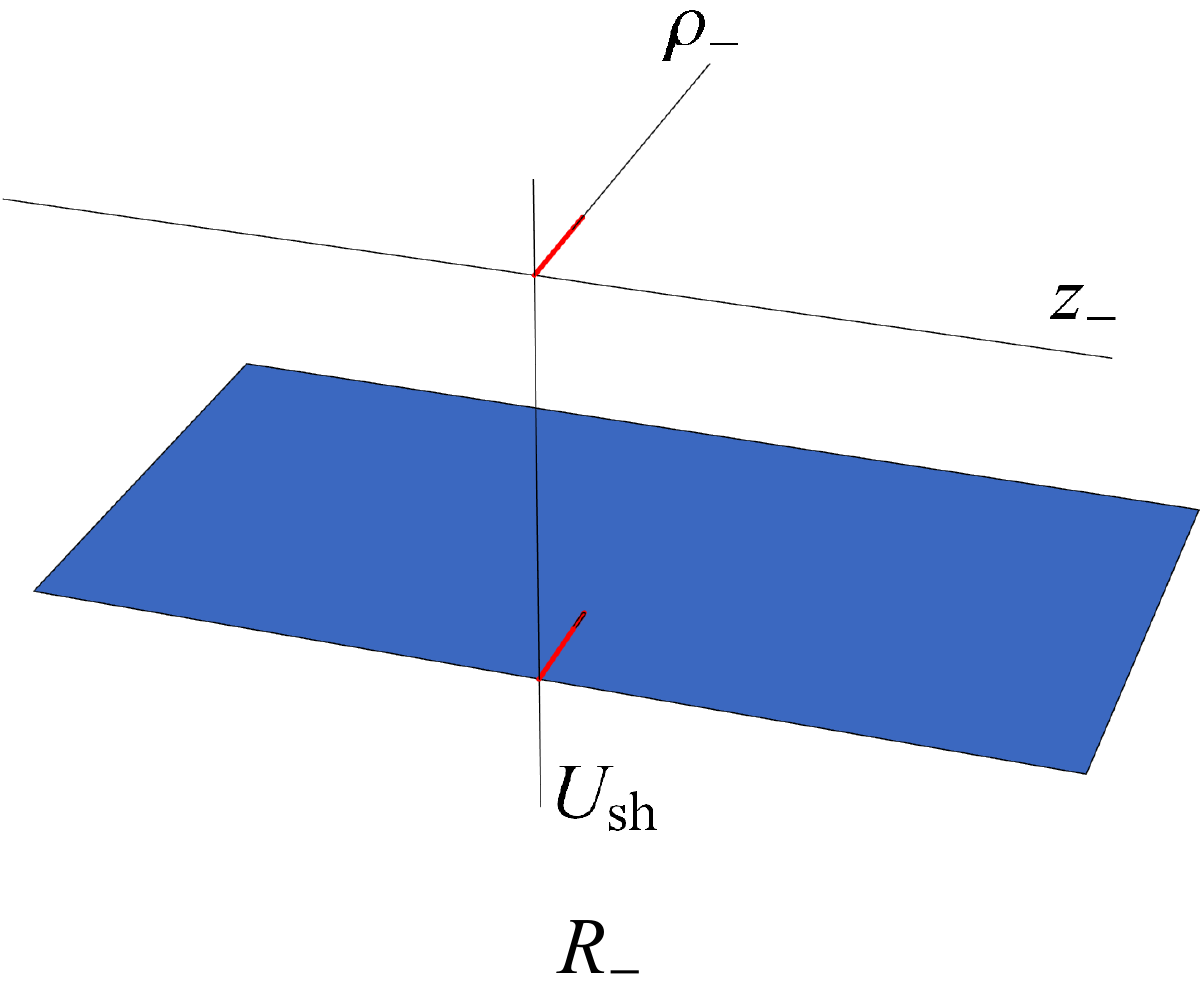}\includegraphics[width=0.5\columnwidth]{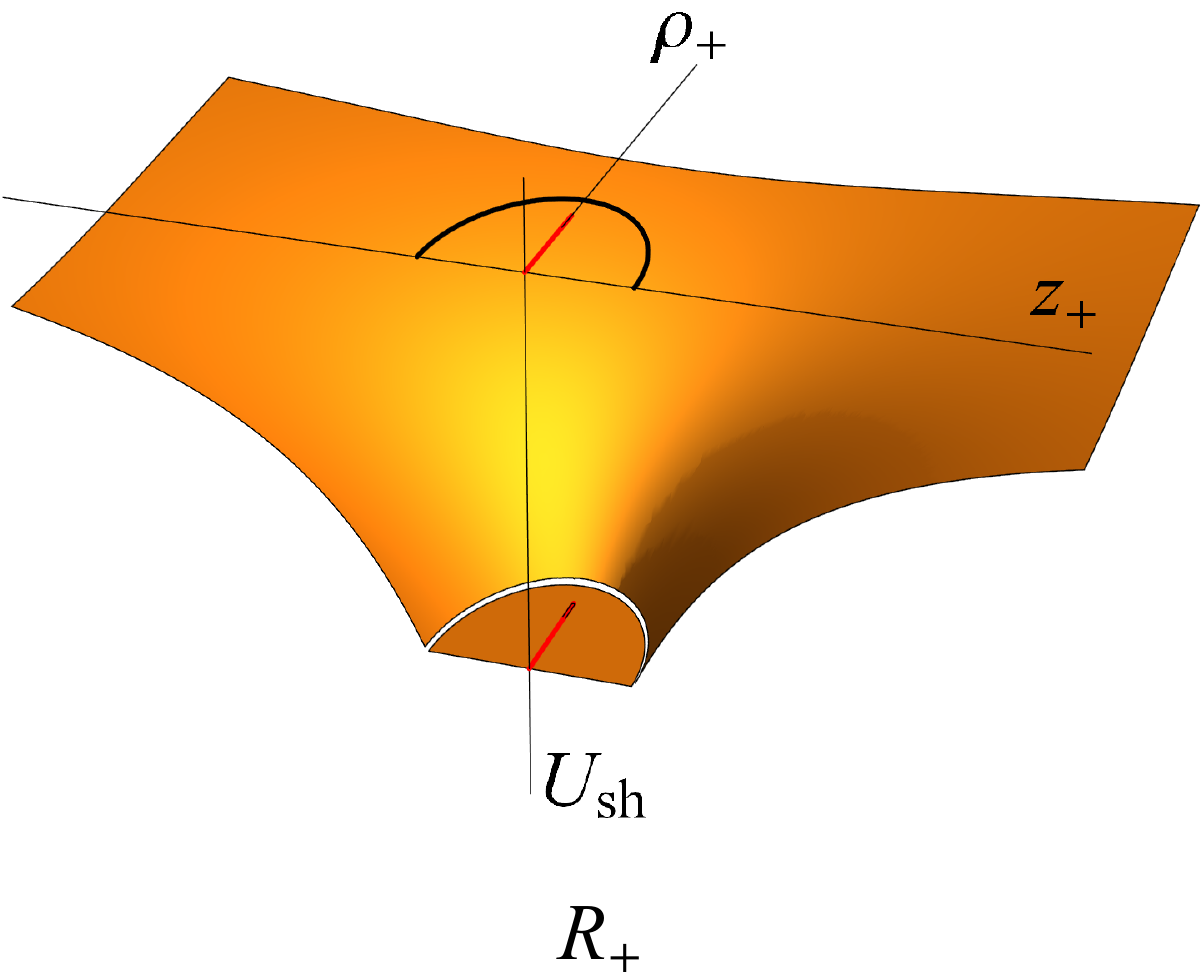}\\[3ex]
      \includegraphics[width=0.5\columnwidth]{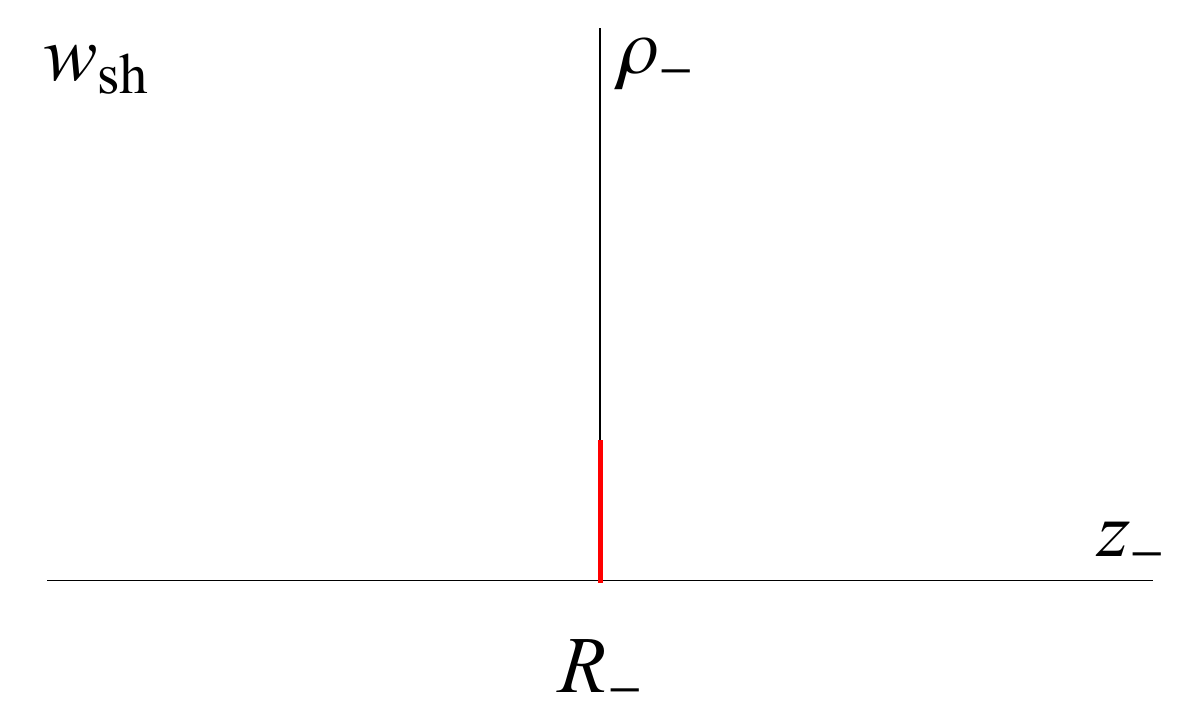}\includegraphics[width=0.5\columnwidth]{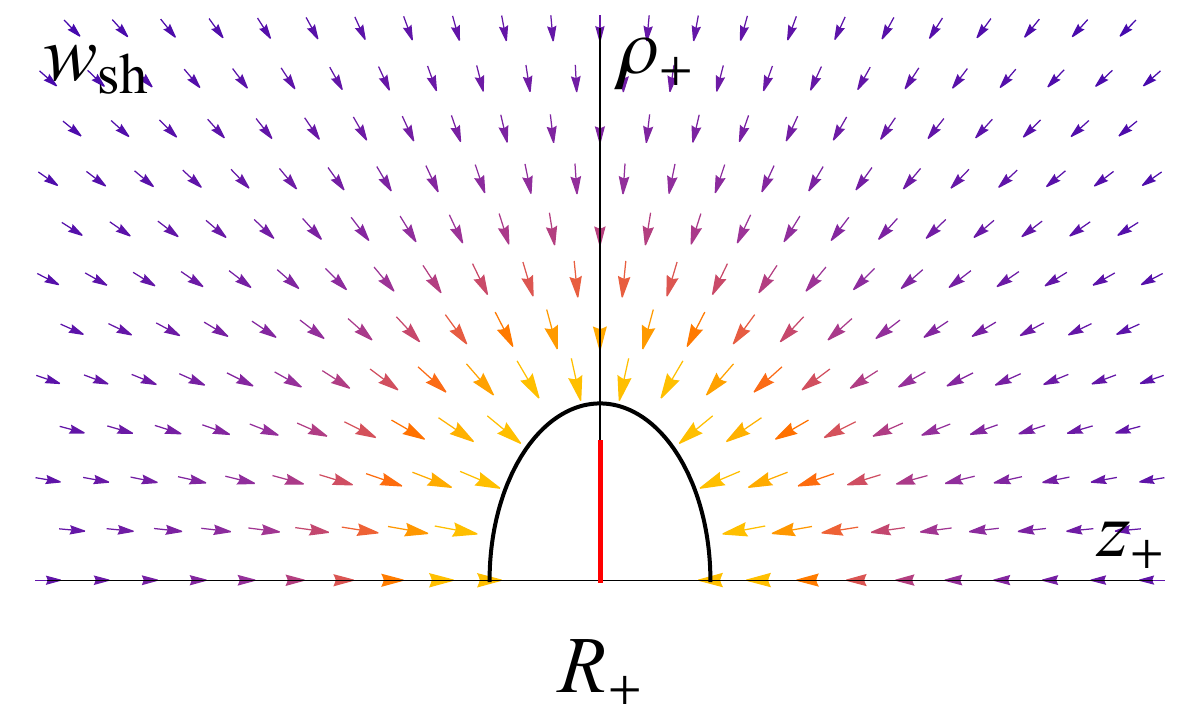}\\[2ex]
    \caption{\label{fig:shUW}Potential $\pot_\sh$ (top) and field strength $w_\sh^j$ (bottom) of a thin massive spheroidal shell located around the wormhole in $R_+$. The left diagram shows the field in space $R_-$,  and the right one in $R_+$. Thick segments on the $\rho_\pm$ axes represent mouths of the wormhole. The shell (depicted by the thick ellipse) is confocal with the wormhole mouth. Angular coordinate $\ph$ is ignored and can be recovered by a rotation around the axes $\rho_\pm=0$.}
\end{figure}

Let us assume now  that there exists a static mass distribution surrounding the wormhole. In particular, we consider a thin massive shell which has a form of an oblate spheroid $S_\oix$ in $R_+$ surrounding the wormhole and confocal to its ring.
In what follows we assume that the this massive shell is formed by massive particles that are brought together from $R_+$ infinity. During this process no zero-mode field would be created and they should be excluded in our solution.

In oblate spherodial coordinates  $(\chi,\tht,\ph)$ the equation of shell is $\chi=\chi_\oix>0$. Let $M$ be the mass of the massive shell, and $m=M/a$ be its dimensionless variant. The mass is not distributed homogeneously over the shell surface and its surface density is
\be\label{thinshsigma}
\sigma = \frac{m}{4\pi\sqrt{\sinh^2\chi+\cos^2\tht}\,\cosh\chi}\, .
\ee
The distributional volume density is\footnote{%
Here, $\delta_S$ is a covariant $\delta$-function localized on a surface $S$,
\[\int f\, \delta_S\, dV = \int_S f\, dS\;,\]
and $\delta(\chi-\chi_\oix)$ is its coordinate representation. They are related
\[\delta_S\,dl = \delta(\chi-\chi_\oix)\,d\chi\;,\]
with the proper length $dl$ orthogonal to $S$
\[dl=\sqrt{\sinh^2\chi+\cos^2\tht}\,d\chi\;.\]}
\be
\mu = \sigma\, \delta_{S_\oix} = \frac{m}{4\pi (\sinh^2\chi+\cos^2\tht)\cosh\chi} \delta(\chi-\chi_\oix)\, .
\ee
Surprisingly, this mass distribution can be obtained \cite{Chandra:1969} as a thin shell limit of the thick homogeneous spheroidal shell discussed in Sec.~\ref{ssc:thickshell}. Nonhomogeneity of the surface mass density arises from the fact, that the thick shell is formed by two similar oblate spheroids and its thickness vary with position on the spheroid.
We will see that for the thin shell the potential $\pot_\sh$ depends only on the $\chi$ coordinate and hence it is
constant on the shell.\footnote{This solution is similar to the well known solution for a static electric potential of a charged conducting oblate spheroid which is discussed, e.g., in \cite{lan84}}

In fact, to obtain a solution of the Poisson equation for the thin shell we can use solutions $\Zo(\chi)$ of the Laplace equation given by (\ref{Z0def}). The field $\pot_\sh$ analogous to that of the thick shell discussed in Sec.~\ref{ssc:thickshell} is
\be \n{thinshell}
\pot_\sh(\chi)=
  \begin{cases}
       m\, \Zo(\chi)   &\text{for}\;  \chi>\chi_\oix\; ,\\
       \pot_\oix = m\, \Zo(\chi_\oix) &\text{for}\;  \chi<\chi_\oix\; .
  \end{cases}
\ee
The potential inside the shell, as well as in $R_-$, is constant. Outside the shell, it is the same as the zero mode, and it vanishes at the infinity of $R_+$. The jump in the field strength at the shell corresponds to the surface mass distribution \eqref{thinshsigma}.

Of course, we can superpose this solution with the zero mode solution
\be\n{Umzm}
\pot = \pot_\sh + \pot_\zm\; .
\ee
At the infinity of $R_+$ the potential $\pot$ again vanishes, while at the infinity of $R_-$ its value is
\be\n{Umzm1}
\pot|_{\chi=-\infty} = m\, \Zo(\chi_\oix) - \mzm\,\pi\, .
\ee

We thus found that a general potential \eqref{Umzm} of the thin shell vanishing in the infinity of $R^+$  contains two constants, $m$ and $\mzm$. Their meaning can be understood using the gravity field flux. As we have seen, cf.~\eqref{PPMM}, the flux may be used to define the mass function $m_S=\Pi_S$ characterizing mass inside a surface $S$. For the equipotential surface $\chi=\text{const}$ the same calculation as in \eqref{fluxzm} gives
\be\n{cchh}
   m_S(\chi)=m\Theta(\chi-\chi_\oix)+\mzm\, .
\ee
It should be emphasized that both in $R_+$ and $R_-$ the normal to the spheroid $\chi=\const$ is chosen in the direction of growing $\chi$.

In the exterior of the shell, the mass function is $m_S=m+\mzm$. Inside the shell (including space $R_-$, where $\chi<0$), the mass function is just $m_S=\mzm$. It correctly jumps by the mass of the shell when crossing the shell. Since $m_S$ does not change when we further decrease $\chi$, we can interpret $\mzm$ as a mass `far behind' the wormhole, located under arbitrary negative $\chi$, i.e., located in the infinity of space $R_-$.

However, this mass had to be brought into the infinity of $R_-$ through the wormhole from the infinity of $R_+$. Because by fixing vanishing potential $\pot$ at the infinity of $R_+$, we have chosen that all matter forming the gravitational field has to be brought from this infinity. The mass $\mzm$ at the infinity of $R_-$ thus had to be moved from the infinity of $R_+$, and the potential difference \eqref{zmpotdif} corresponds to the work needed for that.

We interpreted the potential assuming that it vanishes at infinity of $R_+$. We selected one of the asymptotics as the  origin of all matter. We thus have broken the symmetry of the system. Of course, we can also allow the matter to originate at the infinity of $R_-$. In such a case, we have to recover the constant $C$ in \eqref{zmpotC}.

Then, one can interpret the zero mode as the gravitational field trapped by the wormhole, which holds memory about a previous activity how the matter has been moved through the wormhole.

\subsection{Wormhole in a homogeneous gravitational field}
\label{ssc:whhomfield}

\begin{figure}
    \centering
      \includegraphics[width=0.5\columnwidth]{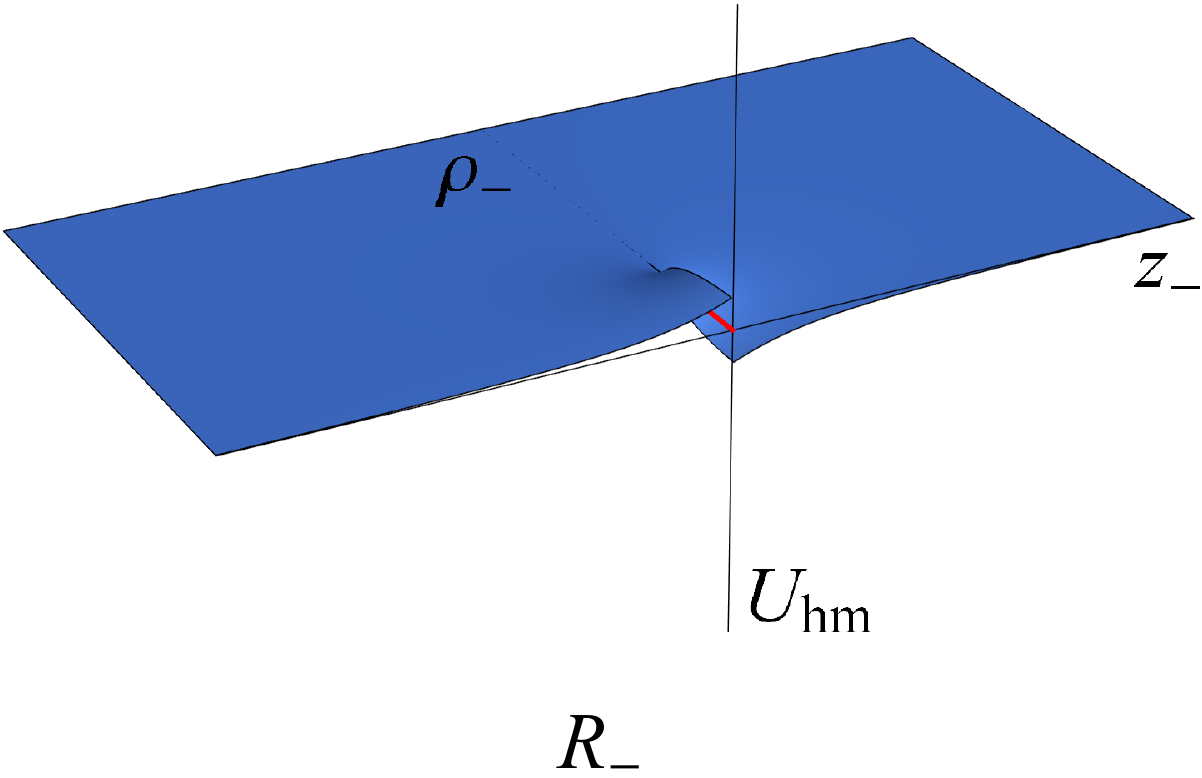}\includegraphics[width=0.5\columnwidth]{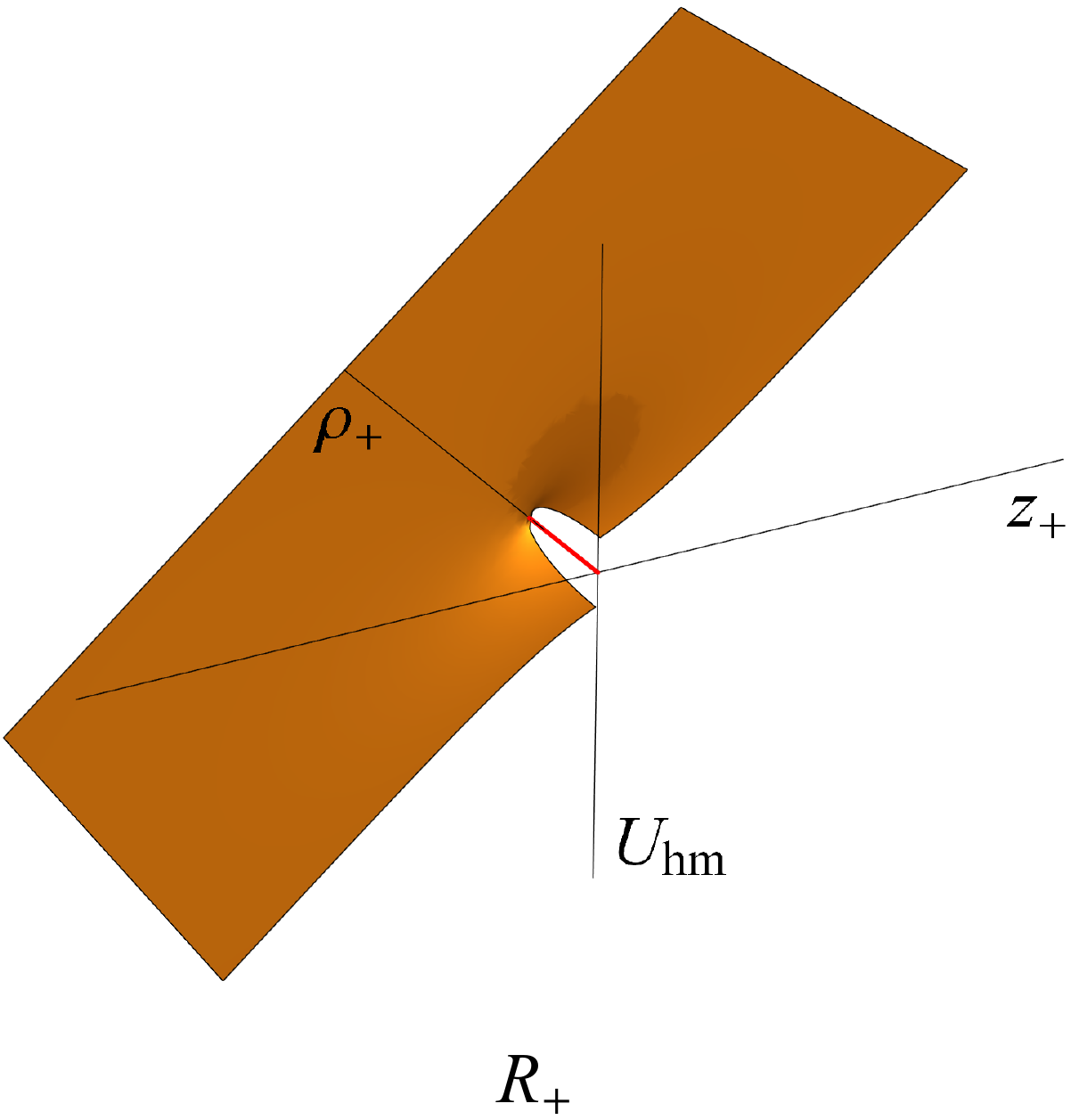}\\[3ex]
      \includegraphics[width=0.5\columnwidth]{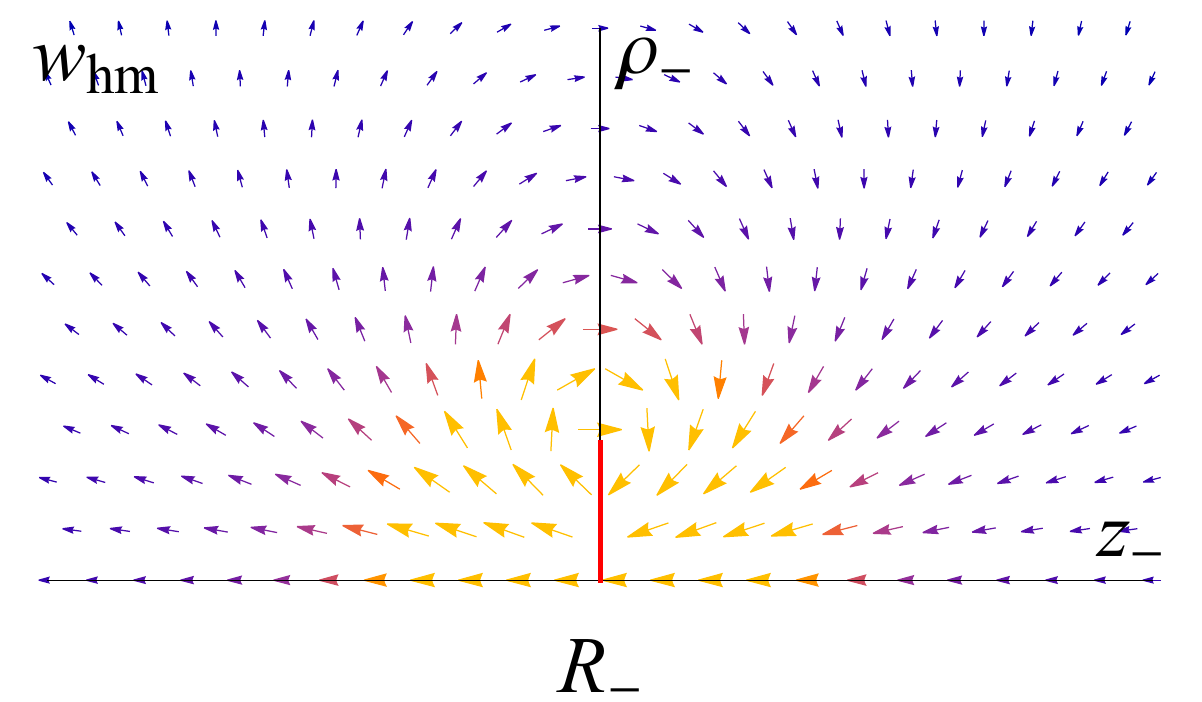}\includegraphics[width=0.5\columnwidth]{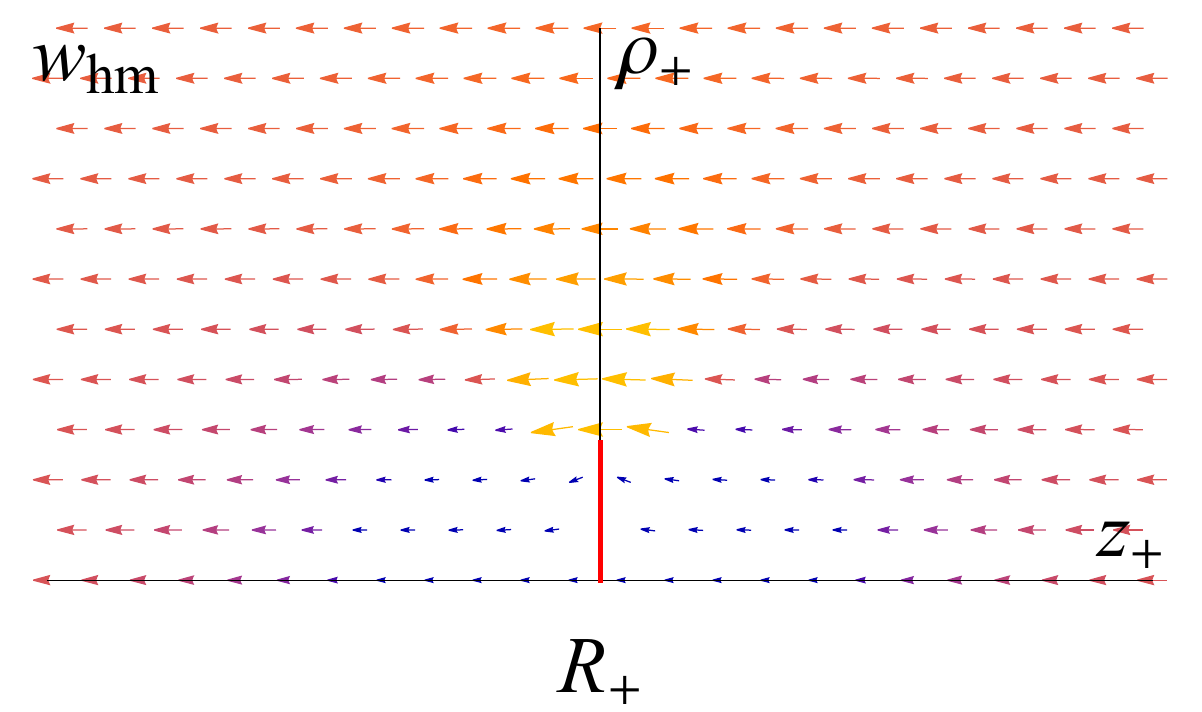}
    \caption{\label{fig:hmUW}Potential $\pot_\hm$ (top) and field strength $w_\hm^j$ (bottom) of asymptotically homogeneous field in $R_+$. The left diagram shows the induced field in space $R_-$,  and the right one the modified homogeneous field in $R_+$. Thick segments on the $\rho_\pm$ axes represent mouths of the wormhole. Angular coordinate $\ph$ is ignored and can be recovered by a rotation around the axes $\rho_\pm=0$.}
\end{figure}

Let us consider a homogeneous gravitational field in a flat space. Its potential is
\be \n{UU00}
\pot = \hmf z\, .
\ee
The acceleration field has constant value $\hmf$ and is directed along $z$-axis
\be
w_i = \nabla_{\!i}\, \pot = \hmf \,\nabla_{\!i}\, z\, .
\ee

Let us assume now that a ring wormhole leading to another space is immersed in this gravitational field. For simplicity, we assume the wormhole is in the origin $z=0$, with its mouth orthogonal to $z$-axis. To find how such a wormhole modifies the field, one needs to solve the Laplace equation in $\rwhspc$ space which has an asymptotic \eqref{UU00} at \mbox{$R_+$-infinity}. The corresponding solution is
\be\n{TTUU}
\pot_\hm = -\hmf\, \Zp(\chi)\, \cos\tht\, .
\ee

Here, we introduce functions\footnote{Functions ${\Zpm}$ are related to Legendre function ${\mathsf{Q}_1}$ evaluated on the imaginary axis, with properly chosen asymptotic, and with a branch cut chosen such that it is analytical on the whole imaginary axis.} $\Zpm(\chi)$
\begin{equation}\label{Zpmdef}
  \Zpm(\chi)= \mp\frac12\sinh\chi-\frac1\pi\bigl(1+\sinh\chi\,\arctan\sinh\chi\bigr)\, .
\end{equation}
We require ${\arctan s\in(-\frac\pi2,\frac\pi2)}$. Clearly
\begin{equation}\label{Z1split}
  \Zp(\chi)=\Zm(-\chi)= -\sinh\chi + \Zm(\chi)\, .
\end{equation}
Function $\Zp(\chi)$ asymptotically behaves as
\begin{equation}\label{Zpprop}
    \Zp|_{\chi\to-\infty} = 0\;,\quad
    \Zp|_{\chi\to+\infty} \approx - \sinh\chi\;,
\end{equation}
and function $\Zm(\chi)$ as
\begin{equation}\label{Zpprop}
   \Zm|_{\chi\to-\infty} \approx \sinh\chi\;,\quad
    \Zm|_{\chi\to+\infty} = 0\;,
\end{equation}
and
\begin{equation}\label{Zpprop}
   \Zpm'(0) = \mp\frac12\;.
\end{equation}
\begin{figure}[h]
    \centering
      \includegraphics[width=0.6\columnwidth]{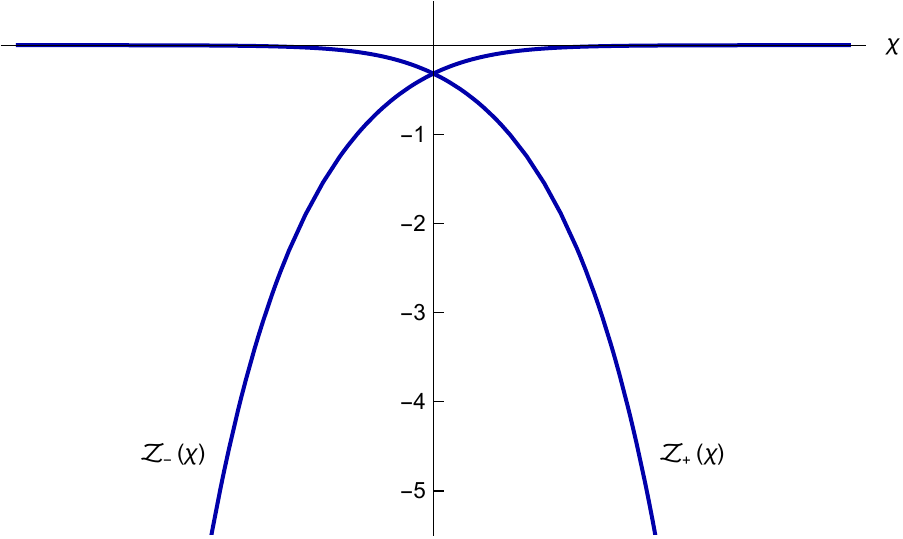}
    \caption{\label{FigZ1}Functions $\Zm(\chi)$ and $\Zp(\chi)$.}
\end{figure}

To check the validity of the boundary condition of the field $\pot_\hm$ at ${z\to\infty}$ in $R_+$, we use relation \eqref{Z1split}
\be\label{homsplit}
  \pot_\hm = \hmf\sinh\chi\cos\tht-\hmf\Zm(\chi)\cos\tht  \;.
\ee
In $R_+$, where $z=\sinh\chi\cos\tht$, the first term correctly reproduces the asymptotic behavior \eqref{UU00}, since the second term falls down at ${\chi\to +\infty}$.

One can interpret this result as follows. The ring wormhole changes the homogeneous gravitational field \eqref{UU00}, which was originally introduced only in ${R_+}$. When we consider it in the full wormhole space ${\rwhspc}$, we call it truncated field ${\pot_\trn}$. In the presence of the wormhole, it will be modified by an induced field $\pot_\ind$. The contribution of the induced field in the region $R_+$ is described by term ${-\hmf\Zm(\chi)\cos\tht}$, while in the region $R_-$ it is given by field \eqref{TTUU} penetrating through the wormhole, i.e., by $-\hmf\Zp(\chi)\cos\tht$. Together,
\begin{equation}\label{Uhomsplit}
    \pot_\hm = \pot_\trn + \pot_\ind\;,
\end{equation}
where the truncated and the induced fields are
\begin{align}
  \pot_\trn &= \hmf\,\Theta(\chi)\,\sinh\chi\,\cos\tht \;, \label{trnpot}\\
  \pot_\ind &= -\hmf\,\Zm(\abs{\chi})\,\cos\tht \;, \label{indpot}
\end{align}
$\Theta(\chi)$ being the Heaviside step function.

These fields have a natural physical interpretation. The truncated potential $\pot_\trn$ vanishes in $R_-$ and describes the homogeneous field in $R_+$. It is continuous everywhere, including the wormhole mouth $\chi=0$, but it is not smooth here. Its field strength is discontinuous at $\chi=0$. In fact, it is discontinuous at both faces $D^<$ and $D^>$ of the ring wormhole. The jump of the normal component of the field strength is $+\hmf$ and $-\hmf$ at $D^<$ and $D^>$, respectively, since the normal is pointing in a negative $z$ direction at $D^<$ and in a positive $z$ direction at $D^>$. Hence, $\pot_\trn$ does not satisfy the homogeneous Laplace equation at the wormhole. The jump corresponds to a massive thin shell at the wormhole mouth $\chi=0$ with a surface density $\mp\frac{\hmf}{4\pi}$.

The potential induced by the wormhole, $\pot_\ind$, is also continuous everywhere and not smooth at the wormhole. Using properties of function $\Zp$ one can find that it satisfies the Laplace equation everywhere except the  wormhole throat $\chi=0$. The jump of the normal component of the field strength at $\chi=0$ is in this case $-\hmf$ and $+\hmf$ at $D^<$ and $D^>$, respectively. So, it also corresponds to a massive shell at the throat of the wormhole with a surface density $\pm\frac{\hmf}{4\pi}$. Potential $\pot_\ind$ has dipole character both in $R_-$ and $R_+$.

Clearly, if one adds both potentials $\pot_\trn$ and $\pot_\ind$, then the sum \eqref{homsplit} is a solution of the source-free Laplace equation everywhere, including the wormhole throat ${\chi=0}$.

Let us note that the potential $\pot_\hm$ defined by \eqref{TTUU} is finite and its value is proportional to $\hmf$ (see Fig.~\ref{fig:hmUW}). Hence, for small $\hmf$ the field strength outside the ring is uniformly small. However, the corresponding field strength diverges at the ring. This divergency is a consequence of the assumption that the corresponding cosmic string located at the ring is infinitely thin. As it was discussed in Sec.\ref{sc:ringyWH}, the smearing of the distribution of the matter of the string makes the background spacetime of the wormhole regular. One can expect that the gradient of the potential $\tilde{U}$ on such a smooth background would be regular in the vicinity of smeared string as well. The adopted weak-field approximation remains valid until a narrow `tube' surrounding the smeared string. The behavior of the gravitational field inside such a `tube' depends on details of the string's matter distribution.

\section{Gravitational field near the wormhole in spacetime with one asymptotic}
\label{sc:oneasympWHspc}

In this section, we will study a wormhole connecting two places in one asymptotically Minkowski spacetime, and we will discuss a gravitational perturbation of such a space. We described the background wormhole spacetime $R_\wh$ in Sec.~\ref{sc:ringyWH}.

Now we consider the gravitational potentials $\pot$ of the massive thin shell located around the wormhole mouth $D_-$. In oblate spheroidal coordinates ${(\chi_-, \tht_-, \ph_-)}$ constructed around the mouth ${D_-}$, the shell is given by $\chi_-=\chi_\oix>0$. It is highly nontrivial to find the full, consistent solution of the Poisson equation in the wormhole spacetime ${R_\wh}$. Namely, one needs to obtain a solution of the Laplace equation with specially chosen boundary conditions. Besides natural conditions of decreasing of the gravitational potential $\pot$ at the infinity  and jump conditions for it on the surface of the shell one needs to impose additionally a condition of the regularity and absence of matter at the discs which are identified in such a space. The latter condition means that the field's strength $\nabla \pot$ should be continuous on the discs. This problem belongs to a class of so called mixed boundary value problem. A general discussion of such problems and references to the papers where they have been studied can be found in the nice book \cite{Sneddon:1966}.

We will solve this problem only approximately, for the wormhole mouths sufficiently distant, ${\ell\gg 1}$.\footnote{We must distinguish two approximate schemes employed here. First, we use the perturbative expansion of the gravitational interaction, where we consider only the first-order approximation in which the gravity description reduces to the Newtonian potential satisfying the Poisson equation. Second, for the particular wormhole spacetime, we look for the gravitational potential of the massive shell using the expansion for large mouths' distance $\ell$. Here, we discuss the later expansion.}

\subsection{Large distance approximation -- 0th order}

Let's start with the zeroth-order approximation. We return to the first-order corrections below. In the zeroth order, we assume that the field of the shell near the second mouth ${D_+}$ is negligible, and we can thus ignore any interaction with the wormhole. The potential  is thus given by field \eqref{thinshell} positioned around ${D_-}$, which is achieved by employing ${(\chi_-,\,\tht_-,\,\ph_-)}$ coordinates,
\begin{equation}\label{thinshellmin}
  \pot =
  \begin{cases}
       m\, \Zo(\chi_-)   &\text{for}\;  \chi_->\chi_\oix\; ,\\
       \pot_\oix = m\, \Zo(\chi_\oix) &\text{for}\;  0<\chi_-<\chi_\oix\; .
  \end{cases}
\end{equation}

\begin{figure}
      \includegraphics[width=0.8\columnwidth]{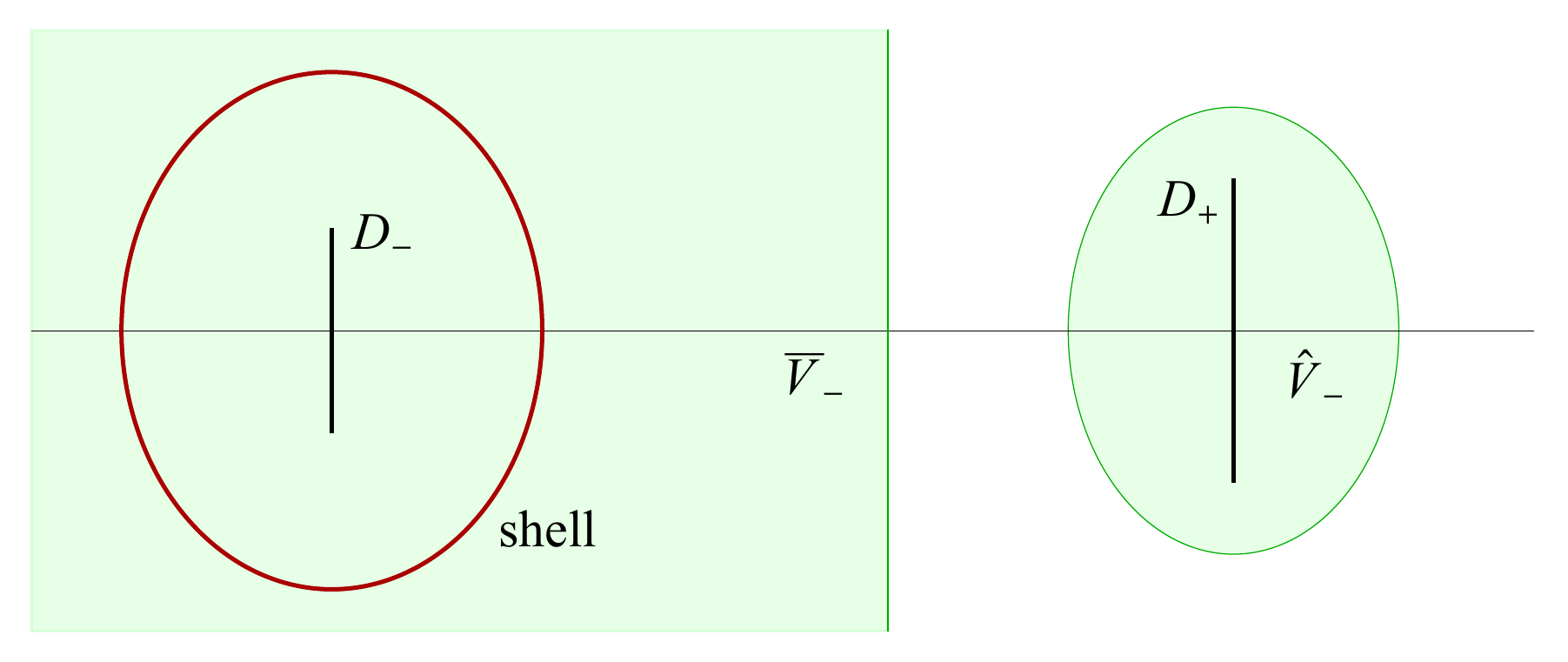}\\[-1ex]
      \includegraphics[width=0.8\columnwidth]{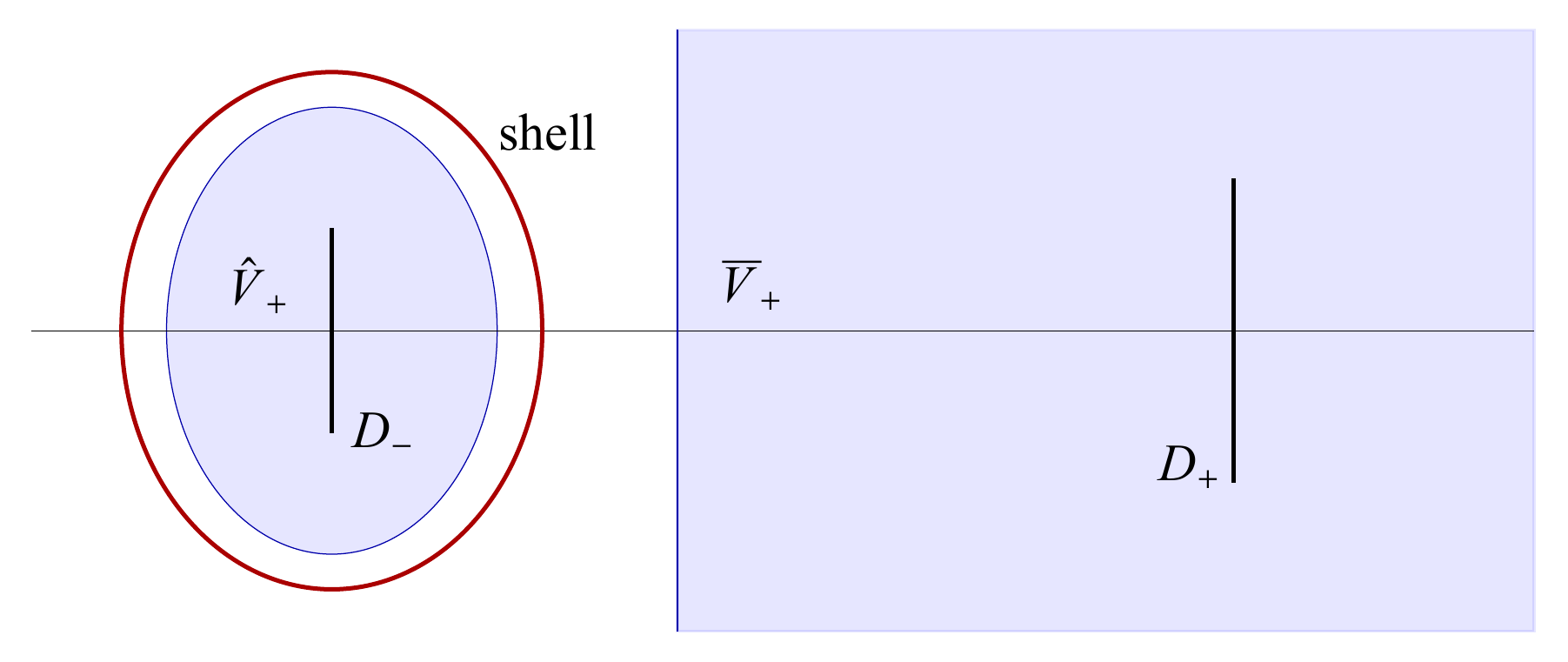}\\[-1ex]
      \includegraphics[width=0.8\columnwidth]{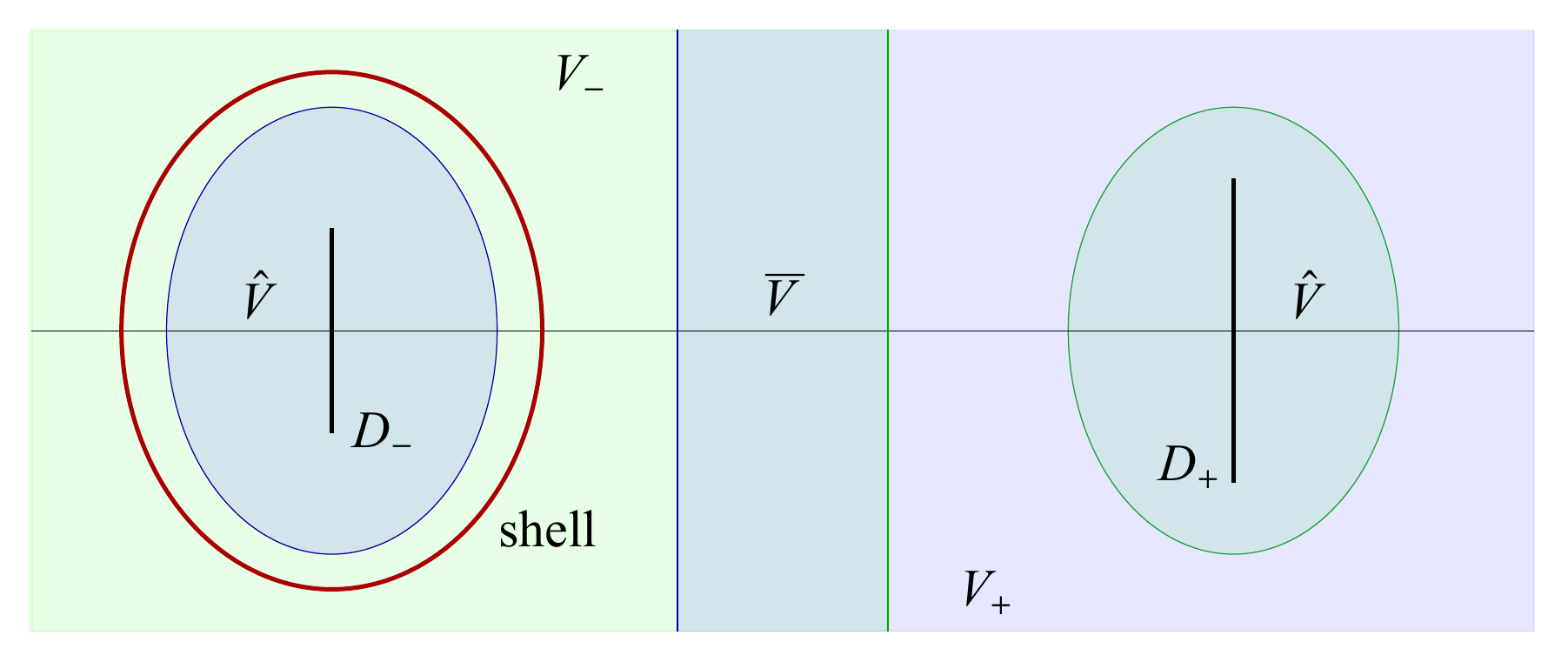}
    \caption{\label{fig:whdom}Coordinate domains around the wormhole in the presence of gravitational field of a thin massive shell near the left wormhole mouth. Domains $V_-$ (top) and $V_+$ (middle) are covering together the whole space. The division into main part $\bar{V}_\pm$ and `leaked' part $\hat{V}_\pm$ is indicated. The bottom diagram shows two intersections $\bar{V}$ and $\hat{V}$ of these two domains. The wormhole mouths $D_\pm$ are represented by the thick vertical lines. The real size of both mouths is the same. However, since the gravitational field of the shell affects the geometry, the global cartesian coordinates used for mapping to the diagram do not represent the geometry properly. The coordinate size of the both mouths is thus different, as can be observed in the diagram. The distance between mouths should be large -- diagrams do not reflect this feature accurately.}
\vspace{4ex}
      \includegraphics[width=\columnwidth]{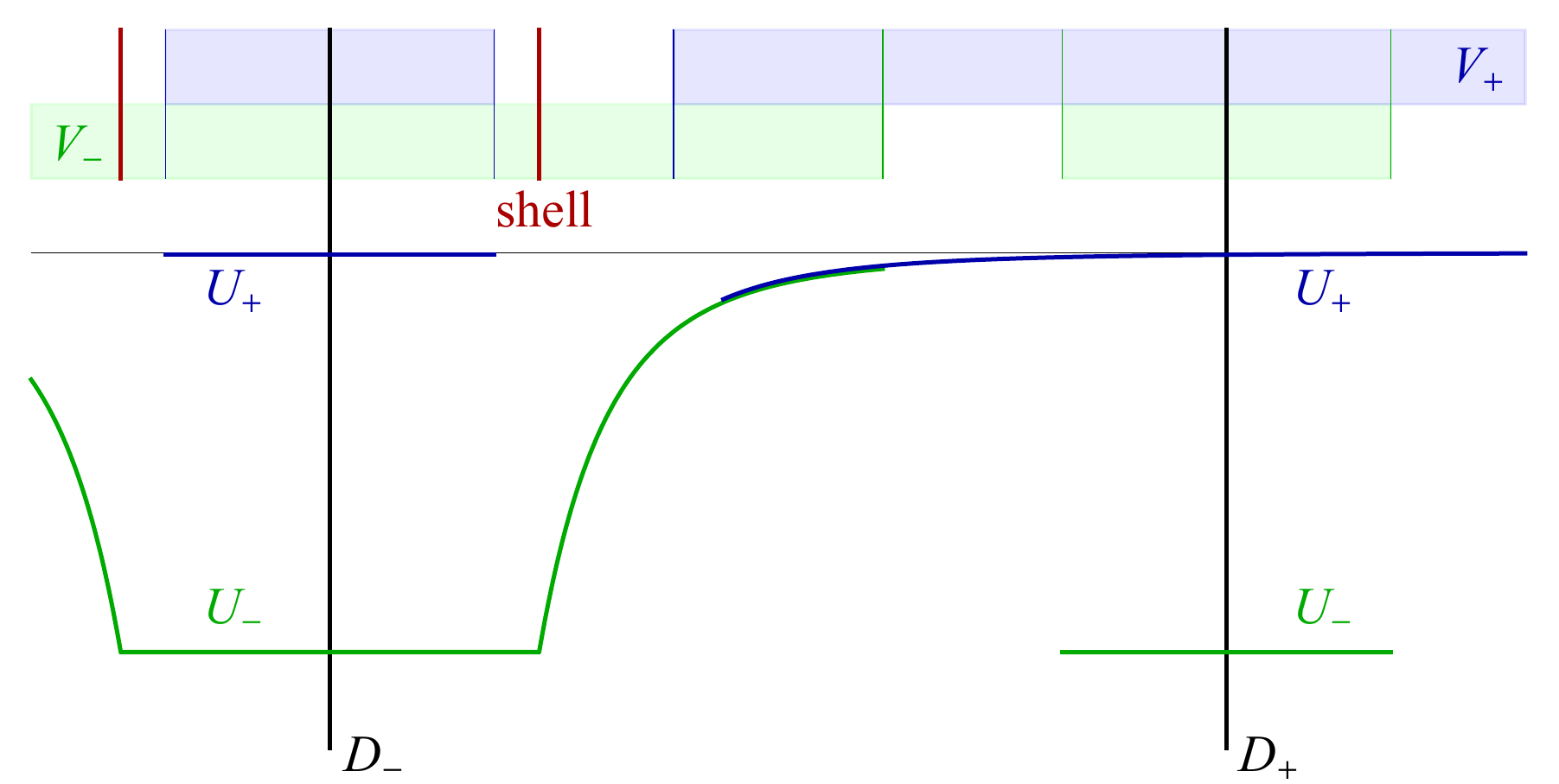}
    \caption{\label{fig:whpot0}Diagram shows the zeroth order approximation of potentials $\pot_\pm$ drawn along the axis $z$. The potentials $\pot_-$ and $\pot_+$ are continuous on the domains $V_-$ and $V_+$, respectively. They coincide on the intersection domain $\bar{V}$ but differ by a constant $\delU$ on the intersection domain $\hat{V}$. The fall-off of the potential is exaggerated to compensate a small distance between mouths. Top part of the diagram indicates the domains $V_-$ and $V_+$ and the position of the wormhole and of the massive shell.}
\end{figure}

However, even in the zeroth approximation, we have to extend the potential through the wormhole. We perform that separately in the domain ${V_-}$ around ${D_-}$ and in ${V_+}$ around ${D_+}$. We thus obtain two potentials, $\pot_-$ and $\pot_+$, each of which satisfies the Laplace equation. On the intersection $\bar{V}$, both potentials agree. In the zeroth-order approximation, the potential $\pot_-$ is constant inside the shell around the mouth ${D_-}$,  and it thus leaks as constant $\pot_\oix$ `behind' the wormhole, into the domain $\hat{V}_-$. On the other hand, the potential $\pot_+$ is negligible around ${D_+}$ in the zeroth order, and thus nothing leaks `behind' the wormhole into the domain $\hat{V}_+$. See Fig.~\ref{fig:whpot0}.

Clearly, the potentials differ around the mouths as
\begin{equation}\label{potdiff}
    \pot_+ = \pot_- + \delU\;,
\end{equation}
and in this order of approximation we obtain
\begin{equation}\label{potdiff0ord}
   \delU = -\pot_\oix = - m\, \Zo(\chi_\oix)\quad \text{in $\hat{V}$}\;.
\end{equation}
It indicates that the global potential is not well defined and we deal with a locally static spacetime. The globally well defined quantity is the geometry.

The perturbed metrics in the domains $V_-$ and $V_+$ are
\begin{equation}\label{pertmet}
    ds^2_\pm =  - e^{2\pot_\pm}\, dt_\pm^2 + e^{-2\pot_\pm}\, dl_\pm^2\;. 
\end{equation}
We want that these two metrics describe one globally well-defined geometry. Therefore, they have to coincide as tensors on both intersections $\bar{V}$ and $\hat{V}$. Their identification will be simple on the intersection $\bar{V}$, since the potentials $\pot_-$ and $\pot_+$ coincide here. We can express the both metrics just in global cylindrical coordinates ${(\rho, z, \ph)}$ on $\bar{V}$, cf.~\eqref{metcylcoor}.

However, the potentials differ on other intersection $\hat{V}$. Therefore, we have to adjust the coordinate identification in this domain to match both metrics. By identifying the proper time of the static observers, we get
\begin{equation}\label{dtpmrel}
  e^{\pot_-}dt_-=e^{\pot_+}dt_+\;.
\end{equation}
Identifying spatial distance, we find
\begin{equation}\label{dlpmrel}
  e^{-\pot_-}dl_-=e^{-\pot_+}dl_+\;.
\end{equation}
Here, the background proper lengths $dl_-$ and $dl_+$ should be expressed using \eqref{mtrcsphroidal} in appropriate spheroidal coordinates ${(\chi_-, \tht_-, \ph_-)}$ and ${(\chi_+, \tht_+, \ph_+)}$, respectively. Relation \eqref{dlpmrel} then gives the identification of coordinates ${(\chi_\pm, \tht_\pm, \ph_\pm)}$ in the domain $\hat{V}$, which guarantees the matching geometry. The resulting relation is, however, rather complicated.

The identification of coordinates is simpler if expressed in cylindrical coordinates. For that, however, we have to introduce gravitationally modified cylindrical coordinates `leaked' behind the wormhole. Let's consider first coordinates ${(\chi_-, \tht_-, \ph_-)}$ around the wormhole mouth $D_-$. In the domain $\bar{V}_-$ (above the mouth), they are related to the original global cylindrical coordinates ${(\rho, z, \ph)}$ as \eqref{rrtt}. However, behind the mouth, in the domain $\hat{V}_-$, they are related by analogous relations \eqref{rrttleak} to new `leaked' cylindrical coordinates ${(\rho_-, z_-, \ph_-)}$. Of course, the flat background metric $dl_-^2$ is given on both sides of the wormhole throat by \eqref{metcylcoor}
\begin{equation}\label{dl2min}
\begin{aligned}
    dl_-^2\big|_{\bar{V}_-} &= d\rho^2 + dz^2 +\rho^2 d\ph^2 \;,\\
    dl_-^2\big|_{\hat{V}_-} &= d\rho_-^2 + dz_-^2 +\rho_-^2 d\ph_-^2\;.
\end{aligned}
\end{equation}
Similarly, $dl_+^2$ is related by \eqref{rrtt} to ${(\rho, z, \ph)}$ in $\bar{V}_+$ and by \eqref{rrttleak} to new `leaked' coordinates ${(\rho_+, z_+, \ph_+)}$ in domain $\hat{V}_+$, and
\begin{equation}\label{dl2pl}
\begin{aligned}
    dl_+^2\big|_{\bar{V}_+} &= d\rho^2 + dz^2 +\rho^2 d\ph^2 \;,\\
    dl_+^2\big|_{\hat{V}_+} &= d\rho_+^2 + dz_+^2 +\rho_+^2 d\ph_-^2\;.
\end{aligned}
\end{equation}

Comparing \eqref{dl2min} and \eqref{dl2pl} using relation \eqref{dlpmrel}, we find
\begin{equation}\label{drhozrelVmin}
  e^{-\pot_-}d\rho_-=e^{-\pot_+}d\rho\;,\quad
  e^{-\pot_-}dz_-=e^{-\pot_+}dz\;
\end{equation}
in domain $\hat{V}_-$, and
\begin{equation}\label{drhozrelVpl}
  e^{-\pot_-}d\rho=e^{-\pot_+}d\rho_+\;,\quad
  e^{-\pot_-}dz=e^{-\pot_+}dz_+\;
\end{equation}
in $\hat{V}_+$.

Assuming, that the mouths of the wormhole are identified at times $t_-=t_{\oix-}$ and $t_+=t_{\oix+}$, the differential relation \eqref{dtpmrel} gives
\begin{equation}\label{tpmrel}
    \bigl(t_+-t_{\oix+}\bigr) = e^{-\delU}\bigl(t_--t_{\oix-}\bigr)\;,
\end{equation}
valid in $\hat{V}$.

\begin{figure}
      \includegraphics[width=\columnwidth]{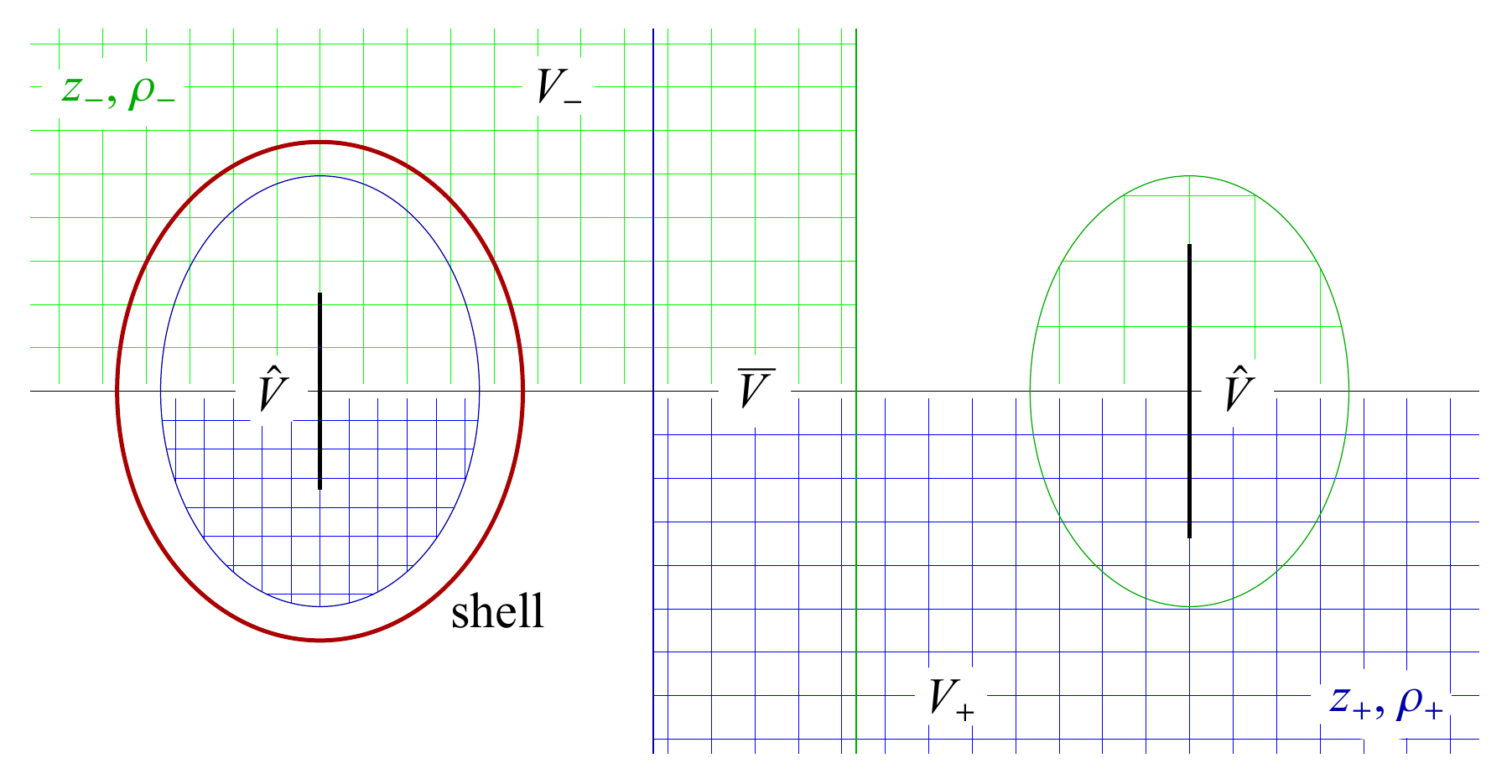}
    \caption{\label{fig:whflatcoor}Finite wormhole with distant mouths. Diagram shows coordinates $x_-^j$ and $x_+^j$ defined on the domains $U_-$ and $U_+$, respectively. Since the potential modifies the geometry, the coordinates continuously extended through the wormhole does not match the original coordinates.}
\end{figure}

For spatial coordinates, we get that in domain $\hat{V}_+$ the `leaked' cylindrical coordinates ${(\rho_+, z_+, \ph_+)}$ are related to the global cylindrical coordinates ${(\rho, z, \ph)}$ as
\begin{equation}\label{cylplrel}
\begin{gathered}
    \rho_+ = e^{\delU} \rho\;,\quad
    z_+ + \frac\ell2 = e^{\delU} \Bigl(z + \frac\ell2\Bigr)\;.
\end{gathered}
\end{equation}
Similarly, in domain $\hat{V}_-$, the `leaked' cylindrical coordinates ${(\rho_-, z_-, \ph_-)}$ are related to the global cylindrical coordinates ${(\rho, z, \ph)}$ as
\begin{equation}\label{cylminrel}
\begin{gathered}
    \rho = e^{\delU} \rho_-\;,\quad
    z - \frac\ell2 = e^{\delU} \Bigl(z_- - \frac\ell2\Bigr)\;.
\end{gathered}
\end{equation}
See Fig.~\ref{fig:whflatcoor}.

We can reformulate these relations in the rule, how the global cylindrical coordinates defined on both side of the wormhole throat should be identified through the throat. We observe that $t_+=t$ in $\bar{V}_+$ and $t_-=t$ in $\bar{V}_-$. Relation  \eqref{tpmrel} thus gives the identification of time coordinate $t= t_{\oix-} + {\Delta t}_*$ in $\bar{V}_-$ with $t=t_{\oix+} + e^{-\delU}{\Delta t}_*$ in $\bar{V}_+$ for any ${\Delta t}_*$. Similarly, we observe that the coordinates  ${(\rho, \ph)}$ approaching the mouth $D_-$ from $\bar{V}_-$, are matched by `leaked' coordinates ${(\rho_-, \ph_-)}$ approaching $D_+$ from $\hat{V}_-$. The result for the time coordinate and relations \eqref{cylminrel} thus give the identification
\begin{equation}\label{cylcoorident}
\begin{gathered}
   \bigl[t,\,\rho,\,z,\,\ph\bigr]\big|_{\bar{V}_-}=
   \Bigl[ t_{\oix-} + {\Delta t}_*,\,\rho_*,\,-\frac\ell2\pm0,\ph_*\Bigr]\\
   \updownarrow\\
   \bigl[t,\,\rho,\,z,\,\ph\bigr]\big|_{\bar{V}_+}=
   \Bigl[t_{\oix+} + e^{-\delU}{\Delta t}_*,\,e^{\Delta\phi} \rho_*,\frac\ell2\mp0,\,\ph_*\Bigr]
\end{gathered}\;,
\end{equation}
for any ${\Delta t}_*$, $\rho_*$ and $\ph_*$. The term $\pm0$ distinguishes the faces of the mouths.

It means that the identification of points through the wormhole throat changes under the gravitational perturbation. It causes a complication with the size of the wormhole. To keep the proper size of the wormhole mouths $a$ the same from both sides, one has to choose a different coordinate radius for each mouth, $a_-$, and $a_+$, such that
\begin{equation}\label{whrad}
    a = e^{-\pot_-}\, a_- = e^{-\pot_+}\, a_+\;,\quad\text{i.e.,}\quad a_+ = e^{\delU}\, a_-\;.
\end{equation}

Notice that the change in the coordinate size of the wormhole mouths is of the first order in the gravity perturbations and it produces only next-order perturbations to the gravitational potential. Thus, we can continue to use the potential calculated on the original wormhole background.

Let us summarize. We have perturbed the background with metric \eqref{flatbackgr} and found the perturbed metrics \eqref{pertmet}. Since the potentials ${\pot_-}$ and ${\pot_+}$ cannot be extended to a global function on the background spacetime, the perturbation has to be performed separately on domains ${\bar{V}_-}$ and ${\bar{V}_+}$. In these domains we can use global cylindrical coordinates ${(t,\rho,z,\ph)}$ to identify the background spacetime with the perturbed spacetimes.

The perturbed metrics  ${ds_-^2}$ and ${ds_+^2}$ can be easily extended to domain ${V_-}$ and ${V_+}$ through the wormhole throat, using, for example, spherodial coordinates ${(t_-,\chi_-,z_-,\ph_-)}$ and  ${(t_+,\chi_+,z_+,\ph_+)}$, respectively. These  metrics match trivially on the intersection ${\bar{V}}$, as can be seen in global coordinates ${(t,\rho,z,\ph)}$, since the potentials are equal here.

However, since the potentials differ on the other intersection ${\hat{V}}$, the metrics cannot be matched without a proper identifications of the coordinates. Because the potentials  ${\pot_-}$ and ${\pot_+}$ differ only by a constant, cf.~\eqref{potdiff}, the metrics ${ds_-^2}$ and ${ds_+^2}$ on ${\bar{V}}$ differs only by a rescaling of time and spatial directions. It can be demonstrated using cylindrical coordinates ${(t_\pm,\rho_\pm,z_\pm,\ph_\pm)}$ introduced `behind' the wormhole. Namely, the global cylindrical coordinates are related to `leaked' cylindrical coordinates by a rescaling \eqref{cylplrel} and \eqref{cylminrel}.

However, this shows that we need to perform perturbation of the background in two domains independently, and only after that we can glue together the global geometry representing a weak gravitational field. As the result, the spacetime is not globally static but only locally static spacetime.

\subsection{Large distance approximation -- 1st order}

Let us now improve our approximation of the shell field in the wormhole spacetime. We now include an interaction of the field with the wormhole in the first order of~$\frac1\ell$. We start with some geometrical preliminaries.

The oblate spheroidal coordinates ${(\chi_-,\,\tht_-,\,\ph_-)}$ and  ${(\chi_+,\,\tht_+,\,\ph_+)}$ introduced around the mouths $D_-$ and $D_+$ are related by \eqref{sphcoorpmrel}. We evaluate it at the mouth $D_+$ where $\chi_+=0$
\be \n{Dm}
\sinh\chi_- \cos\tht_-= \ell   \;,\quad \cosh\chi_-\sin\tht_- =\sin\tht_+ \;.
\ee
Solving these equations one finds
\be
\sin\tht_+=\sqrt{ \frac{\ell^2}{\cos^2\tht_-}+1}\,\sin\tht_-\ge \ell\,\sin\tht_-\, .
\ee
The value of $\sin\tht_+$ is restricted by 1, thus the angle $\tht_-$ evaluated at $D_+$ is small, $\tht_-\big|_{D_+} \ll 1/\ell$.  Substituting this result back to \eqref{Dm} we obtain that up to the quantity of order $O(\ell^{-2})$ the value of $\sinh\chi_-$ on the mouth $D_+$ is constant and equal to~$\ell$.

From the discussion of the first order approximation, we expect that the potential $\pot$ of the shell around the wormhole mouth $D_-$ is not globally defined. As we discussed in section \ref{sc:weakgrav}, we can define the potential $\pot$ using integration \eqref{UINF} along a path from infinity not crossing the wormhole throat. It will not be continuous at the throat of the wormhole, but it is smooth everywhere else. And although the potential is not continuous through the wormhole, the field strength $-\nabla\pot$ is globally defined and smooth. We want to find an approximation of this potential.

In the zeroth approximation of the potential we assumed that the field \eqref{thinshellmin} is negligible at the mouth~$D_+$. Now, we take into account, that it is not negligible here. We denote this potential as truncated, $\pot_\sht$. Equipotentials of $\pot_\sht$ are surfaces of $\chi_-=\const.$ Therefore, it follows from the discussion above, that the field \eqref{thinshellmin} can be treated as homogeneous near the mouth $D_+$ in the next order of approximation.

We can thus use our discussion from Sec.~\ref{ssc:whhomfield}. Here we learnt that the truncated homogeneous field around the mouth of the wormhole has to be compensated by the field \eqref{indpot} adjusted to the mouth at $z=\frac\ell2$.

This field is nonvanishing on both sides of the wormhole, it thus changes the field not only in domain $\bar{V}_+$ `above' the mouth $D_+$, but also in domain $\hat{V}_+$ `below' the mouth. Above the mouth $D_+$ we use expression \eqref{indpot} employing spheroidal coordinates ${(\chi_+,\,\tht_+,\,\ph_+)}$. Below the mouth, we transfer the induced field into spheroidal coordinates ${(\chi_-,\,\tht_-,\,\ph_-)}$, using relations\footnote{We do not employ here the corrections of the coordinate matching discussed after \eqref{dlpmrel} in the previous section (which we also expressed using cylindrical coordinates in \eqref{cylcoorident}) since it would produce corrections of higher order.} \eqref{sphcoorpmrel}. In these domains, values of relevant spheroidal coordinate $\chi_\pm$ are positive, and both sets of coordinates can be understood as functions of the global coordinates ${(\rho,\,z,\,\ph)}$, cf.~\eqref{rrtt}.

We can thus write the induced field $\pot_\ind$ outside the wormhole throat as
\begin{equation}\label{indpotm}
    \pot_\shi = \hmf\,\bigl(\Zm(\chi_-)\,\cos\tht_- - \Zm(\chi_+)\,\cos\tht_+ \bigr) \;.
\end{equation}
The corrected field of the shell thus reads
\begin{equation}\label{thinshel1ord}
    \pot = \pot_\sht + \pot_\shi
\end{equation}
with $\pot_\sht$ given by \eqref{thinshellmin} and $\pot_\shi$ given by \eqref{indpotm}.

The constant $\hmf$ in the induced field has to be fixed by the condition, that the field strength should by continuous through the wormhole. Namely, we require
\be \n{CONT}
\begin{aligned}
\frac{dU}{dz}\bigg|_{D_+^<} &= \frac{dU}{dz}\bigg|_{D_-^>}\, ,\\
\frac{dU}{dz}\bigg|_{D_+^>} &= \frac{dU}{dz}\bigg|_{D_-^<}\, .
\end{aligned}
\ee

In our approximation, we assume that the field is homogeneous and thus, it is sufficient to check these conditions just along the $z$-axis. Relation \eqref{rrttleak} gives
\be
   z\mp \ell/2=\sinh\chi_{\pm}\, \cos\tht_\pm\;,
\ee
where along the axis, values of $\cos\tht_\pm$ are just $-1$ or $+1$ depending on a position with respect of the wormhole mouths. Hence
\be \n{chiz}
{d\chi_{\pm}\over dz}=\frac{\cos\tht_\pm}{\cosh\chi_{\pm}}\; .
\ee
Using these relations, we can calculate $z$-derivatives of the potential $\pot$ at the mouths of the wormhole. Actually, it turns out that it is the same on both faces $D^<$ and $D^>$ of the mouth. Using relations \eqref{thinshellmin}, \eqref{indpotm}, \eqref{Z0prop} and \eqref{Zpmdef} one finds
\begin{align}\label{Uders}
\begin{aligned}
\frac{dU}{dz}\bigg|_{D_+} &= \frac{m}{\ell^2+1}-\frac{\hmf}{2}\bigl(1+V_{\ell}\bigr)\, ,\\
\frac{dU}{dz}\bigg|_{D_-} &= \frac{\hmf}{2}\bigl(1+V_{\ell}\bigr)\, .
\end{aligned}
\end{align}
Here $V_\ell$ is coming from the derivative of $\Zm'(\ell)$, i.e., form contribution of the induced field at distant wormhole mouth
\be
V_{\ell}={2\over \pi}\left( \arctan\ell -\frac{\pi}{2} +{\ell\over \ell^2+1}\right)=-\frac{4}{3\pi \ell^3}+\OO(\ell^{-5}) .
\ee

Now, it is easy to check that the continuity conditions \eqref{CONT} are satisfied provided
\be\label{ww}
\hmf={m\over (\ell^2+1)(1+V_{\ell})}=\frac{m}{\ell^2}+\OO(\ell^{-4}) .
\ee
We see that the contribution of the induced field is rather small.

\begin{figure}
  \includegraphics[width=0.8\columnwidth]{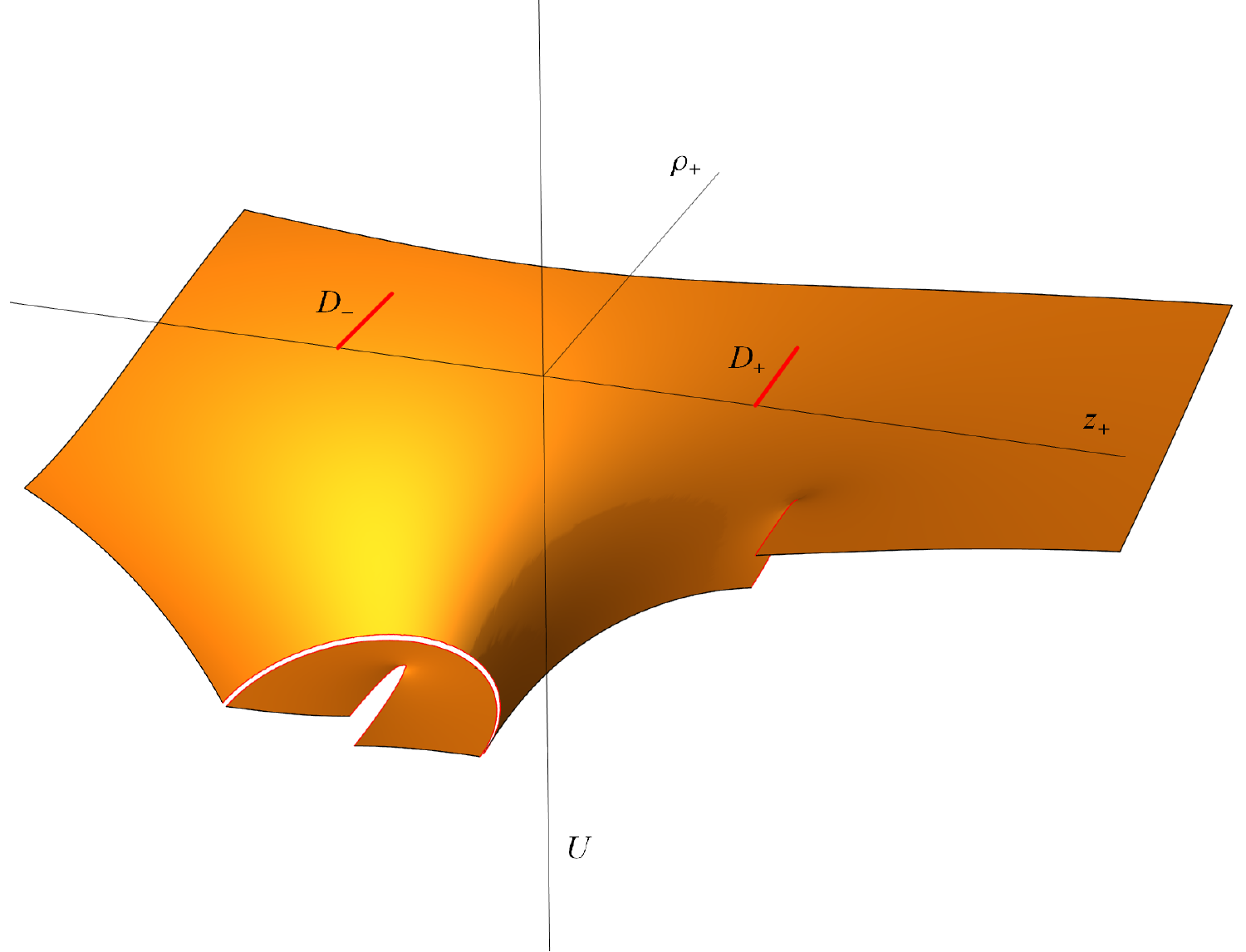}\\
  \includegraphics[width=0.8\columnwidth]{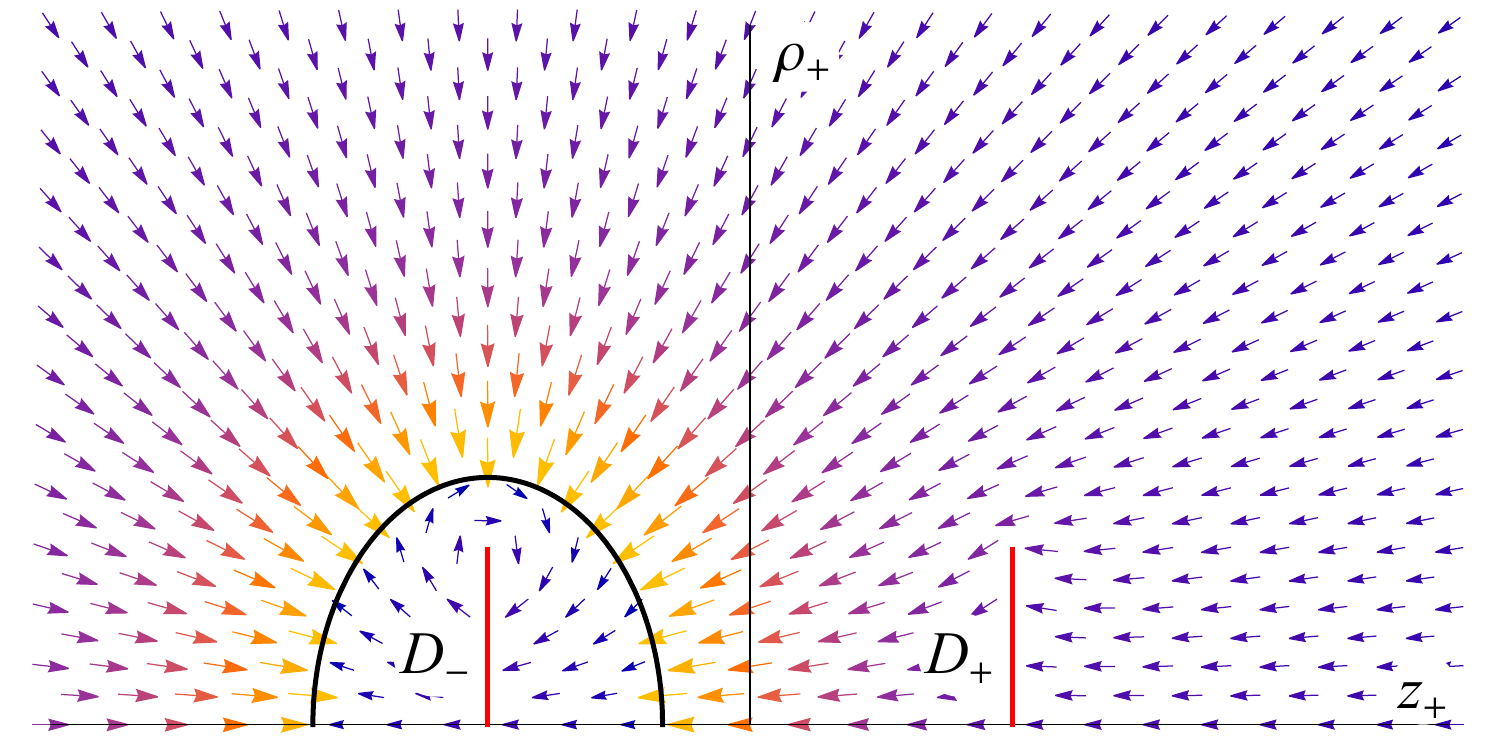}
    \caption{\label{fig:shell1ord}Potential $\pot$ [top] and field strength [bottom] of the thin massive shell localized around the wormhole mouth $D_-$. The potential depicted here is smooth outside the wormhole throat, but it is discontinuous at the throat. The zeroth order value $\pot_\sht$ of the potential corresponds to the shell in space without a wormhole. The correction given by induced field $\pot_\shi$ makes the field strength smooth through the wormhole throat.}
\end{figure}

The potential \eqref{thinshel1ord} is shown in figure \ref{fig:shell1ord}. As we emphasized, it is not globally smooth, it has discontinuity across the wormhole throat.

Similarly as we did in the zeroth order approximation, it can be extended as a solution of the homogeneous Laplace equation through each wormhole mouth, obtaining thus two potentials $\pot_-$ and $\pot_+$ defined in domains $V_-$ and $V_+$, respectively. However, in this case we have to adjust the induced field $\pot_\shi$ to guarantee also the continuity of the potential. It can be achieved by adding a suitable constant to the extended field `behind' the mouth. We obtain
\begin{equation}\label{shellwhextp}
  \pot_+ =
  \begin{cases}
    \pot_\sht + \pot_\shi \quad&\text{in $\bar{V}_+$}\,,\\
    \pot_\sht\big|_{z=\frac{\ell}{2}} + \pot_\shi \quad&\text{in $\hat{V}_+$}\,,\\
  \end{cases}
\end{equation}
and
\begin{equation}\label{shellwhextm}
  \pot_- =
  \begin{cases}
    \pot_\sht + \pot_\shi \quad&\text{in $\bar{V}_-$}\,,\\
    \pot_\oix + \pot_\shi \quad&\text{in $\hat{V}_-$}\,.\\
  \end{cases}
\end{equation}
Here $\pot_\oix=\pot_\sht|_{z=-\frac{\ell}{2}}=m\Z(\chi_\oix)$ is the value of the truncated potential inside the shell `leaking' behind the mouth $D_-$ -- the effect which we encountered already in the zeroth order approximation..

The constant
\begin{equation}\label{constleakingDp}
   \pot_\sht\big|_{z=\frac{\ell}{2}} = -m\arccot\ell = - \frac{m}{\ell} + \OO(\ell^{-3})
\end{equation}
`leaking' through the mouth $D_+$ into $\hat{V}_+$ appears only in the first order of approximation. Comparing with \eqref{ww}, we observe that it is actually more significant than the induced field $\pot_\shi$ itself.

\begin{figure}
  \includegraphics[width=\columnwidth]{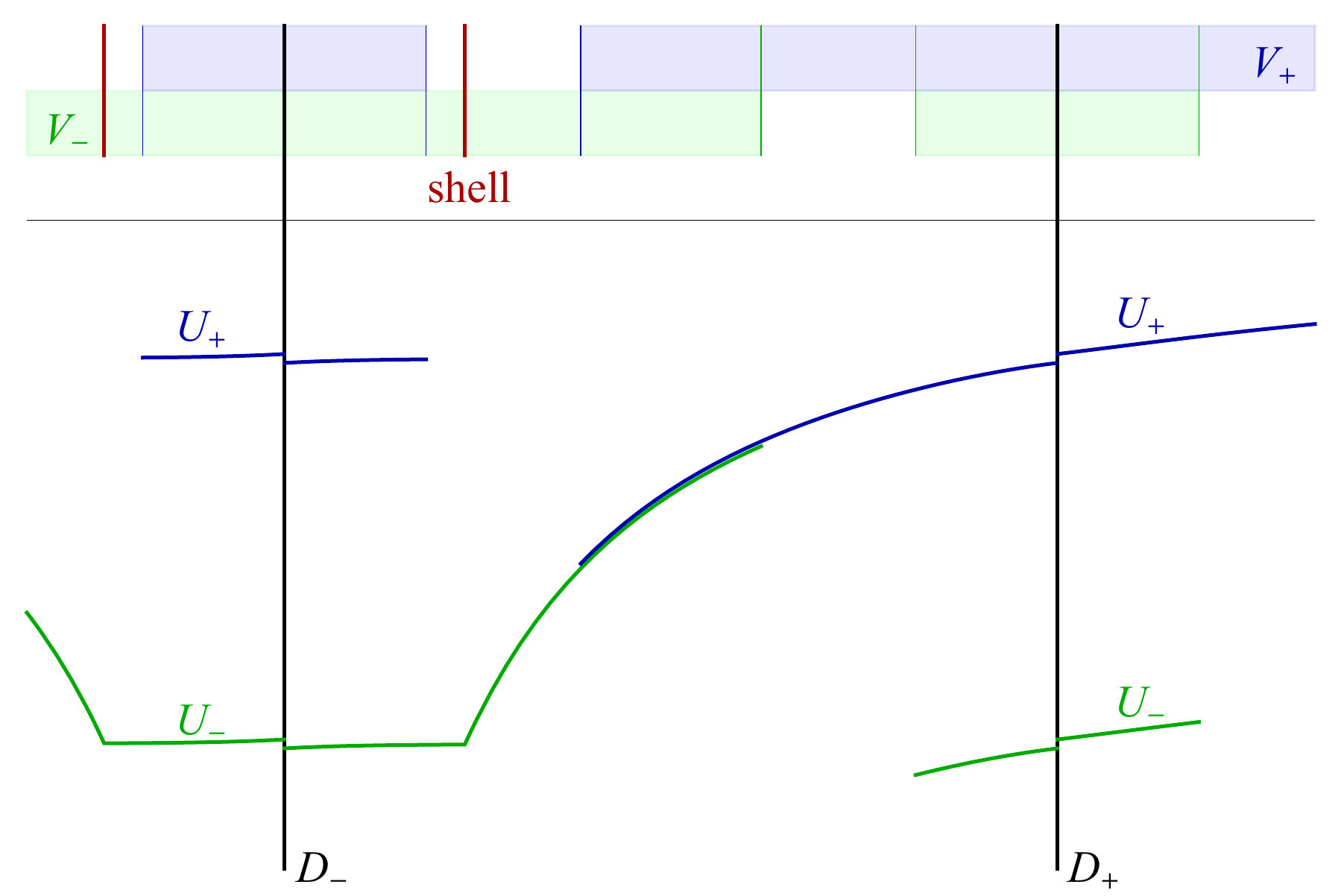}
    \caption{\label{fig:whpot1}Diagram shows the first order approximation of potentials $\pot_\pm$ drawn along the axis $z$. The potentials $\pot_-$ and $\pot_+$ are continuous on the domains $V_-$ and $V_+$, respectively. The potentials include the induced field correcting the zeroth-order contribution on the wormhole throat. The both potentials coincide on the intersection domain $\bar{V}$ but differ by a constant $\delU$ on the intersection domain $\hat{V}$. Top part of the diagram indicates the domains $V_-$ and $V_+$ and the position of the wormhole and of the massive shell.}
\end{figure}

The both potentials $\pot_-$ and $\pot_+$ along $z$-axis are depicted in Fig.~\ref{fig:whpot1}.

We can estimate the difference between both potentials. Analogically to \eqref{potdiff} we obtain the correction to \eqref{potdiff0ord}
\begin{equation}\label{potdiff1ord}
\begin{split}
  \delU &= 
    \pot_\sht\big|_{z=\frac{\ell}{2}} - \pot_\oix
    = - \pot_\oix -\frac{m}{\ell} +\OO(\ell^{-2})
    \\&= m \Bigl(\arccot\sinh\chi_\oix - \frac1\ell + \OO(\ell^{-2})\Bigr)
    \;.
\end{split}
\end{equation}

Since the jump $\delU$ plays a role in the identifications of the geometry through the wormhole \eqref{cylcoorident}, namely in the identification of the time coordinate \eqref{tpmrel}, it is worth to investigate it more from a different point of view.

\subsection{Invariant $I_C$ calculation}

In a multiply connected space with a non-potential locally static gravitational field one can define the following integral
\be\n{IINN}
I_C=\oint_C w_i dx^i\;.
\ee
Here $C$ is a closed contour, and $w_i$ is a three-dimensional co-vector of the acceleration. This integral is a topological invariant in the following sense. The integral \eqref{IINN} is the same for any two closed paths, which can be obtained one from another by a continuous transformation.

Clearly, for a globally potential field ${w_i=U_{,i}}$, this integral vanishes. For a locally potential field which is not potential globally, it estimates the non-potentiality of the field ${w_i}$. If evaluated along a path between different points which belongs to a domain, where the potentiality holds, it gives a potential difference. Therefore, we can conclude, that for a closed path it estimates a potential difference which is an obstacle for a global definition of the potential,
\begin{equation}\label{ICdU}
    \delU = I_C\;.
\end{equation}

\begin{figure}[!hbt]
    \centering
      \includegraphics[width=0.3\textwidth]{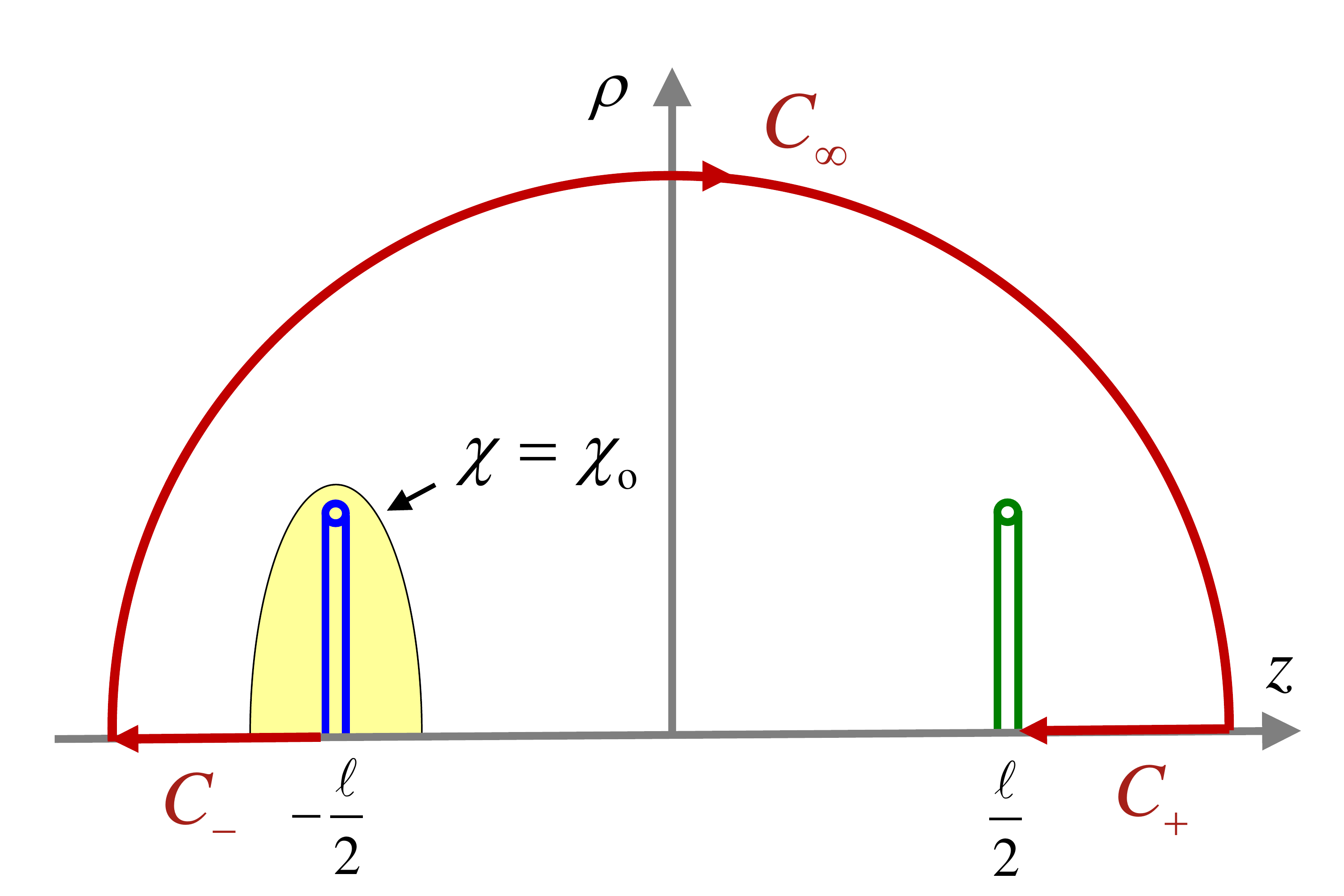}  
    \caption{Contour $C$}
    \label{INV}
\end{figure}

Let us calculate the invariant $I_C$ for situation discussed in this section: the wormhole space $R_\wh$ with one of its mouths surrounded by a massive thin shell. We choose a closed $C$ which passes through the wormhole. Since its value does not depend on a special choice of such a path we use for $C$ the contour shown at Fig.~\ref{INV}. It consists of three parts $C_-$, $C_+$ and $C_{\infty}$. We choose $C_-$ and  $C_+$ to be intervals along $z-$ axes and
$C_{\infty}$ to be a circle $R\equiv \sqrt{\rho^2+z^2}=$const connecting the end points of these intervals. In the limit $R\to\infty$ the contribution of the part $C_{\infty}$ to  $I_C$ vanishes.

For the other parts of the contour $C$ one has
\ba
I_- =&\int_{0}^{\infty} w_{\chi_-} d\chi_-=U|_{D_-^<}^{\chi_-=\infty}\\
&=-\pot_\sht|_{\chi_-=\chi_\oix} -\pot_\shi|_{D_-^<},\\
I_+ =-&\int^{\infty}_{0} w_{\chi_+} d\chi_+  =-U|_{D_+^>}^{\chi_+=\infty}\\
=&\pot_\sht|_ {D_+^>} +\pot_\shi|_{D_+^>},
\ea
The induced potential $\pot_\shi$ (\ref{indpotm}) is a symmetric functions with respect to reflection $\chi_- \rightleftarrows \chi_+, ~\tht_- \rightleftarrows \pi-\tht_+$ and, hence,  satisfies the condition
\begin{equation}
\pot_\shi|_{D_-^<}=\pot_\shi|_{D_+^>}\;,
\end{equation}
therefore it cancels in the sum of $I_-$ and $I_+$. Then we obtain
\ba
I_C=& -m[\Z(\chi_\oix)+\arccot\ell ] \;.
\ea
For $\ell\gg 1$ we finally get
\ba
I_C=& m\Big[\arccot\sinh\chi_\oix -\frac{1}{\ell}\Big]+\OO(\ell^{-2}) \, .
\ea
One can check that when $\sinh\chi_\oix\ll \ell$ the integral $I_C$ is positive. The main contribution comes from the first term.
The work along the closed path passing through the wormhole does not vanish and the gravitational field is non-potential.

\section{Time machine formation}

\label{sec:TMformation}

\begin{figure}
    \centering
      \includegraphics[width=0.25\textwidth]{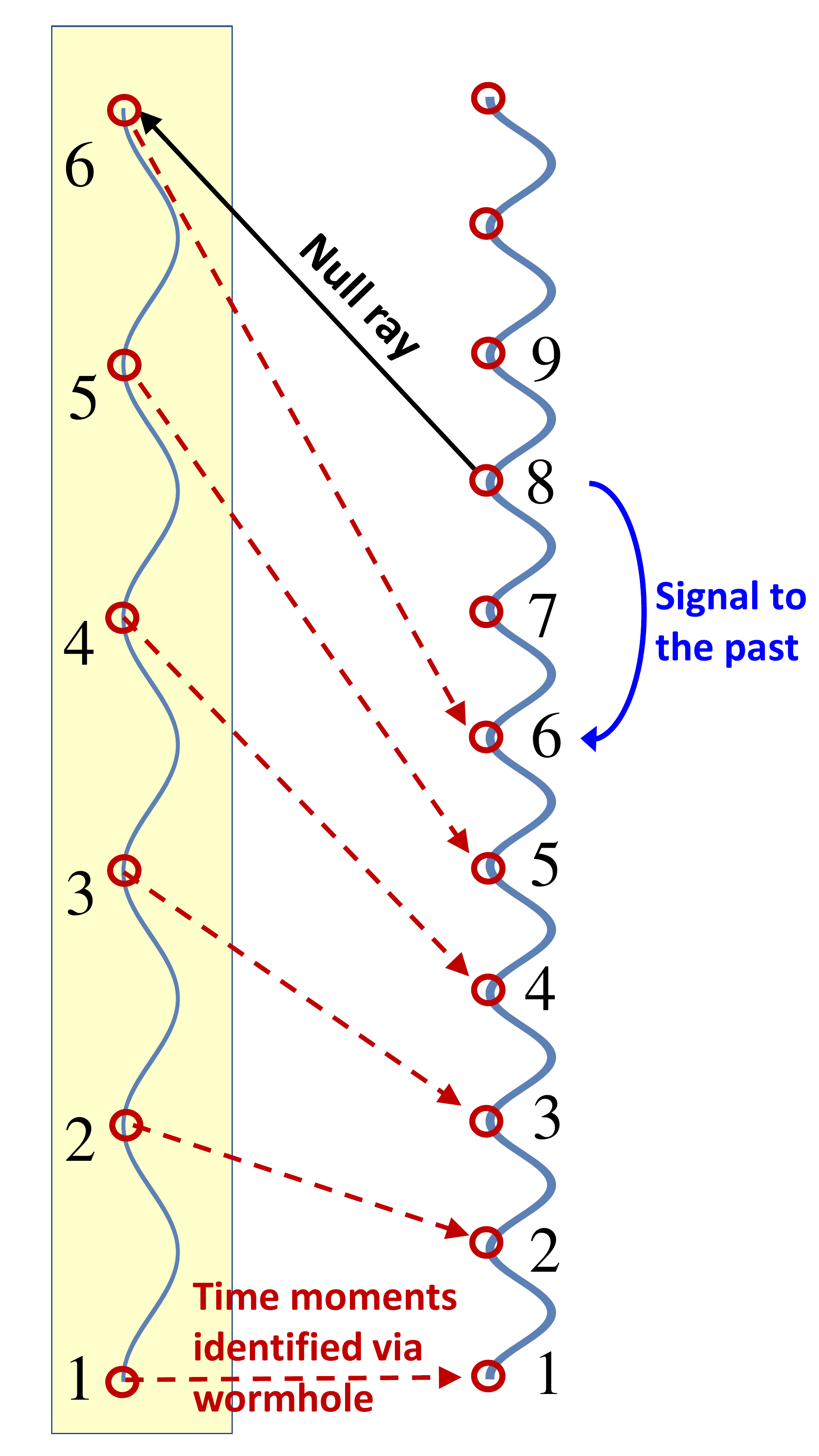}
    \caption{Red dashed lines mark the connection of events via wormhole. Before the massive shell was turned on around $D_-$ the events are globally synchronized. After an adiabatic formation of a massive shell the pace of time at  $D_-$ slows down as seen by an observer at infinity and the mouths of the wormhole become desynchronised from his/her point of view. When the time difference accedes $\ell/c$ the chronology horizon forms. After that the time machine appears. }
    \label{Fig17}
\end{figure}

Let us show that in the spacetime with a ring wormhole, where one of its mouthes is surrounded by matter, closed timelike curves are created. Assume that $D_-$ mouth of the wormhole is inside the massive shell while $D_+$ mouth is located far away outside the shell (see Fig.~\ref{Fig17}) where its gravitational field is small. Because these mouthes are identified through the wormhole the proper times at  $D_-$ and $D_+$ are the same. Hence, the coordinate times $t_-$  at $D_-$ and $t_+$ at $D_+$ are related as (\ref{tpmrel}). In this setup $\Delta U=I_C>0$. It means that for the same proper time interval the coordinate time
at the disk $D_-$ which is inside the massive shell spans more than that at $D_+$. Let us assume that clocks at the disks are synchronized  at the moment $t=0$. If a particle enters $D_-$ at time $t_1$  it appears from $D_+$ at slightly different time $t_2$. Denote the time gap $\Delta t =t_2 - t_1$. Then in such a process the time gap is negative
\begin{equation}\label{Dt}
\Delta t=-\big[e^{I_C}-1\big] t_2 <0\;.
\end{equation}

Let us consider the following experiment. Suppose a null ray emitted at some time $t_0$ at the point with coordinate $z=0$  propagates along $z-$axis to the left, enters the disc $D_-$ at time $t_1$ and after it appears from $D_+$ at the time $t_2$ returns to the point of its emission at the time $t_3$. Let us find how the corresponding time $t_3$ (as measured by an observer at infinity) depends on $t_0$.
For the metric (\ref{MMUU}) the equation of the null ray moving along $z-$axis is
\be  \n{ZZUU}
{dz\over dt}=- e^{2U(z)}\, ,
\ee
where $U(z)$ is a value of the potential $U$ on $z$-axis. A sign minus in the right-hand side of this equation indicates that  at both stages (motion to $D_-$ and motion from $D_+$) the ray moves in the direction opposite to the direction of $z$-axis. Using (\ref{ZZUU}) one finds that the time of the arrival of the ray to $D_-$ is
\be \n{TTM}
t_1=t_0+\int_{-\ell/2}^0 e^{-2U} dz\, .
\ee
Similarly, time $t_3$ of arrival to the original point $z=0$ after the ray appears from $D_-$ at time $t_2$ is
\be \n{TTP}
t_3=t_2 +\int_0^{\ell/2} e^{-2U} dz\, .
\ee
Using equations (\ref{Dt}), (\ref{TTM}) and (\ref{TTP}) one finds
\ba
&t_3-t_0=-\big[e^{I_C}-1\big] t_2+B\; ,\\
&B=\int_{-\ell/2}^{\ell/2} e^{-2U} dz>0  \, .
\ea
For a chosen direction of null ray motion $I_C>0$ . Hence, $t_3-t_0$ is a decreasing function of the time $t_2$ at $D_+$ which for large $\ell$ is close to the time of an observer at infinity. This equation shows that  $t_3$ becomes lesser than $t_0$ when
\be
t_2\ge T, \hskip 1cm T= \frac{B}{e^{I_C}-1}\, .
\ee
Because $B>0$ and $I_C>0$ this means that for sufficiently late time $t_2$ the null ray arrives at the initial point $z=0$ earlier than it was emitted and a closed time curve is formed.

 In the weak field approximation adopted in this paper $I_C$ is small and $B\approx \ell$. Then formation of the time machine happens after time
 \be
 T\approx \frac{\ell}{I_C}\; .
 \ee
Let $r=\sinh\chi_\oix$ be a characteristic  size of the massive thin shell surrounding the $D_-$ mouth of the ring wormhole.
Then the characteristic time of the transformation of the wormhole into the time machine is
\be
T\approx {r\ell/m}\; .
\ee
Restoring the dimensions one can write this expression in the form
\be
T\approx {R L c\over GM}\; .
\ee
Here $T$ is the characteristic time of closed timelike curves formation in the initially traversable ring wormhole surrounded by a massive thin shell of radius $R$ and mass $M$. $L$ is the distance between mouths.

Let us emphasize, that the main conclusion concerning a closed time-like curves formation in the wormhole spacetime is a rather robust property. Such curves arise as a result of the asymmetry of the mass distribution surrounding its two mouths.

For more details concerning non-potential gravitational field and time machine formation in such fields see \cite{Frolov:1990si,Visser:1995cc,Frolov:1998wf,FrolovZelnikov:2011}.

\section{Discussion}

\n{S6}

In the present paper we study the gravitational field in a spacetime with a traversable ring wormhole. We considered two types of such wormholes: (i) a wormhole connecting to flat spaces, and (ii) a wormhole connecting two distant domains in the same space. We focused on study solutions of the gravity equations in the weak field approximation in the presence of matter in these spaces. For the wormhole connecting two flat spaces we demonstrated that there exist zero mode solutions describing the gravitational field trapped by the wormhole. Such a field obeys the homogeneous  Laplace equation. One can relate them to a proper choice of the matter distribution located at the infinity either of $R_+$, or/and $R_-$  spaces. We also obtained a solution for the gravitational field of the massive thin oblate spheroidal shell confocal to the disc representing the ring wormhole. A main important feature of such a solution is that one cannot impose a condition that the corresponding gravitational potential $U$ vanishes at both infinities in $R_{\pm}$. There always exists a non-vanishing difference of these potentials which is proportional to the mass $m$ of the shell and depending on its size. We also found an exact solution for the case when a ring wormhole is immersed in the gravitational field which is homogeneous at $R_+$ infinity. Such a solution contains the dipole type component which describes the modification of the homogeneous field by the wormhole.

In section~\ref{sc:oneasympWHspc} we study the gravitational field of a massive thin shell surrounding one of the mouths of the ring wormhole connecting two distant domains in a single space. We solved the gravity equation in the approximation when the distance $L$ between the mouths is much larger that the ring radius $a$. A main property of this solution that the gravitational potential $U$ becomes a multi-valued function. This happens because the space is multi-connected.  While the strength of the gravitational force, $\vec{w}$ is well defined and unique, a solution of the equation $\nabla U=-\vec{w}$ does not posses this property. One can integrate this equation and find the function $U$ along any chosen curve connecting two points. But the value of $U$ may be different for two different paths which cannot be transformed one to the other by means of a continuous transformation. As a result the gravitational field is non-potential. In such a case there exists a growing  time gap for clocks synchronized along specially chosen non-contractible closed paths. This implies that a closed timelike curves formation occurs in such a space. This is a mechanism which transforms an original traversable wormhole without a  time-gap  into a time machine.

Let us emphasize that above described properties of the ring wormholes are similar to well known properties of the `standard' wormholes with a spherical topology of their throat. A main difference is that for the ring wormhole an observer passing through it moves in a flat (or practically flat spacetime), while in the case of `standard' wormholes he/she should pass a domain filled with the matter violating the null energy condition.  In this paper we focused
on the classical aspects of the ring wormhole model. It is well known that for the `standard' wormholes the quantum effects are important and may dramatically change their property in the regime of the time-machine formation. Namely, it was shown \cite{Kim:1991mc,Frolov:1991nv} that at the moment close to the time when closed timelike curves are formed, the renormalized quantum average of the stress-energy tensor of a quantum field infinitely grows. Hawking formulated a chronology protection conjecture \cite{Hawking:1991nk}, according to which the back-reaction  of quantum effects should always forbid  formation of closed timelike curves.  There exists quite a lot of papers where different aspects of the role of quantum effects in wormholes and time machines are discussed (see e.g.
\cite{Visser:1992py,Politzer:1992zm,Visser:1992tx,Krasnikov:1995jn,Kay:1996hj,Gonzalez-Diaz:1996iea,Visser:2002ua,Emparan:2021yon}). It would be interesting to study quantum effects in the ring wormholes and time-machines. These models are rather simple from the mathematical point of view and they may provide one with a simple analytical tools for study such a quantum mechanical problem.

\section*{Acknowledgments}

The authors V.F. and A.Z. thank the Natural Sciences and Engineering Research Council of Canada and the Killam Trust for their financial support.
P.K.\ was supported by Czech Science Foundation Grant GA\v{C}R~22-14791S.

\appendix

\section{Static spacetime}\label{apx:statST}

\subsection*{Metric}

Let us consider a static spacetime. It is locally described by metric
\begin{equation}\label{axeqstaticmetric}
\begin{gathered}
  ds^2=g_{\mu\nu}dx^{\mu} dx^{\nu} = -e^{2\pot} dt^2 +dq^2\;,\\
  dq^2=q_{ij}dx^i dx^j\;,\\
  \pa_t \pot= 0 \;,\quad \pa_t q_{ij}=0 \;.
\end{gathered}
\end{equation}
We denote $\xi^\mu=\delta_t^\mu$  a time-like Killing vector and $u^{\mu}$ a four-velocity of static observers moving along this Killing vector,
\begin{equation}\label{axequxirel}
    u^\alpha = e^{-\pot} \xi^\alpha\;,\quad u^\mu u^\nu g_{\mu\nu}=-1\;.
\end{equation}
Obviously,
\begin{equation}\label{axequdtrel}
    u_\alpha = -e^{\pot} t_{,\alpha}\;.
\end{equation}
The four-acceleration of these static observers,
\begin{equation}\label{axeqwdef}
     w^{\mu}=u^{\mu}_{\ ;\nu}u^{\nu}\;,
\end{equation}
is orthogonal to $u^\alpha$. Using the orthogonality, the Killing equation $\xi_{(\mu;\nu)}=0$, and staticity ${u^\alpha \pot_{,\alpha}=0}$, one finds that $U$ is a local potential of $w_\mu$
\begin{equation}\label{axeqwdurel}
    w_\alpha = \pot_{,\alpha}\;.
\end{equation}

We call $\pot$ the gravitational potential, $w^\alpha$ the acceleration field, and $-\pot_{,\alpha}$ the gravitational field strength. Since $w^\alpha$ is orthogonal to the time direction $u^\alpha$, it can be restricted to spatial components. The field strength is a fictitious force that one assumes in the static frame to explain a tendency of free observers to move with respect to the frame. Non-moving static observers have to `compensate' this force by a real force equal to $w_j$ per unit mass, cf.~\eqref{wdurel}. Of course, in the spacetime description, the static observers move along non-geodesic trajectories with four-acceleration $w^\alpha$ caused by the real force.

\subsection*{Weak static gravitational field}

Now we want to study a weak gravitational field on a flat background. We assume that the resulting spacetime is static, as described above, and we want to formulate the field equation in an approximation of a weak field. Since the metric \eqref{axeqstaticmetric} is invariant under time reflection ${t\to -t}$, the extrinsic curvature of a surface $t=\const$ vanishes, and Gauss-Codazzi equations imply that
\be \n{GGG}
G_t^t=-{1\over 2}{\cal R}\, .
\ee
Here, $G_{\mu\nu}$ is Einstein tensor of the spacetime metric $ds^2$ and ${\cal R}$ is a scalar curvature of the spatial metric $dq^2$ (see, e.g., (B.2.6) in \cite{FrolovZelnikov:2011}).

We choose a stress-energy tensor for a static distribution of matter in the form
\be\label{Tmn}
T^{\mu\nu}=\eps\, u^{\mu}u^{\nu}\;.
\ee
Here $\eps$ is the mass density. In the metric (\ref{axeqstaticmetric}) one has
\be\label{dm1}
dm=\eps\sqrt{q}\, d^3x\, .
\ee

Using Einstein equations and relation \eqref{GGG} one gets
\be\label{Rrhorel}
{\cal R}=16\pi \eps\, .
\ee

Let us consider a special case when the spatial geometry is conformally flat with a factor given by $e^{-2\pot}$
\be\label{dqconfflat}
dq^2=e^{-2\pot} dl^2\, ,
\ee
$dl^2=\delta_{ij} dx^i dx^j$ being a flat spatial metric. Then one has
\be \label{RRFF}
{\cal R}=2e^{2\pot}\bigl(2\lap\pot-(\nabla\pot)^2\bigr)\, .
\ee
Here, $\nabla$, $\lap$ and $(\dots)^2$ are the covariant derivative, the Laplace operator, and the square with respect to the flat spatial metric~$dl^2$.

We say that the gravitational field is weak if
\be
\abs{\nabla\pot \nabla\pot}\ll \abs{\nabla\nabla\pot}\, .
\ee
In such a case the second term in the brackets of (\ref{RRFF}) can be omitted and one has ${\cal R}=4e^{2\pot} \lap\pot$.
Equation (\ref{Rrhorel}) implies that in the weak field approximation the mass density $\varepsilon$ is of the same first-order as ${\cal R}$.

The second term in the bracket can be neglected in our approximation, but the prefactor is not negligible. From \eqref{Rrhorel} we obtain the equation for potential $\pot$
\be\label{LAPfulleps}
\lap\pot = 4\pi e^{-2\pot} \eps\, .
\ee

In fact, in the full theory, for a static distribution of the matter the stress-energy tensor should be slightly modified by adding a contribution responsible for repulsive forces keeping the matter at rest.
Indeed, for the stress-energy tensor \eqref{Tmn} one has
\begin{equation}
  T^{\mu\nu}{\!}_{;\nu}=\eps\, w^{\mu}\;,
\end{equation}
This means that a static distribution of matter is possible only when some repulsive forces are present that compensate the gravity attraction. This can be achieved by including the pressure $p$ and writing $T_{\mu\nu}$ in the form
\begin{equation}
T^{\mu\nu}=({\eps} +p)u^{\mu}u^{\nu}+p g^{\mu\nu}\, .
\end{equation}
For this tensor
\begin{equation}
T^{\mu\nu}{\!}_{;\nu}=({\eps}+p) w^{\mu}+p^{;\mu}\, .
\end{equation}

In the weak field approximation, both ${\eps}$ and $w^\alpha$ are small.  We say that each of this quantities is of the first order in the weak field approximation.
To satisfy the conservation law $T^{\mu\nu}{\!}_{;\nu}=0$,  the pressure should be of the second order. Such pressure does not contribute to the Einstein equations in the leading first-order approximation.

Let us discuss now the other Einstein equations (besides (\ref{RRFF})) in the same leading order approximation.
It is easy to check that all time-spatial components of the Einstein tensor $G_{tj}$ vanish identically, and the spatial components $G_{ij}$ are of the second order in $\nabla\pot$  (see, e.g., (B.2.6) in \cite{FrolovZelnikov:2011}).
This means that in the leading order of the weak field approximation  the spatial components of the Einstein equations are also satisfied.

When working in the flat space, it is natural to use a mass density in (\ref{dm1}) normalized on the flat volume element $dV=d^3x$. The relation to $\eps$ follows from
\be
dm = \mu\, dV \, .
\ee
Thanks to rescaling \eqref{dqconfflat} we obtain
\begin{equation}\label{muepsfullrel}
    \mu = e^{-3\pot} \eps\;.
\end{equation}
Then, equation \eqref{LAPfulleps} for the potential takes the form
\be\label{LAPfullmu}
\lap \pot=4\pi e^{\pot} \mu\, .
\ee

Let us note that   metric \eqref{axeqstaticmetric}, \eqref{dqconfflat} and the field equation \eqref{LAPfullmu} are  invariant under the following scaling transformation
\be \n{axeqUUCC}
\begin{gathered}
\pot=\hat{\pot}+\delU\; ,\\
t= e^{-\delU} \hat{t}\; ,\quad dl^2=e^{2\delU} d\hat{l}^2\;,\\
\mu = e^{-3\delU} \hat{\mu}\; .
\end{gathered}
\ee

\subsection*{Small value of potential $U$ case}

Let us emphasize, that in the above discussion we assume that the potential $U$ is slowly changing in space, but we do not assume that  its value is small.

To simplify the presentation we make an additional assumption that the potential $U$ is uniformly small, that is its value satisfies the relation
\begin{equation}\label{weakgrfield}
   \abs{\pot}\ll 1\, .
\end{equation}
everywhere in space.

In this approximation equation \eqref{LAPfullmu} takes the form
\be\n{axeqPoisson}
\lap \pot = 4\pi \mu\; .
\ee

In our perturbative approach it is the same as $\eps$ in the leading order
\begin{equation}\label{mueps}
    \mu = \eps\;,
\end{equation}
but for $\pot$ that are not small, they differ by the factor $e^{-3\pot}$, cf.~\eqref{muepsfullrel}..

\section{Locally static spacetimes}\label{apx:locstaticst}

In this appendix we demonstrate how to describe a locally static spacetime without introducing initially a static time coordinate ${t}$ and a gravitational potential ${\pot}$ as we did in Eq.~\eqref{axeqstaticmetric}  in the previous Appendix.

Let us consider a  spacetime $M$ with metric $\ts{g}$ and assume that it admits two vector fields, a future directed unit vector $\ts{u}$ and a vector $\ts{w}$ which obey the equations
\bea \n{uw}
& u_{\mu}u^{\mu}=-1\, ,\nonumber\\
&u_{\mu ;\nu}=-w_{\mu} u_{\nu}\, ,\\
&w_{[\mu ;\nu]}=0\, .\nonumber
\eea
We call such a spacetime locally static. The  relations (\ref{uw})  imply that
\be\n{uwa}
w^{\mu}=u^{\nu} u^{\mu}_{\ ;\nu}\hh w_{\mu}u^{\mu}=0\, .
\ee
Denote by $\gamma$ integral lines of $\ts{u}$
\be
{dx^{\mu}\over d\tau}=u^{\mu}\, .
\ee
Normalization condition $\ts{u}^2=-1$ implies that $\tau$ is a proper time along a worldline $\gamma$.

The last of the relations in (\ref{uw}) implies that the 1-form $w_{\mu}dx^{\mu}$ is closed. Let $p_0$ be a point in $M$ and $\Omega_{p_0}$ be a simply connected region which contains this point. Then there exist such a function $U$ in $\Omega_{p_0}$ that
\be
w_{\mu}=U_{,\mu}\, .
\ee
Let $p$ be a point in $\Omega_{p_0}$ and $C_{p_0 p}$ be a path in  $\Omega_{p_0}$ connecting points $p_0$ and $p$.
Then
\be
U(p)=U(p_0)+\int_{C_{p_0 p}} w_{\mu}dx^{\mu}\, .
\ee
The function $U$ called the gravitational potential is defined in $\Omega_{p_0}$ up to a constant $U_0\equiv U({p_0})$.
Let us denote
\be
z^{\mu}=e^{\epsilon U} u^{\mu}\hh \epsilon=\pm 1\, .
\ee
Using relations (\ref{uw}) one gets
\be
z_{\mu;\nu}=e^{\epsilon U}(-u_{\nu}w_{\mu}+\epsilon u_{\mu}w_{\nu})\, .
\ee
For $\epsilon=-1$ one has $z_{[\mu;\nu]}=0$, while for $\epsilon=1$ one has $z_{(\mu;\nu)}=0$.
We denote
\be
\xi^{\mu}=e^{ U} u^{\mu}\hh \eta^{\mu}=-e^{- U} u^{\mu}\, .
\ee
Thus one has
\be \n{xeU}
\xi_{(\mu;\nu)}=0\hh \eta_{[\mu;\nu]}=0\hh
\ts{\xi}^2=-e^{2U} \, .
\ee
The first of these equations shows that $\ts{\xi}$ is a Killing vector,  while the second equations implies that there exists such a scalar function $t$ in $\Omega_{p_0}$ that
\be
\eta_{\mu}=t_{,\mu}\, .
\ee
It is also easy to check that
\be
\xi^{\mu}\pa_{\mu}U=0\, .
\ee

Consider displacement $dx^{\mu}$ along a worldline $\gamma$. Then one has
\ba\n{ttau}
dt&=\eta_{\mu}dx^{\mu}=-e^{-U} u_{\mu}dx^{\mu}\\
&=-e^{-U} u_{\mu}u^{\mu}d\tau=e^{-U}d\tau\, .
\ea
One also has
\be
{dx^{\mu}\over dt}=e^{U}{dx^{\mu}\over d\tau}=e^{U}u^{\mu}=\xi^{\mu}\, .
\ee
Hence
\be
\xi^{\mu}\pa_{\mu}=\pa_t\, ,
\ee
and $t$ is the Killing time parameter. Let us emphasize that the gravitational potential $U$ in $\Omega_{p_0}$ is defined up to a constant.  A shift $U\to U+c$ results in the rescaling both the Killing vector and Killing time
\be
\xi^{\mu}\to e^{-c}\xi^{\mu}\hh t\to e^{c} t\, .
\ee

The integral lines $\gamma$ provide a foliation  of the region $\Omega_{p_0}$. Each of the Killing trajectory can be specified by giving 3 numbers $y^i$, $i=1,2,3$, and a point $p$ on a given trajectory can be specified by giving its Killing time value $t$. This defines coordinates $x^{\mu}=(t,x^i)$ in $\Omega_{p_0}$. In these coordinates the metric takes the form
\be
ds^2=-e^{-2U} dt^2+h_{ij}dy^i dy^j\hh \pa_t U=\pa_t h_{ij}=0\, .
\ee

In what follows we assume that $\ts{\xi}^2$ does not vanish and remains negative everywhere in $M$. This property excludes a case of black holes. But in the spacetime considered in the present paper this property is valid.

\begin{figure}
    \centering
      \includegraphics[width=0.3\textwidth]{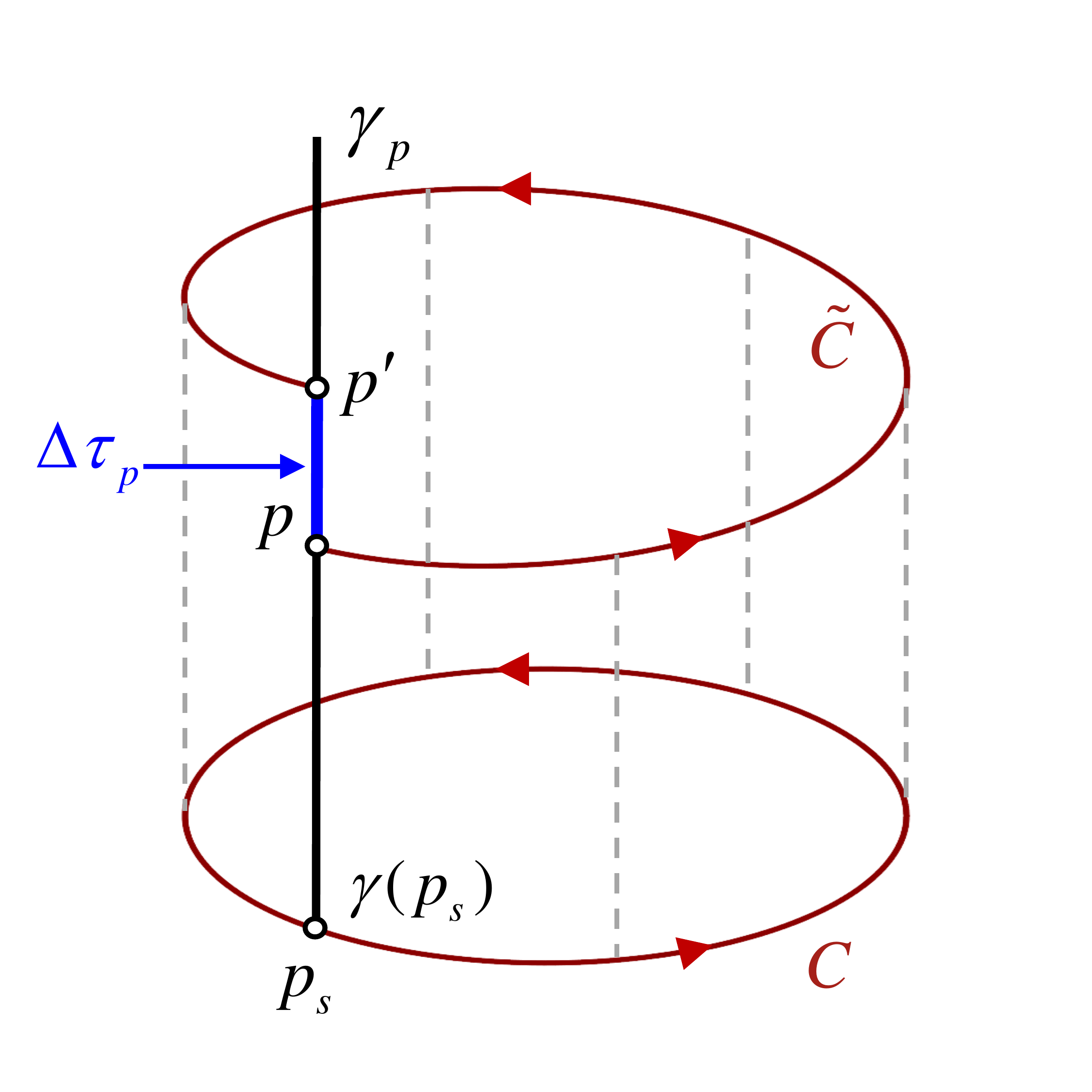} 
    \caption{Projection of $M$ onto the space of Killing trajectories $S$.}
    \label{CURVE_S}
\end{figure}

Following \cite{Geroch:1970nt} we denote by $S$ a collection of all trajectories of $\ts{\xi}$ in $M$. There exists a natural map $\Psi$ from $M$ to $S$ which is defined as follows: For any point in $M$ there exists a Killing trajectory passing through it which determines a point in $S$ (see Fig.~\ref{CURVE_S}). Three-dimensional space $S$ has the induced metric
\be
dh^2=h_{ij}dy^i dy^j\, .
\ee

Since ${\cal L}_{\xi}\ts{w}=0$ and $\ts{\xi}\cdot\ts{w}=0$, the 4-vector of acceleration $\ts{w}$ has a natural projection $\vec{w}$ on $S$. In $(t,y^i)$ coordinates one has $\ts{w}=(0,\vec{w})$. Let us consider a path $C^S_{p_0 p}$ connecting two points $p_{S 0}$ and $p_S$ in $S$. We define a 3D potential $U(p_S)$ as follows
\be\n{UUSS}
U(p_S)=U_0+\int_{C^S_{p_0 p}} w_i dy^i\, ,
\ee
where $U_0$ is a constant which is chosen as a value of the potential at $p_{S 0}$.
The quantity $U(p_S)-U_0$ can be interpreted as a work done by the gravitational field on a particle of unit mass for its motion along the path $C^S$.

Consider a closed path $C^S$ which starts at some point $p_0$ and returns to it again and denote
\be
I_{C^S}=\oint_{C^S} w_i dy^i\, .
\ee
Since 3D 1-form $w$ is closed, the Stokes' theorem implies that $I_{C^S}$ depends only on the  cohomology class of the path $C^S$.
If $S$ is a simply connected manifold the invariant $I_{C^S}$ vanishes. If $S$ is not simply connected there may exist paths
for which $I_{C^S}\ne 0$. For such a closed path work done by the gravitational field does not vanish. We call such a field non-potential. For this field the potential $U$ is a multi-valued function.

A path $C^S$  connecting $p_S$ and $p'_S$ in $S$ can be lifted to $M$ as follows. Consider a Killing trajectory corresponding to $p_S$ and choose a point $p$ on it. If in local coordinates $y^i$ the equation of $C^S$ is $y^i=y^i(\lambda)$, then the equation of the path $S$ in $(t,y^i)$ coordinates in $M$  is $x^{\mu}=(t_p,y^i(\lambda))$, where $t_p$ is the time coordinate of  point $p$.
We require that the path $C$ has a property that its tangent vector is orthogonal to $\ts{u^{\mu}}$. For a given initial point $p$ and chosen path $C^S$ in $S$ this property uniquely define $C$. It is easy to see that
\be
\int_{C^S} w_i dy^i=\int_C w_{\mu}dx^{\mu}\, .
\ee

Let $C^S$ be a closed path in multiply-connected $S$ passing through a point $p_S$ and $\tilde{C}$ be its lift to $M$ passing through $p$. The end point $p'$ of $\tilde{C}$ lies on the same Killing trajectory as the initial point $p$ but in a general case it does not coincide with it. Denote by $\Delta C$ a path from $p'$ to $p$ along their common Killing trajectory. The path $C=\tilde{C}+\Delta C$ is closed. Since $\ts{w}$ is orthogonal to a Killing trajectory, $\int_{\Delta C} w_{\mu}dx^{\mu}=0$, and one has
\be\n{CCSW}
I_C=\oint_{C^S} w_i dy^i=\oint_C w_{\mu}dx^{\mu}\, .
\ee
The value of this integral does not depend on a choice of the initial point $p$ on the path, and  relation (\ref{CCSW}) is valid not only for a chosen lift of the path $C^S$ but for any closed path $C$ in $M$ such that its projection on $S$ coincides with $C^S$.

Let us consider a closed path $C^S$ with a starting point $p_S$ for which $I_C=0$ and chose two close points $p$ and $\tilde{p}$ on the Killing trajectory determined by $p_S$. They have the same  coordinates $y^i$. Let us denote by $d t$ time difference between  $p$ and $\tilde{p}$ and assume that $d t$ is small. By construction, $d t$ remains the same for the paths $C$ and $\tilde{C}$ with the initial points $\tilde{p}$ and $p$, respectively. In particular this is true for the end points $p'$ and $\tilde{p}'$ of these curves. Since $I_C=0$, the value of the potential $U$ defined along the paths by equation (\ref{UUSS}) at the end point coincides with its value at the initial point, one concludes that the initial proper time interval $d\tau$ between $p$ and $\tilde{p}$ is the same as the final proper time interval $d\tau'$ between $p'$ and $\tilde{p}'$. We call the proper time interval $\Delta\tau_p$ between the initial point $p$ of $C$ and its final point $p'$ the proper time gap (see figure \ref{CURVE_S}). The above discussion implies that the proper time gap is constant for closed paths for which the invariant $I_C$ vanishes.

If a locally static spacetime $M$ is simply connected so that $I_C=0$ for any closed path in it, then
\begin{enumerate}[label=(\roman*)]
\item gravitational potential is globally uniquely defined, up to a constant;
\item gravitational field is potential;
\item Killing vector field is globally uniquely defined after choice of its norm in some point.
\end{enumerate}
In other words, this spacetime is (globally) static.

\begin{figure}
    \centering
      \includegraphics[width=0.25\textwidth]{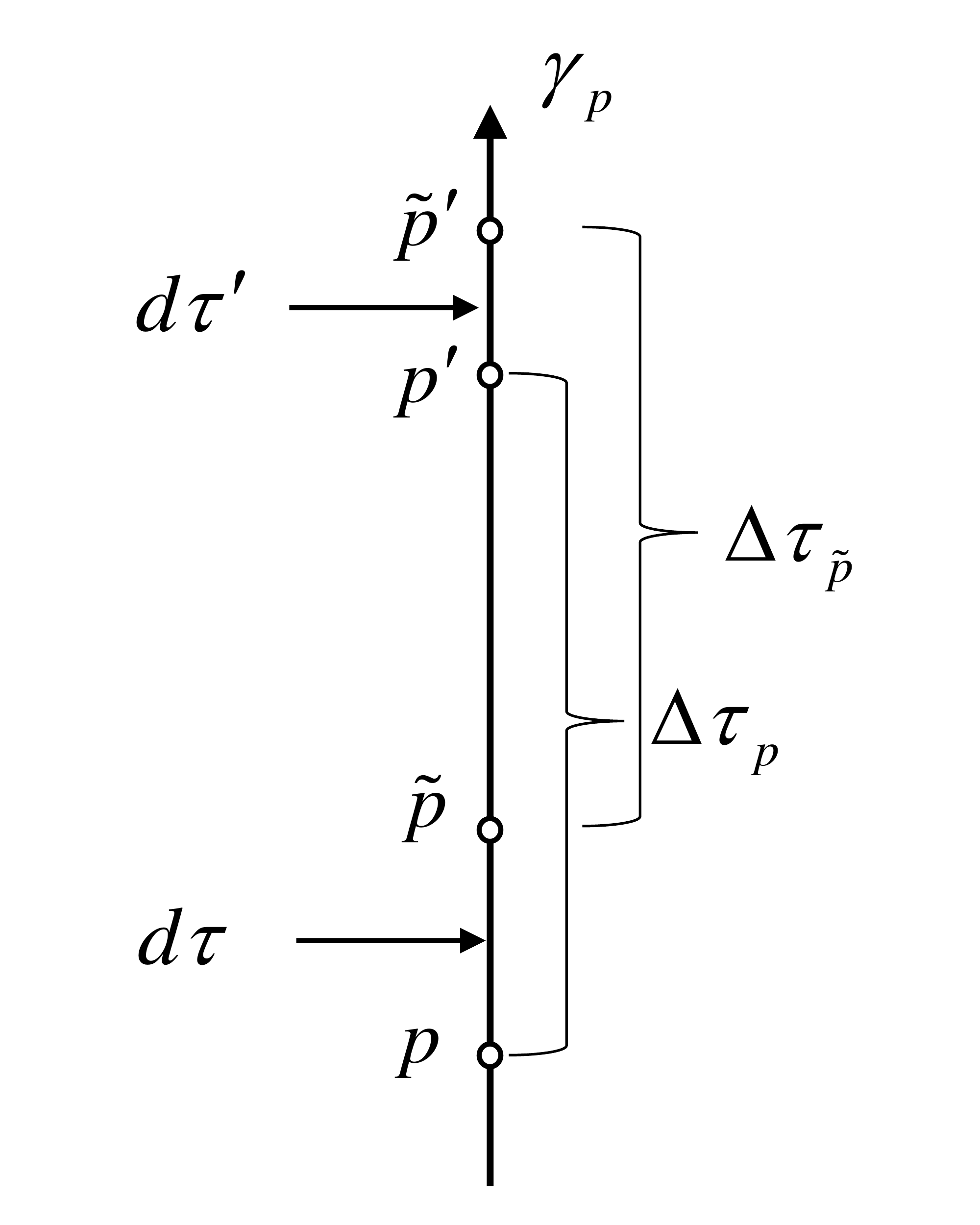}
    \caption{Time gaps.}
    \label{PROP}
\end{figure}

Let us now repeat the above consideration, but do not assume that the invariant $I_C$ for a chosen closed path vanishes. Once again the Killing time interval $d t$ for two close lifts of $C^S$ with starting points at  $p$ and $\tilde{p}$ remains constant along the path. Denote by $d\tau$  proper time interval between initial points $p$ and $\tilde{p}$, and by $d\tau'$ proper time interval between final points $p'$ and $\tilde{p}'$ (see figure~\ref{PROP}).
Then using (\ref{ttau}) one gets
\be
d\tau=e^{U(p)} dt\hh
d\tau'=e^{U(p')} d t\, .
\ee
Thus
\be
d\tau'=e^{U(p')-U(p)}d\tau=\exp{(I_C)} d\tau\, .
\, .
\ee
Denote by $\Delta\tau_p$ the proper time gap for the point $p$ and by $\Delta\tau_{\tilde{p}}$ the proper time gap for $\tilde{p}$. Then one has (see figure \ref{PROP})
\be
d\Delta\tau\equiv \Delta\tau_{\tilde{p}}-\Delta\tau_p=d\tau'-d\tau=
(\exp{(I_C)} -1) d\tau
\, .
\ee
Hence, the proper time gap $d\Delta\tau$ linearly grows with time $\tau$ so that
\be
{d\Delta\tau\over d\tau}=\exp{(I_C)} -1\, .
\ee
For more details see \cite{Frolov:1990si,Frolov:1998wf,FrolovZelnikov:2011}.





%

\end{document}